\documentclass[pdflatex, sn-basic]{sn-jnl}

\usepackage{graphicx}%
\usepackage{multirow}%
\usepackage{amsmath,amssymb,amsfonts}%
\usepackage{amsthm}%
\usepackage{mathrsfs}%
\usepackage[title]{appendix}%
\usepackage{xcolor}%
\usepackage{textcomp}%
\usepackage{manyfoot}%
\usepackage{booktabs}%
\usepackage{algorithm}%
\usepackage{algorithmicx}%
\usepackage{algpseudocode}%
\usepackage{listings}%
\usepackage{gensymb}
\usepackage{pdflscape}

\usepackage{aas_macros}

\begin{document}

\title[The CCOR Compact Coronagraphs for the Geostationary Operational Environmental Satellite-19 (GOES-19) and the Space Weather Follow On (SWFO) Missions]{The CCOR Compact Coronagraphs for the Geostationary Operational Environmental Satellite-19 (GOES-19) and the Space Weather Follow On (SWFO) Missions}


\author*[1]{\fnm{Thernisien} \sur{Arnaud F.}}\email{arnaud.f.thernisien.civ@us.navy.mil}

\author*[1]{\fnm{Chua} \sur{Damien H.}}\email{damien.h.chua.civ@us.navy.mil}

\author[1]{\fnm{Carter} \sur{Michael T.}} 
\author[1]{\fnm{Rich} \sur{Nathan B.}}

\author[1]{\fnm{Noya} \sur{Mario}}

\author[1]{\fnm{Babich} \sur{Timothy A.}}

\author[1]{\fnm{Crippa} \sur{Cameron E.}}

\author[2]{\fnm{Baugh} \sur{Becky}} 
\author[18]{\fnm{Bordlemay} \sur{Yadira}}
\author[1]{\fnm{Socker} \sur{Dennis}}
\author[12]{\fnm{Biesecker} \sur{Douglas}}
\author[2]{\fnm{Korendyke} \sur{Clarence}}
\author[2]{\fnm{Wang} \sur{Dennis}}
\author[17]{\fnm{Wang} \sur{Nai-Yu}}
\author[17]{\fnm{Vassiliadis} \sur{Dimitrios}}
\author[1]{\fnm{Abbay} \sur{Samuel}}
\author[1]{\fnm{Bagnall} \sur{Sean}}
\author[7]{\fnm{Balmaceda} \sur{Laura}}
\author[3]{\fnm{Brown} \sur{Samuel}}
\author[4]{\fnm{Bonafede} \sur{Joseph}}
\author[9]{\fnm{Boyer} \sur{Darrell}}
\author[21]{\fnm{Declet} \sur{John}}
\author[9]{\fnm{Cheng} \sur{Peter}}
\author[4]{\fnm{Corsi} \sur{Keith}}
\author[9]{\fnm{Cremerius} \sur{Luke}}
\author[4]{\fnm{Chavis} \sur{Isabella}}
\author[2]{\fnm{Chiralo} \sur{Joel}}
\author[10]{\fnm{Clifford} \sur{Greg}}
\author[4]{\fnm{Dancheck} \sur{John}}
\author[1]{\fnm{Davis} \sur{Joseph}}
\author[13, 14]{\fnm{Dima} \sur{Gabriel}}
\author[20]{\fnm{Dudley} \sur{Renee}}
\author[9]{\fnm{Gardner} \sur{Don}}
\author[2]{\fnm{Gardner} \sur{Larry}}
\author[4]{\fnm{Hagood} \sur{Britta}}
\author[4]{\fnm{Hagood} \sur{Robert}}
\author[4]{\fnm{Hohl} \sur{Bruce}}
\author[2]{\fnm{Hunt} \sur{Tonia}}
\author[11]{\fnm{Jenkins} \sur{Frank}}
\author[13, 15]{\fnm{Johnson} \sur{Jeffrey M.}}
\author[3]{\fnm{Koehler} \sur{Matt}}
\author[7]{\fnm{Kuroda} \sur{Natsuha}}
\author[2]{\fnm{Lanagan} \sur{Andrew}}
\author[2]{\fnm{Laut} \sur{Sophie}}
\author[4]{\fnm{Lynch} \sur{Brian}}
\author[8]{\fnm{Mallory} \sur{Tyrell}}
\author[4]{\fnm{Mechel} \sur{David}}
\author[13, 15]{\fnm{Miles} \sur{Nathan D.}}
\author[4]{\fnm{Miranda} \sur{Anthony}}
\author[8]{\fnm{Newman} \sur{Miles}}
\author[1]{\fnm{Nguyen} \sur{Bang}}
\author[4]{\fnm{Ogindo} \sur{Moses}}
\author[4]{\fnm{Pellak} \sur{Ken}}
\author[6]{\fnm{Podgurski} \sur{Robert}}
\author[16]{\fnm{Ragan} \sur{Tai}}
\author[4]{\fnm{Richards} \sur{Victoria}}
\author[10]{\fnm{Silver} \sur{Dennis}}
\author[2]{\fnm{Simmons} \sur{Jeff}}
\author[13, 14]{\fnm{Donald} \sur{Schmit J.}}
\author[4]{\fnm{Smith} \sur{Linda}}
\author[5]{\fnm{Spitzak} \sur{John}}
\author[19]{\fnm{Tadikonda} \sur{Sivakumara K.}}
\author[4]{\fnm{Tanner} \sur{Steve}}
\author[4]{\fnm{Uhl} \sur{Drew}}
\author[1]{\fnm{Verzosa} \sur{Julius}}
\author[5]{\fnm{Walker} \sur{Peter}}
\author[4]{\fnm{Wiggins} \sur{Grayson}}
\author[6]{\fnm{Williams} \sur{Eric}}
\author[4]{\fnm{Wilson} \sur{Courtni}}
\author[4]{\fnm{Zurcher} \sur{Dallas}}

\affil*[1]{\orgname{U.S. Naval Research Laboratory}, \orgaddress{\street{4555 Overlook Ave, SW}, \city{Washington}, \state{DC}, \postcode{20375}, \country{USA}}}

\affil[2]{\orgname{Space Systems Research Corporation} \orgaddress{\street{2000 Duke Street, Suite 300} \city{Alexandria}, \state{VA}, \postcode{22314}, \country{USA}}}

\affil[18]{\orgname{TSC} \orgaddress{\street{251 18th Street South, Suite 705}, \city{Arlington}, \state{VA}, \postcode{22202}, \country{USA}}}

\affil[4]{\orgname{KBR} \orgaddress{\street{8120 Maple Lawn Blvd}, \city{Fulton}, \state{MD}, \postcode{20759}, \country{USA}}}

\affil[5]{\orgname{Computational Physics Inc.} \orgaddress{\street{8001 Braddock Road, Suite 210}, \city{Springfield}, \state{VA}, \postcode{22151}, \country{USA}}}

\affil[6]{\orgname{National Air and Space Administration/Goddard Space Flight Center}, \orgdiv{Optics Branch, Code 551}, \orgaddress{\city{Greenbelt}, \state{MD}, \postcode{20771}, \country{USA}}}

\affil[7]{\orgname{George Mason University} \orgaddress{\street{4400 University Drive}, \city{Fairfax}, \state{VA}, \postcode{22030}, \country{USA}}}

\affil[8]{\orgname{Peraton}, \orgaddress{\street{1875 Explorer Street}, \city{Reston}, \state{VA}, \postcode{20190}, \country{USA}}}

\affil[9]{\orgname{Software Control Solutions, LLC}, \orgaddress{\street{1962 Dairy Rd}, \city{Melbourne}, \state{FL}, \postcode{32904}, \country{USA}}}

\affil[10]{\orgname{Silver Engineering Inc.}, \orgaddress{\street{255 East Dr}, \city{Melbourne} \postcode{32904}, \state{FL}, \country{USA}}}

\affil[11]{\orgname{Aerotemp Consulting},
\orgaddress{\street{11611 Whittier Rd}, 
\city{Bowie}, \state{MD}, \postcode{20721}, \country{USA}
}}

\affil[12]{\orgname{McCallie Associates}, \orgaddress{\city{Boulder}, \state{CO}, \country{USA}}}

\affil[13]{\orgname{Cooperative Institute for Research in Environmental Sciences, University of Colorado}, \orgaddress{\city{Boulder}, \state{CO}, \country{USA}}}

\affil[14]{\orgname{National Centers for Environmental Information, National Oceanographic and Atmospheric Administration}, \orgaddress{\city{Boulder}, \state{CO}, \country{USA}}}

\affil[15]{\orgname{Space Weather Prediction Center, National Oceanographic and Atmospheric
Administration}, \orgaddress{\city{Boulder}, \state{CO}, \country{USA}}}

\affil[16]{\orgname{Research Support Instruments, Inc.}, \orgaddress{\city{Pasadena}, \state{MD}, \country{USA}}}

\affil[17]{\orgname{National Oceanographic and Atmospheric
Administration NESDIS Office of Space Weather Observations (SWO)}, \orgaddress{\city{Lanham}, \postcode{20706} \state{MD}, \country{USA}}}

\affil[18]{\orgname{National Aeronautics and Space Administration, Earth Science Division}, \orgaddress{\city{Washington}, \state{DC}, \country{USA}}}

\affil[19]{\orgname{Columbus Technologies and Services, Inc.}, \orgaddress{\city{Greenbelt}, \state{MD}, \country{USA}}}

\affil[20]{\orgname{Aerospace Corporation}, \orgaddress{\street{14745 Lee Road}, \city{Chantilly}, \state{VA}, \postcode{20151}, \country{USA}}}

\affil[21]{\orgname{ASRC Federal System Solutions}, \orgaddress{\city{Reston}, \state{VA}, \postcode{20190}, \country{USA}}}

\abstract{The CCOR Compact Coronagraph is a series of two operational solar coronagraphs sponsored by the National Oceanic and Atmospheric Administration (NOAA). They were designed, built, and tested by the U.S. Naval Research Laboratory (NRL). The CCORs will be used by NOAA's Space Weather Prediction Center (SWPC) to detect and track Coronal Mass Ejections (CMEs) and predict space weather at Earth and everywhere in the solar system. CCOR-1 is on board the Geostationary Operational Environmental Satellite -U (GOES-U, now GOES-19/GOES-East). GOES-U was launched from Kennedy Space Flight Center (KSFC), Florida, on 25 June 2024. CCOR-2 is on board the Space Weather Follow On at Lagrange point 1 (SWFO-L1). SWFO-L1 is scheduled to launch in the fall of 2025. SWFO will be renamed SOLAR-1 once it reaches L1.

The CCORs are white-light coronagraphs that have a field of view and performance similar to the SOHO LASCO C3 coronagraph. CCOR-1 FOV spans from 4 to 22 Rsun, while CCOR-2 spans from 3.5 to 26 Rsun. The spatial resolution is 39 arcsec for CCOR-1 and 65 arcsec for CCOR-2. They both operate in a band-pass of approximately 470 - 740 nm. Polarizers are not used. They use the same 2048 x 1920 pixels Active Pixel Sensor (APS) as the Wide-Field Imager for Parker Solar Probe (WISPR) and Solar Orbiter Heliospheric Imager (SoloHI). The synoptic cadence is 15 min and the latency from image capture to the forecaster on the ground is less than 30 min.

Compared to past generation coronagraphs such as the Large Angle and Spectrometric Coronagraph (LASCO), CCOR uses a compact design; all the solar occultation is done with a single multi-disk external occulter. No internal occulter is used. This allowed a substantial reduction in size and mass compared to SECCHI COR-2, for example, but with slightly lower signal-to-noise ratio.

In this article, we review the science that the CCORs will capitalize on for the purpose of operational space weather prediction and how they fit in the collection of sensors used for that purpose. We give a description of the driving requirements and accommodations, and then we provide some details on the instrument design. In the end, information on ground processing and data levels is also provided.  }


\maketitle

\section{Introduction}\label{sec1}
The CCOR Compact Coronagraph is a series of two operational solar coronagraphs for the National Oceanic and Atmospheric Administration (NOAA). The compact coronagraph is a new class of coronagraphs designed, built, and tested by the U.S. Naval Research Laboratory (NRL). The CCOR series are the first coronagraphs designed for operational use by NOAA's Space Weather Prediction Center (SWPC) to detect and track Coronal Mass Ejections (CMEs) to predict space weather at Earth. CCOR-1 coronagraph was launched on board the Geostationary Operational Environmental Satellite version U (GOES-U, now GOES-19/GOES-East) from Kennedy Space Flight Center (KSFC), Florida, on 25 June 2024. CCOR-2 is on board the Space Weather Follow On at Lagrange point 1 (SWFO-L1) scheduled to launch in the fall of 2025. Note that SWFO will be renamed SOLAR-1 once t reaches L1 and becomes operational.

NOAA National Environmental Satellite, Data, and Information Service (NESDIS) Office of Space Weather Observations (SWO) develops, deploys, and supports NOAA operational satellite systems that study space weather and safeguard society. SWO provides real-time CCOR data to NOAA National Weather Service's Space Weather Prediction Center (SWPC), which in turn generate space weather products and services. SWPC is the United States' authority for monitoring solar activity and providing real-time space weather forecasts. The SWPC forecasters monitor solar activity and provide forecasts and warnings to, among others, power grid operators, satellite operators, and airline companies and make them available to the general public. Observations from coronagraphs are cornerstones of the data that SWPC uses to forecast space weather and geomagnetic storms. Coronagraph observations are used to monitor and track the primary drivers of geomagnetic storms, coronal mass ejections (CMEs). These transients can be traced from the surface of the Sun out to the solar corona, and all the way to Earth in the same way a weather satellite would track the formation, strengthening, and projected landfall of a terrestrial hurricane. The direction, mass, and speed of the CMEs obtained from the coronagraph data are fed into solar wind models such as the Wang-Sheeley-Arge (WSA-Enlil) \citep{Pizzo_WSA_2011SpWea...9.3004P} which allow one to predict the time of arrival at Earth and the strength of the storm.

CMEs are sudden releases of plasma and magnetic field from the Sun that can travel through the heliosphere and disrupt Earth's magnetic field causing a geomagnetic storm. CME driven magnetic storms can have a significant impact on human assets on Earth and in space, and affect a broad range of industries. The main effects include induced currents on the power grid, GPS errors, communication outages, increased satellite drag, and radiation concerns for astronauts, airline crew, and satellites. An extreme example is known as the Carrington event of 1859, which was strong enough to induce direct current on telegraph lines, causing damaging effects on the primary communications system of the time. Another event in 1989, driven by two consecutive flares and CMEs caused a widespread blackout of the Hydro-Quebec power system \citep{Boteler_2019SpWea..17.1427B}. Today, with our increased reliance on technological systems, such an event could cause billions of dollars in damage to satellites and the power grid, with obvious impacts on human lives \citep{Impact_of_Space_Weather_NOAA}. Forecasting these disturbances is critical to take actions to protect assets and lives, and react accordingly if critical systems such as communication or Global Positioning Systems (GPS) are degraded.

The CCORs are the first operational coronagraphs designed to provide observations for space weather forecasting at NOAA. The necessity of coronagraph observations for space weather forecasting has been demonstrated by the observations from Large Angle and Spectrometric Coronagraph (LASCO; \citealp{Brueckner_1995}) on board the Solar and Heliospheric Observatory (SOHO; \citealp{Domingo_1995SoPh..162....1D}). LASCO is a package of three coronagraphs, C1, C2, and C3. The FOV of each coronagraph are nested like an onion: C1 1.1 to 3 R$_\odot$, C2 2.2 to 6 R$_\odot$, C3 3.6 to 30 R$_\odot$. Since the launch of the SOHO mission in December 1995, LASCO has been observing the corona almost continuously, except during a 4-months period in 1998, when SOHO was lost and then recovered. Unfortunately, C1 suffered permanent damage during that interruption and was never recovered. However, C2 and C3 have been observing the corona since then, a period of almost 30 years at the time this article was written. LASCO was designed for scientific research of the solar corona and has become central for geomagnetic storm forecasting. Among contributions, statistical studies with LASCO observations have allowed better modeling and prediction of CME arrival time at Earth \citep{Gopal_2001JGR...10629207G}. Advances from LASCO and the entire SOHO mission in understanding CMEs and their contribution to space weather in general would be too long to summarize here. Please refer to \cite{Howard_2023FrASS..1064226H} and the references therein.

Coronagraphs use the effects of Thomson scattering \citep{Billings_1966} of sunlight by free electrons to image the plasma of the solar corona. For a comprehensive history of coronagraphs and evolution of understanding of CMEs we invite the reader to refer to \cite{Howard_2015}, and \cite{Howard_2023FrASS..1064226H}; we only recall some major milestones. The white-light coronagraph was invented as a ground-based telescope by Lyot in 1936 \citep{Lyot_1933}. The Lyot coronagraph was internally occulted; the direct sunlight is blocked after the first optical element, and the Field of View (FOV) extends only a fraction of a solar radius above the solar limb. \cite{Evans_1948JOSA...38.1083E} improved the design by adding an external occulter, prior to the first optical element. Because of the air scattering, ground-based observations were limited to a few solar radii above the solar limb. In the 1960s, the first balloon-borne and sounding rocket-born coronagraphs were developed, which allowed for an extended view of the corona, compared to ground observation, although only for a few minutes at a time. In 1971, the Orbital Solar Observatory 7 (OSO-7) was launched with a white-light coronagraph \citep{Koomen_1975} and provided day-to-day observations of the corona. Coronal Mass Ejections (CMEs) were first detected in OSO-7 coronagraph images \citep{Tousey_1973}, although their existence was suspected before then \citep{2006GMS...165....7H_Howard}. In the next two decades, successive missions included coronagraphs: Skylab (1973-74; \citealp{MacQueen_1974}), Solwind/P-78 (1979-85; \citealp{Michels_1980}), Solar Maximum Mission (SMM, 1980-1989) \citep{MacQueen_1980}. They allowed for almost daily continuous observation of the corona and allowed the first statistical and morphological studies of CMEs \citep{Howard_1985JGR....90.8173H}. In particular, the first halo CME was observed by Solwind on 27 November 1979 \citep{Howard_1982ApJ...263L.101H}. The geometrical effects due to the direction of the CMEs related to their initiation region in the lower corona and the position of the observer were studied and begun to be understood \citep{Hundhausen_1993JGR....9813177H}. Note also that in April 1993, the SPARTAN 201-01 payload, which had a white-light channel, \citep{spartan_1994SSRv...70..267F} was launched into orbit from NASA's Discovery Space Shuttle OV-103.

In 2006, the Solar TErrestrial RElations Observatory (STEREO) mission was launched, with a suite of two coronagraphs on board, Sun-Earth Connection Coronal and Heliospheric Investigation (SECCHI) COR-1 and COR-2 \citep{Howard_2008}. The two STEREO spacecraft, STEREO-A and STEREO-B, have a SECCHI suite of instruments on board each of them, allowing the corona to be observed from two different vantage points and permitting the localization by triangulation of solar wind transients \citep{Thernisien_2009SoPh..256..111T}. STEREO-B was lost in 2014, but STEREO-A is still running. Heliospheric Imagers (HI) are also part of the SECCHI suite. HIs are cousins of the coronagraph. They are not pointed straight at the Sun center but look on the side, imaging the corona between the Sun and the Earth, in the case of STEREO. Co-rotating Interaction Regions (CIRs) appear at the interface between the slow solar wind (streamer) and the fast solar wind (coronal hole) \citep{Pizzo_1978JGR....83.5563P}. CIRs can be observed in HIs \citep{Wood_2010ApJ...708L..89W}. The transition between slow and fast solar wind, due to the presence of a coronal hole at low latitude, can induce geomagnetic effects/disturbances of the Earth's magnetosphere.

In an attempt to make an inventory of all the current active coronagraphs, we list the following.
\begin{itemize}
    \item Metis, on board Solar Orbiter \citep{METIS_2020A&A...642A..10A}, launched on 10 February 2020.
    
    \item The Lyman-alpha Solar Telescope - Solar Corona Imager (LST-SCI) \citep{Li_ASO_2019RAA....19..158L, Feng_ASO_2019RAA....19..162F} on board the Advanced Space-based Solar Observatory (ASO-S) \citep{Gan_2023SoPh..298...68G}, launched on 9 October 2022.

    \item The Visible Emission Line Coronagraph (VELC) \citep{velc_2017CSci..113..613R, velc_2019AdSpR..64.1455S} on board the Aditya-L1 \citep{seetha_aditya_2017CSci..113..610S}, launched on 26 August 2023.
    
    \item Coronal Diagnostic Experiment (CODEX), on board the International Space Station \citep{codex_2024SPIE13092E..7OC}, launched on 4 November 2024.
  
    \item Association of Spacecraft for Polarimetric and Imaging Investigation of the Corona of the Sun (ASPIICS), a formation flying coronagraph \citep{Lamy_ASPIICS_2017SPIE10565E..0TL, Zhukov_2020AGUFMSH031..02Z, shestov_aspiics_2021A&A...652A...4S}, launched on 5 December 2024.

    \item Polarimeter to UNify the Corona and Heliosphere (PUNCH) Narrow Field Imager (NFI) \citep{DeForest_2024, Colaninno_2025SoPh..300..104C}, launched on 11 March 2025.

\end{itemize}

At this time, NOAA's Space Weather Prediction Center relies on LASCO and STEREO SECCHI COR-2 data to monitor the CME activity. Although these instruments were designed as science instruments, they fulfill SWPC needs: they provide an almost uninterrupted stream of images of the corona, from $> 2$Rsun to 30Rsun, at a regular cadence. The CCOR design requirements were based on these two coronagraphs.

From a design point of view, the coronagraphs looking beyond 2Rsun flown in the past were all based on the so-called two-stage Lyot design. The first stage has an external occulter that casts a shadow on the entrance pupil of the telescope. For the second stage, an objective lens creates the image of the external occulter on an internal occulter that apodizes the diffraction. Then a relay lens and a field lens create the image of the corona on the detector. The Lyot design and the externally occulted coronagraph design are extensively discussed in \cite{1988SSRv...47...95K_koutchmy}. The LASCO design is discussed in \cite{Brueckner_1995}. The CCOR design is a slight departure from the Lyot design; the sunlight occultation is done solely by a single multidisk external occulter. In the optical train, there are no internal occulter or relay-lenses. This reduces the length of the instrument roughly by two, saving a good deal of mass and volume. The challenge of this design is to obtain stray light performances similar to those of a traditional two-stage Lyot design. 

In this paper, we start with science objectives and requirements as well as spacecraft orbits and accommodations. From these driving requirements, we describe the design of CCOR, including details specific to CCOR-1 and CCOR-2. Optical specifications are discussed as well as optomechanical design and stray light performances. The specificity of the earthshine, or earthlight stray light on CCOR-1 is addressed. Details on thermal, mechanical, electrical, flight, and ground software are also addressed. Finally, information on the data and products is provided before the conclusion.

\section{Science Objectives and Requirements}\label{sec:scienceandrequirements}
The basic science objective is to detect CMEs, measure their speed, acceleration, direction, and mass. The detection must be made early, as close as possible to the Sun, with a low data latency, and the instrument must remain robust to the effects of Solar Energetic Particles (SEPs) and the space radiation environment. The CMEs are to be modeled by the Space Weather Prediction Center forecasters using the CME Analysis Tool (CAT) \citep{2013SpWea..11...57M_Millward}. The modeled CMEs are then fed into the Wang-Sheeley-Arge (WSA-Enlil) heliospheric model, which models the solar wind between the Sun and Earth \citep{Pizzo_WSA_2011SpWea...9.3004P}. The CMEs are propagated in the solar wind, and the time of arrival at Earth and the strength of the storm are predicted \citep{2003AdSpR..32..497O_Odstrcil}.

The following bullet list gives a summary of the main driving requirements. The CCOR-1 and CCOR-2 specifications as designed are shown in Table \ref{tab:optical_param}.

\begin{itemize}
    \item SWFO GOES-U shall make the Coronal White Light intensity data available to the Space Weather Prediction Center Forecasters within 30 min after observations are made.
    \item CCOR shall be capable of meeting all mission and data availability requirements when observing a solar storm through an intensity of S4. 
    \item CCOR shall be capable of meeting all mission and data availability requirements when observing a solar flare up to an intensity of X50.
    \item SWFO GOES-U mission shall consist of 5 years of operations, with an additional 5 years of resources. 
    \item CCOR shall be capable of 5 years of on-orbit storage prior to the start of operations.
    \item Bandpass: white light
    \item Inner FOV geometric cutoff: CCOR-1: $3.7R_\odot$, CCOR-2: $3.0R_\odot$
    \item Outer FOV: CCOR-1: $\geq 17R_\odot$, CCOR-2: $\geq 20R_\odot$
    \item Resolution: CCOR-1: $\leq50$ arcsec, CCOR-2: $\leq70$ arcsec
    \item Scene coverage where the resolution and Signal to Noise is required to be met: CCOR-1: $1.36\degree$ (5.1 $R_\odot$) to $5.6\degree$ (21.0 $R_\odot$), CCOR-2 $1.17\degree$ (4.4 $R_\odot$) to $6.05\degree$ (22.7 $R_\odot$)
    \item Signal to Noise Ratio: $\ge 10$
    \item Coronal brightness measurement: 10\% absolute accuracy
    \item Image cadence: 15 minutes (5 minutes if images are 2x2 binned)
    \item Coronal signal minimum intensity (outer field of view): $\geq 1.0e-11 B_\odot$ 
    \item Coronal signal maximum intensity (inner field of view): $\leq 1.0e-8 B_\odot$ 
    \item CME velocity: $\leq 5$\% accuracy for CME velocities from 200 km/s to 3400 km/s
    \item CME mass: $\leq 50$\% accuracy for CME masses from 1.0e7 to 5.0e14 kg
    \item Re-closable door required for the possible 5 year GOES on-orbit storage.
\end{itemize}

\begin{table}[h]
    \centering
    \begin{tabular}{|p{0.32\textwidth}|p{0.05\textwidth}|p{0.24\textwidth}|p{0.24\textwidth}|}
        \hline
        Parameter & BoE & CCOR-1 & CCOR-2 \\
        \hline
        Mass & M & 19.1 kg & 20.1 kg\\
        Mass of PSB & M & 2.3 kg & 2.2 kg \\
        Volume L x W x H & M & 900 x 520 x 390 mm & 800 x 540 x 390 mm \\
        Power & M & 25 W & 25 W \\
        H $\times$ V Detector Pixels & M & $1920 \times 2048$ pixels & $1920 \times 2048$ pixels\\
        Imaging Pixels & M & $1910 \times 2038$ pixels & $1910 \times 2038$ pixels\\
        Pixel size & M & $10\mu m$ & $10\mu m$\\ 
        H $\times$ V, diag FOV & M & $5\degree \times 5\degree$, $6\degree$ & $6\degree \times 6\degree$, $7\degree$ \\ 
        H $\times$ V, diag FOV, at 1 AU & M & $18.8 R_\odot \times 18.8 R_\odot$, $22.5 R_\odot$ & $22.5 R_\odot \times 22.5 R_\odot$, $26.3 R_\odot$ \\ 
        Focal Length & D & 109.53 mm & 85.17 mm \\
        F-Number & D & 6.84 & 5.32 \\
        Plate Scale & M & 19.33 arcsec/pix & 24.27 arcsec/pix \\
        F-Tan Distortion at 6deg Field & D & 0.584\% & 0.299\% \\
        Bandpass FWHM & M & 470 - 740 nm & 469 - 755 nm\\
        Occulter Number of Disks & M & 19 & 24\\
        Geometric Inner FOV Cutoff & D & 3.7 $R_\odot$ ($0.984\degree$) & 3.0 $R_\odot$ ($0.798\degree$)\\
        Photometric Inner FOV Cutoff (S/N $> 1$) & M & 4.0 $R_\odot$ & 3.5 $R_\odot$\\
        Resolution & M & 39 arcsec  & 65 arcsec \\
        Resolution FOV & M & $5R_\odot$ to $20R_\odot$ & $4R_\odot$ to $22R_\odot$ \\
        Calibration Factor (High Gain Mode, Gain Setting = 8) & M & 1.66e-13 $B_\odot/(DN/s) \pm 8\%$ & 1.07e-13 $B_\odot/(DN/s) \pm 8\%$\\
        Baseline Image Cadence in Full Resolution & D & 15 minutes & 15 minutes\\
        Latency to the Ground & D & $\leq$ 15 minutes & $\leq$ 15 minutes\\
        \hline

    \end{tabular}
    \caption{CCOR-1 and CCOR-2 instrument specifications. In the second column BoE means Base of Estimate; M indicates a measured value and D an as-designed target value. The mass includes harnesses and Multi-Layer Insulation (MLI). The volume is the whole envelope plus some margin. The power is on average during nominal operations. The resolution is the averaged tangential and sagittal measurements at 10\% Modulation Transfer Function (MTF). The platescale is given at the center of the field of view. The Rsun values are given for a distance of 1 Astronomical Unit (AU): $1R_\odot = 0.266\degree$. Error values are 1-sigma.}
    \label{tab:optical_param}
\end{table}

\section{Accommodations}
\subsection{GOES-U/19 (CCOR-1)}\label{sec::GOES_accommodations}
CCOR-1 is onboard the Geostationary Operational Environmental Satellite - R Series, version U satellite, or GOES-U \citep{GOES-U_Databook}. After its launch on 25 June 2024, GOES-U was renamed GOES-19. The primary purpose of GOES is Earth imagery for weather forecasts. The first GOES satellite was launched in 1974, and GOES-U is the 19's iteration of the series. It is part of the GOES-R series, which includes instruments dedicated to observing the Sun for monitoring space weather hazards. In addition to the Earth-observing instruments, GOES-19 carries two in situ instruments and three remote sensing instruments that observe the Sun and its corona.

\paragraph{In Situ}
\begin{itemize}
    \item Goddard Magnetometer (GMAG)
    \item Space Environment In Situ Suite (SEISS) 
\end{itemize}

\paragraph{Remote Sensing}
\begin{itemize}
    \item Extreme Ultraviolet and X-ray Irradiance Sensors (EXIS)
    \item Solar Ultraviolet Imager (SUVI)
    \item Compact Coronagraph 1 (CCOR-1)
\end{itemize}

\begin{figure}[h]
\centering
\includegraphics[angle=0, width=0.7\textwidth]{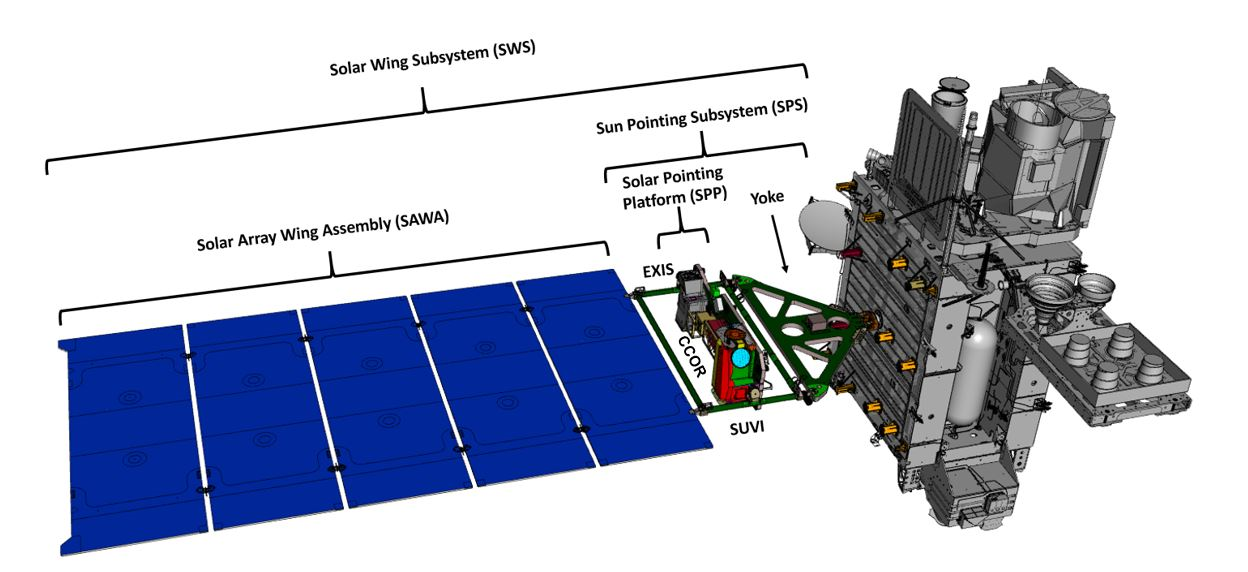}
\caption{GOES-19 spacecraft.}\label{fig_goes19}
\end{figure}

\begin{figure}[h]
\centering
\includegraphics[width=0.6\textwidth]{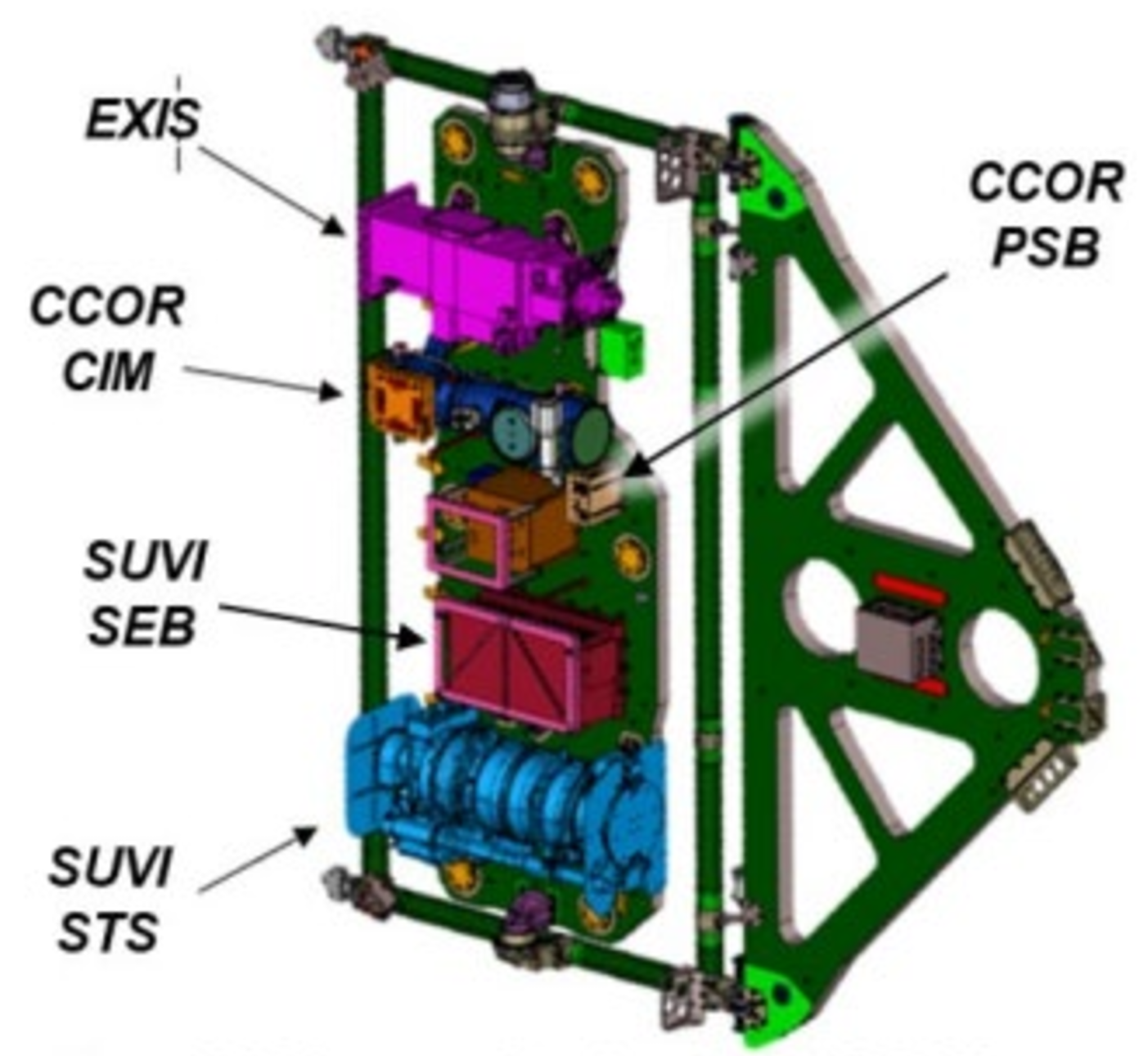}
\caption{GOES-19 Solar Pointing Platform instrument layout.}\label{fig_SPPLayout}
\end{figure}

The GOES-19 spacecraft is shown in Figure \ref{fig_goes19}. The suite of space weather remote sensing imagers is installed on the Solar Pointing Platform (SPP) (Figure \ref{fig_SPPLayout}), a platform that is located just below the solar array. The SPP is articulated around two axes: the solar array yoke axis (Solar Array Drive Assembly or SADA) and the axis transverse to the SPP providing a pitch motion of that SPP (SPP Elevation Gimbal Assembly, or SEGA). These two axes allow fine tracking and sun pointing, based on a Fine Sun Sensor (FSS) and the SUVI Guide Telescope (SUVI-GT), both mounted on the SPP. The SUVI-GT provides the most accurate pointing, but the FSS also provides the two-axis position error when the SUVI-GT data are not available. The CCOR-1 boresight was co-aligned with the SUVI-GT on the ground. 

The CCOR-1 field of view was designed to the NOAA threshold (minimum) requirements. The installation on GOES required the addition of the end cap to reduce the scattered light from the structures on the satellite and earthshine.

GOES-19 is a geostationary satellite, so it sits above the equator, at latitude $0\degree$. The distance from the Sun ranges from 0.984 AU to 1.016 AU. During the Post-Launch Testing (PLT), GOES-19 was located at $\approx 90\degree$ west longitude. In April 2025, GOES-19 was placed at $75.2\degree$ west longitude, where it replaced the current GOES-16, also called GOES-East.

\begin{figure}[h]
\centering
\includegraphics[width=0.8\textwidth]{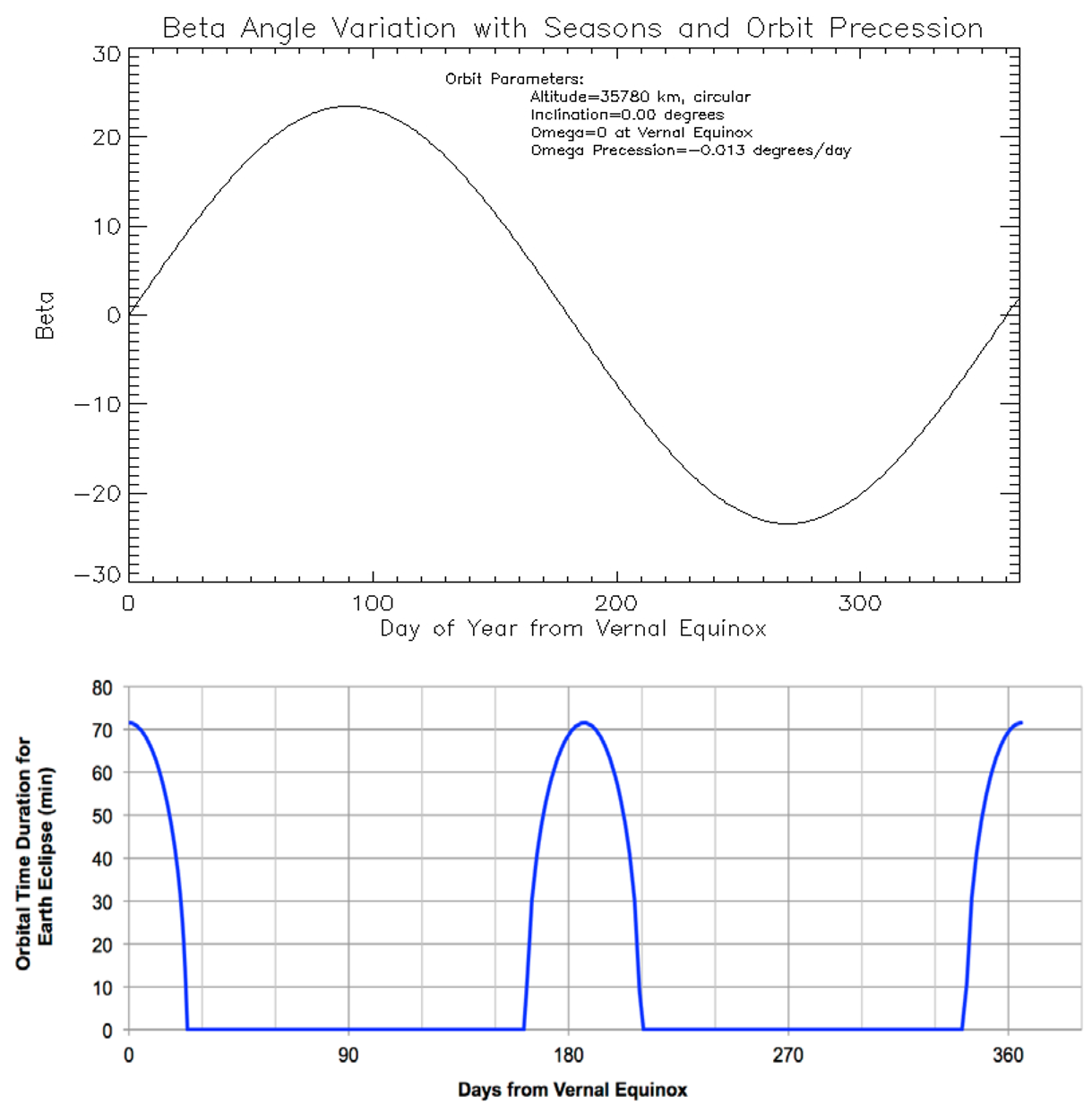}
\caption{Beta angle and total eclipse duration over the solar year, from the vernal equinox ($\simeq$ March 20).}\label{fig_EclipseTiming}
\end{figure}

The geostationary orbit has some consequences on the sun observing instruments. Earth will transit or be in the vicinity of the CCOR FOV on a daily basis. It will not only block the view of the solar corona, but also create strong and highly variable stray light patterns. From a thermal point of view, the detector temperature will also have seasonal variations. GOES makes a full revolution around the center of the globe during a synodic day. The Solar Array Angle (or SADA, as used in the FITS header) needs to do a full 360 degrees to keep on tracking the Sun. As the rotation axis of the Earth is tilted by approximately $23.5 \degree$ with respect to the ecliptic plane, this results in seasonal effects. The angle between the Sun and the Earth equatorial plane, which we call $\beta$, varies from $+23.5 \degree$ on the northern hemisphere summer solstice to $-23.5 \degree$ on the northern hemisphere winter solstice (see Figure \ref{fig_EclipseTiming}). At the equinoxes, the Earth transits the CCOR-1 FOV and even fully eclipses the spacecraft. This is the so-called eclipse season for GOES, which typically starts and ends 25 days before and after the equinoxes. During that time, a special CCOR observing program is run, called eclipse mode. The duration of the eclipses and the angle between the Sun and the Earth equator are given in Figure \ref{fig_EclipseTiming}. After the eclipse season, the Earth will still transit the CCOR FOV for $\pm 50$ days around the equinox date and induce strong stray light signatures in the images, especially when the Earth limb is less than $15\degree$ from the CCOR boresight. More details on earthshine-induced stray light are given in Section \ref{sec::earthshine}.

To avoid having CCOR imaging the spacecraft and being overflowed by sunlight reflections on the deck, GOES performs a yaw flip at each equinox. The orientation of the spacecraft is such that the SPP SEGA gimbal always tilts (or pitch) away from the deck when tracking the Sun. During the northern hemisphere fall and winter season, the GOES-19 solar array points towards the Earth south pole. During spring and summer in the northern hemisphere, the solar array points towards the north pole of the Earth.

Because there are 3 instruments looking at the Sun on the SPP, there are strong co-alignment requirements between them. SUVI drives the pointing requirements, as it has the smallest field of view of the three, and the tightest requirement in terms of sun pointing. As a result, CCOR was tightly co-aligned with the SUVI Guide Telescope (SUVI-GT). The co-alignment budget and SPP pointing accuracy yield a pointing accuracy of 4 arc minutes, 3-sigma for CCOR. Sun pointing stability is 25.2 arc sec, 3-sigma, during 60 seconds, which corresponds to the maximum integration time for the acquisition of a CCOR image.

The radiation environment in the geostationary orbit is harsh, especially in terms of electron radiation \citep{JUN2024_radiationEnviroReview}. CCOR was designed to withstand that environment and keep its performance at the end of its life, which is 5 years of operations + 5 years of possible on-orbit storage. 

The CCOR-1 to spacecraft image transmission latency is less than 12 min. GOES-U is in constant contact with the ground. The images are delivered to the forecasters in about 15 minutes, in addition to the CCOR-1 to spacecraft latency.

\subsection{SWFO-L1}
The Space Weather Follow On L1 (SWFO-L1) is a spacecraft dedicated to space weather monitoring and forecasts. It will be in a lissajous orbit around L1 Lagrange point, following the European Space Agency (ESA) and NASA’s Solar and Heliospheric Observatory (SOHO) \citep{Domingo_1995SoPh..162....1D}, NASA's Advanced Composition Explorer (ACE) \citep{Stone_ACE_1998SSRv...86....1S}, NOAA's Deep Space Climate Observatory (DSCOVR) \citep{DSCOVR}, and Aditya-L1 of the Indian Space Research Organization (ISRO) \citep{seetha_aditya_2017CSci..113..610S}. From the L1 vantage point, there is a continuous and uninterrupted view of the Sun and its corona. SWFO will operate at a distance to the Sun ranging from 0.974 AU to 1.006 AU.

\begin{figure}[h]
\centering

\includegraphics[width=0.9\textwidth]{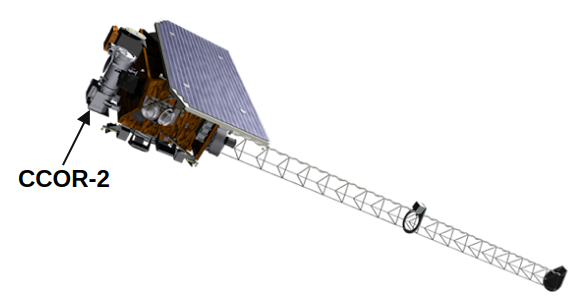}
\caption{SWFO-L1 spacecraft overview.}\label{fig_SWFO}
\end{figure}

SWFO-L1 carries the following 4 in situ instruments and 1 remote sensing imager.

\paragraph{In Situ}
\begin{itemize}
    \item Solar Wind Plasma Sensor (SWiPS)
    \item Supra-Thermal Ion Sensor (STIS)
    \item Suite of two magnetometer sensors (MAG)
\end{itemize}
The SWiPS, STIS, and MAG are collectively known as the Solar Wind Instrument Suite (SWIS).

\paragraph{Remote Sensing}
\begin{itemize}
    \item Compact Coronagraph 2 (CCOR-2)
\end{itemize}

SWFO-L1 is scheduled to launch in the last quarter of 2025. After a cruise phase of approximately 3 months, SWFO will reach L1. The operational phase will start after a 6 months commissioning of the spacecraft and its payloads. The 6 months commissioning include the 3 months of cruise to L1.

Unlike on GOES-U, where there is a platform dedicated to sun-observing instruments, the SWFO spacecraft is simpler and smaller. The whole spacecraft points to the Sun and CCOR-2 is placed in a way where there is little spacecraft-induced stray light compared to GOES (Figure \ref{fig_SWFO}). There will be no Earth eclipses nor Earth-induced stray light since it is at L1. As a result, the thermal environment is also simpler, as there will be no night and day variations and no seasonal variations. Finally, since CCOR-2 is the only sun-pointed remote sensing imager on board, there are no strong co-alignment requirements with the other instruments, and the spacecraft pointing can be adjusted in flight to optimize the CCOR-2 occulter stray light rejection. 

The spacecraft pointing accuracy is 2 arc minutes, 3-sigma in the 3 directions of space. The pointing stability is less than 5 arc seconds 1-sigma over a 60-second period, also in the 3 directions of space. As a result of the smaller spacecraft and tighter pointing accuracy and stability, the CCOR-2 outer field of view, compared to CCOR-1, could be slightly increased and the inner field of view cutoff could be decreased. See Table \ref{tab:optical_param} for a summary of the as-built specifications.

The CCOR-2 to spacecraft image transmission latency requirement is less than 12 min. The rest of the latency to the ground and forecasters is on the spacecraft side and on the ground pipeline side. Due to the L1 location, antenna stations scattered across the globe will allow constant contact with SWFO and ensure data downlinking at the required latency. 

\section{Overall Instrument Design}

\begin{figure}[h]
\centering
\includegraphics[width=0.9\textwidth]{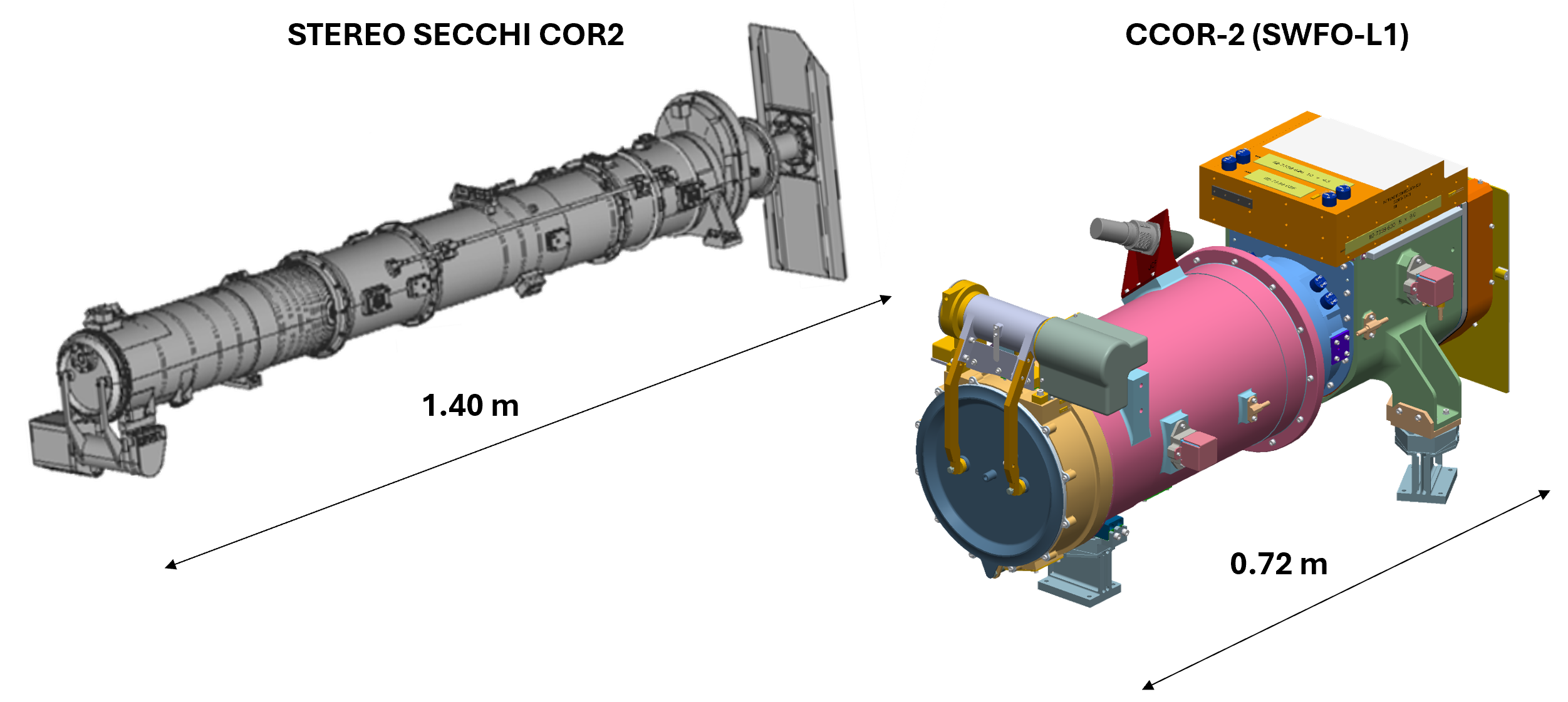}
\caption{Length comparison between SECCHI COR-2 and CCOR-2.}\label{fig_cor2compare}
\end{figure}

\begin{figure}[h]
\centering
\includegraphics[width=0.9\textwidth]{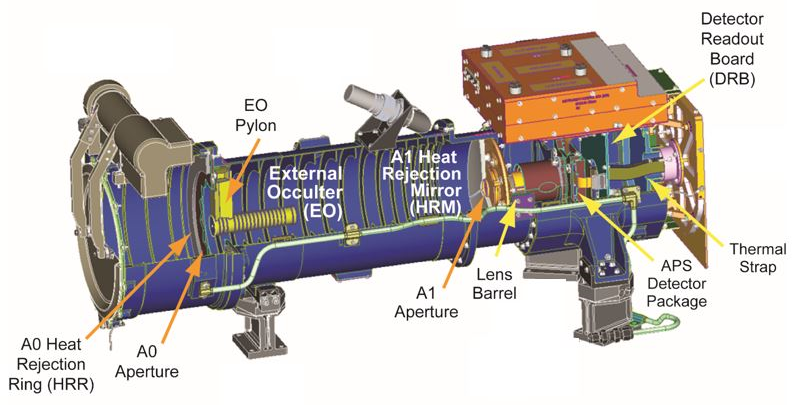}
\caption{CCOR-1 overview.}\label{fig_ccor1cad}
\end{figure}

\begin{figure}[h]
\centering
\includegraphics[width=0.9\textwidth]{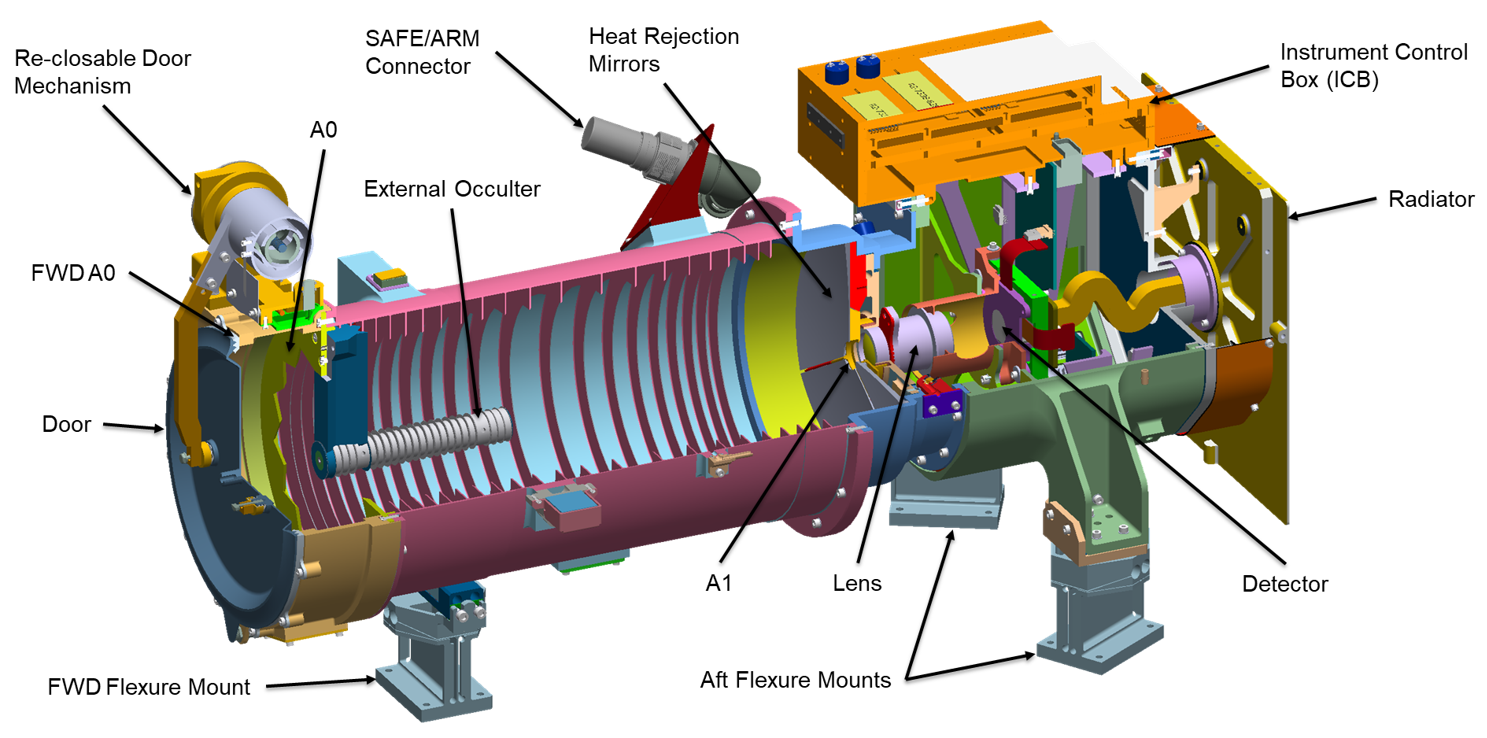}
\caption{CCOR-2 overview.}\label{fig_ccor2cad}
\end{figure}

\begin{figure}[h]
\centering
\includegraphics[width=0.9\textwidth]{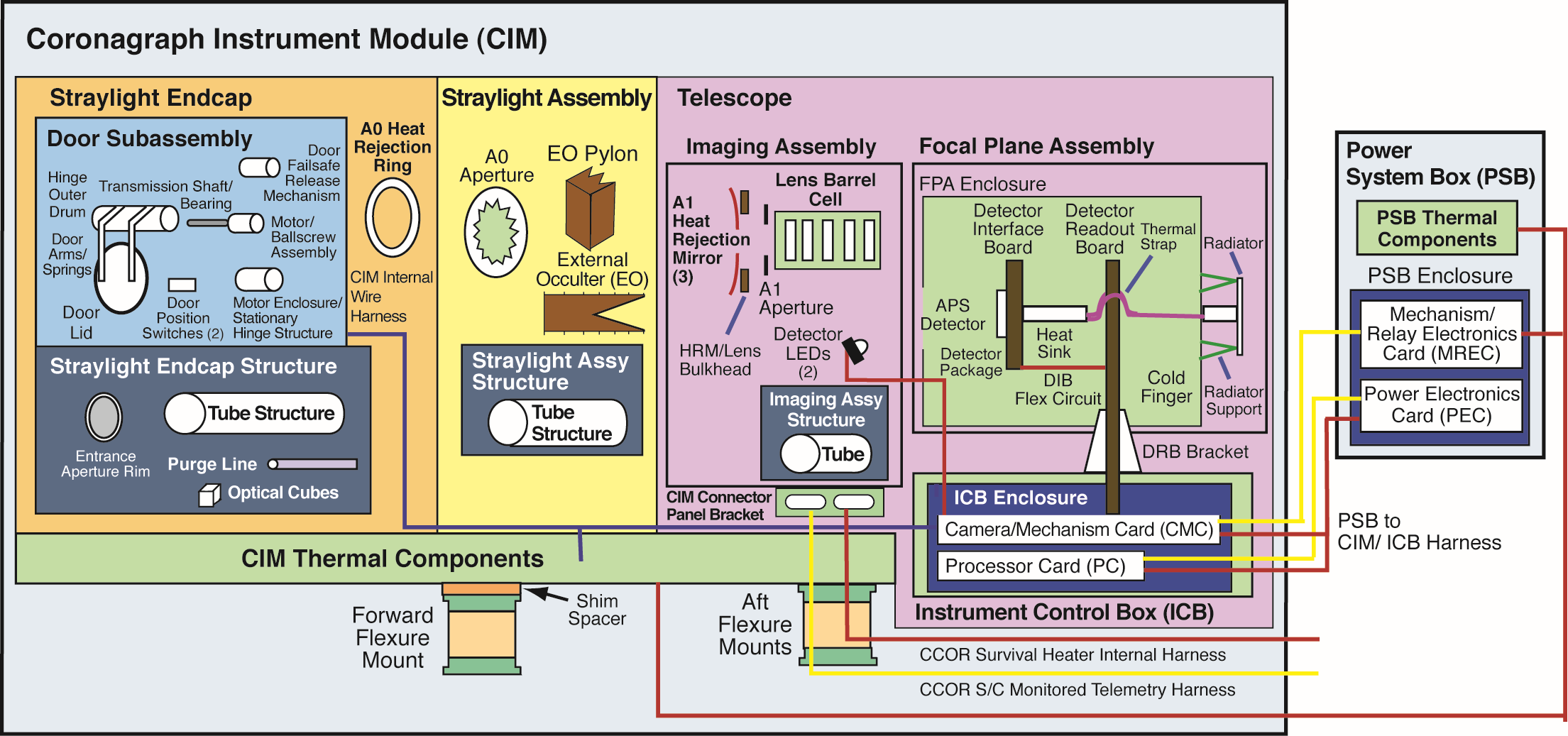}
\caption{CCOR-1 block diagram.}\label{fig_blockdiagram}
\end{figure}

At the start of the GOES-R series, NOAA planned to include a coronagraph based on SECCHI COR-2. However, it was too large to fit and it was descoped during the GOES-R formulation phase. This led NRL to investigate the design for smaller coronagraphs, leading to CCOR. Figure \ref{fig_cor2compare} shows the length of SECCHI COR-2 (1.40 m) compared to that of CCOR-2 (0.72 m). The compact design cuts the length of the instrument by half, but still allowed a sufficient level of signal-to-noise ratio to be useful for space weather application.

The 3D views of the CCOR-1 and CCOR-2 CAD models are shown in Figures \ref{fig_ccor1cad} and \ref{fig_ccor2cad}, respectively. In addition, the instrument block diagram is shown in Figure \ref{fig_blockdiagram}. Both instruments are similar; the main geometric differences are related to the different fields of view. From left to right, a re-closable door is attached on the forefront of the instrument, called the end cap. Behind that is the A0 entrance aperture, a serrated aperture on which a pylon that maintains an external occulter is mounted. A stray light tube equipped with baffles connects the forefront of the instrument to the Focal Plane Assembly (FPA). A set of 3 Heat Rejecting Mirrors (HRMs) at the bottom of the stray light tube reflects the direct sunlight outside the instrument (see Section \ref{subsubsec:HRM}). On the axis, at the center of the HRMs, the A1 entrance aperture of the camera sits in the shadow of the external occulter. Behind the A1, an objective lens creates an image of the corona on an Active Pixel Sensor (APS) detector that is passively cooled; it is connected with a thermal strap to a radiator that mostly sees deep space. The image formed on the APS detector is electronically collected by the Detector Readout Board (DRB) and then transferred to the Instrument Control Box (ICB) for further processing and compression before being sent to the spacecraft through a Space Wire. The ICB includes the Processor Card (PC) and the Camera/Mechanism Card (CMC). A set of 2 LEDs, the brightness of which can be adjusted by the flight software, directly illuminate the detector for ground and in-flight calibration.

\begin{figure}[h]
\centering
\includegraphics[width=0.9\textwidth]{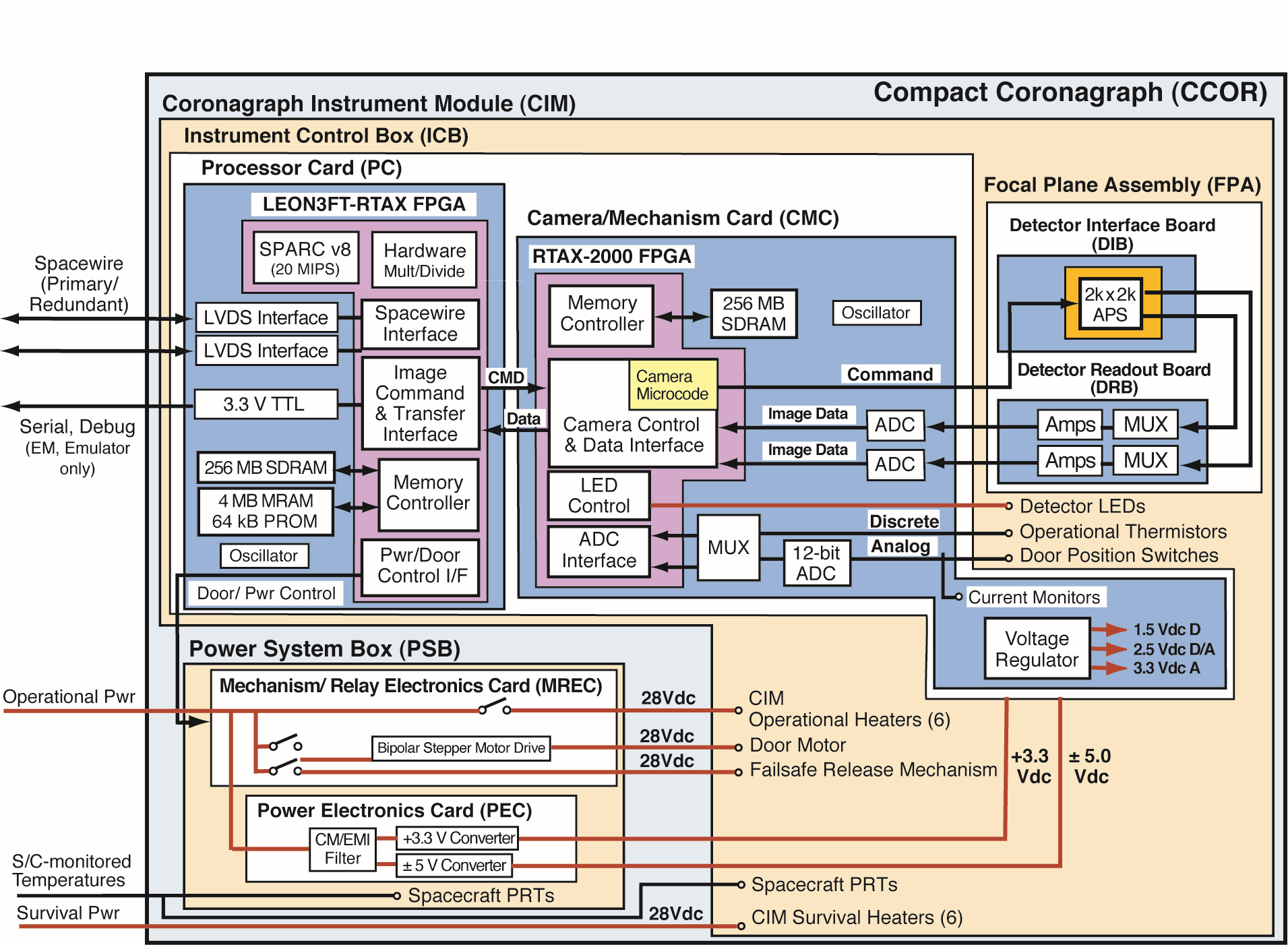}
\caption{CCOR-1 electrical block diagram.}\label{fig_CCOR1ElecDiag}
\end{figure}

The Power System Box (PSB) provides regulated power to CCOR. The PSB includes the Mechanism/Relay Electronics Card (MREC), the Power Electronics Card (PEC), the PSB enclosure, and the PSB thermal components. It also provides eight telemetry points on the CCOR temperatures, bus current, and bus voltages. The PSB telemetry points are accessed from a Serial Protocol Interface (SPI) to the ICB Processor Card. The heater power is controlled by the flight software and provided by the PSB.

The CCORs are equipped with a fully re-closable door. The motor mechanism is a stepper motor that is commanded by setting the direction of the door parameter, the step rate, the dead band, the number of steps, and the use of open / closed limit switches. The CCOR door also has a one-shot paraffin actuator pin puller that will open the door in the event of a door motor failure. After the door is opened using the fail-safe pin puller on-orbit, the door cannot be closed using the Door Motor anymore. See Section \ref{subsec:door} for more details about the door design.

The CCORs are installed on both the GOES-19 SPP and SWFO-L1 deck with 3 flexure mounts. These mounts thermally isolate the CCORs from the spacecraft and dampen vibration loads during launch. They are also designed to maintain the alignment of the CCORs with the other sun-pointed instruments on board each spacecraft. Co-alignment is done on the ground by shims that are implemented in the flexure mounts. Co-aligning and maintaining this alignment in flight is especially critical for CCOR-1 as the Sun pointing is driven by the SUVI field of view. Optimal sun pointing minimizes the diffracted stray light around the occulter. The CCOR-1 optimal pointing towards the Sun needs to be achieved to within $\pm 4$ arcmin during flight. For CCOR-2, the spacecraft is set to optimally point it at the Sun, as it is the primary solar pointed instrument on board. The Sun pointing requirement for CCOR-2 is $\pm 2$ arcmin.

CCOR images and housekeeping data are transmitted through Spacewire. For CCOR-1, the Spacewire connection has an A/B port for redundancy. For CCOR-2, there is only one port, so there is no redundancy.

\section{Coronal Signals}\label{sec_coronalesignals}
The signal-to-noise ratio (S/N) requirement and the field-of-view coverage drive the optical design of the CCORs. The signal of interest is the one from the CMEs, which are transients of the K-corona. We used the \cite{Saito_1977SoPh...55..121S} coronal background electron density model as a proxy of the CME signal. We computed the brightness of that signal using Thomson scattering \citep{Billings_1966} and ray tracing along line of sight that encompass the CCOR FOV elongation range.

For the F-corona, we use the \cite{Koutchmy_Lamy_1985} model . We used the ecliptic plane trace as it is the strongest signal and therefore represents the worst case since the F-corona is considered as a background in the computation of the S/N.

\begin{figure}[h]
\centering
\includegraphics[width=0.8\textwidth]{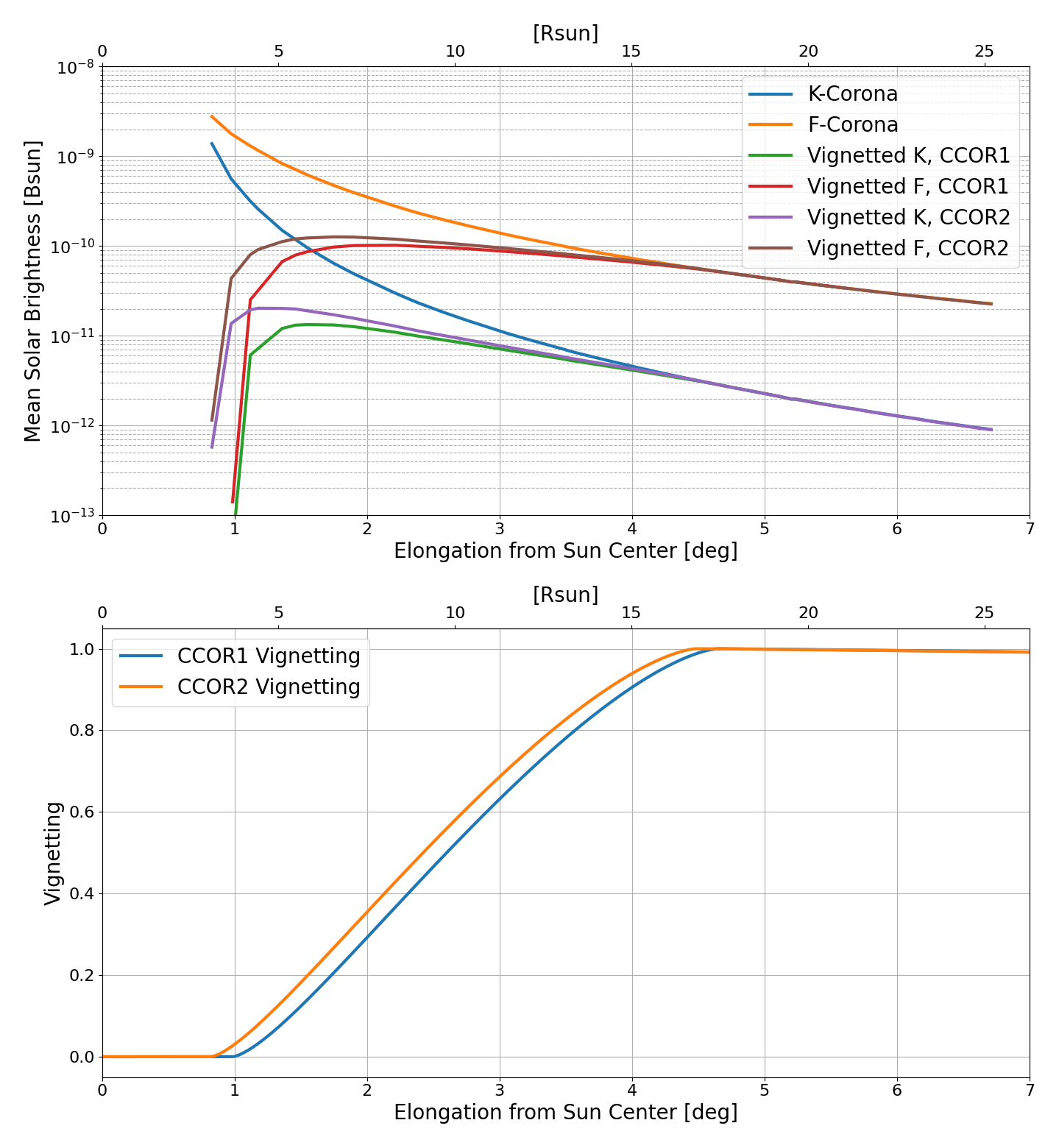}
\caption{Top: K and F coronae signals, un-vignetted and vignetted for both CCOR-1 and CCOR-2. Bottom: vignetting profile for CCOR-1 and CCOR-2, away from the pylon region.}\label{fig_KFsignalsVig}
\end{figure}

\section{Optical Design}

\subsection{External Occulter, Inner Field of View Cutoff, and Vignetting}\label{subsec:IFOVCO}

\begin{figure}[h]
\centering
\includegraphics[width=0.6\textwidth]{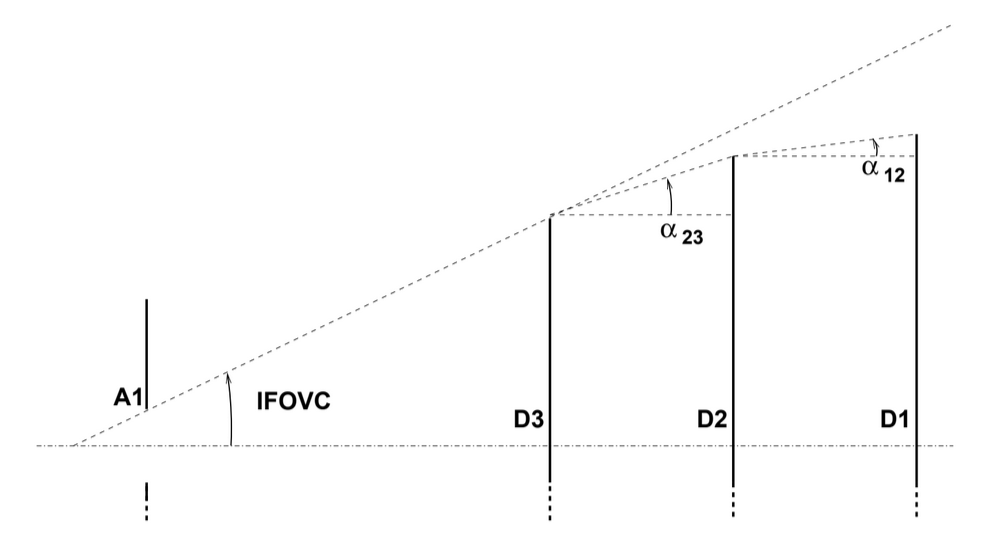}
\caption{Definition of the Inner Field of View Cutoff (IFOVCO). A1 is the telescope entrance pupil. D3 is the last disk of the occulter, in this case. We use only 3 disks here for the sake of clarity of the figure.}\label{fig_IFOVCO}
\end{figure}

The external occulter uses multiple disks; 19 for CCOR-1, 24 for CCOR-2. The disks are sized so that each of them sits in the shadow of the disk in front of it, except for the first one, which sits in full sunlight. This is shown in Figure \ref{fig_IFOVCO}, but only for 3 disks for the sake of clarity of the figure.

The size and distance of the last external occulter disk to the A1 entrance pupil of the telescope determine the inner field of view cutoff (IFOVCO) of the coronagraph. This is shown in Figure \ref{fig_IFOVCO}, which is from \cite{Thernisien_2005}. The baseline values of the IFOVCO for both CCORs are given in Table \ref{tab:optical_param}. As the external occulter is right in the middle of the field of view, it introduces a strong vignetting on the image plane. It fully occults the external scene at the center of the field of view to prevent direct sunlight from entering the telescope entrance pupil, A1. A profile of the vignetting is given in the bottom of Figure \ref{fig_KFsignalsVig}. Right at the IFOVCO, there are still no photons coming from the corona, but the signal starts to ramp up from that point with increasing elongation. The field becomes fully unvignetted at $4.64\degree$ ($17.4 R_\odot$) for CCOR-1 and $4.47\degree$ ($16.8 R_\odot$) for CCOR-2. Although the vignetting attenuates the signal and degrades spatial resolution, it is actually suitable because it flattens the high dynamic range of the corona, which would otherwise saturate the detector in the inner FOV and/or render the signal too weak to be detected in the outer FOV. The effect of vignetting on the signal is shown in the top Figure \ref{fig_KFsignalsVig}. 

\begin{figure}[h]
\centering
\includegraphics[width=0.49\textwidth, trim={50 0 50 20}, clip]{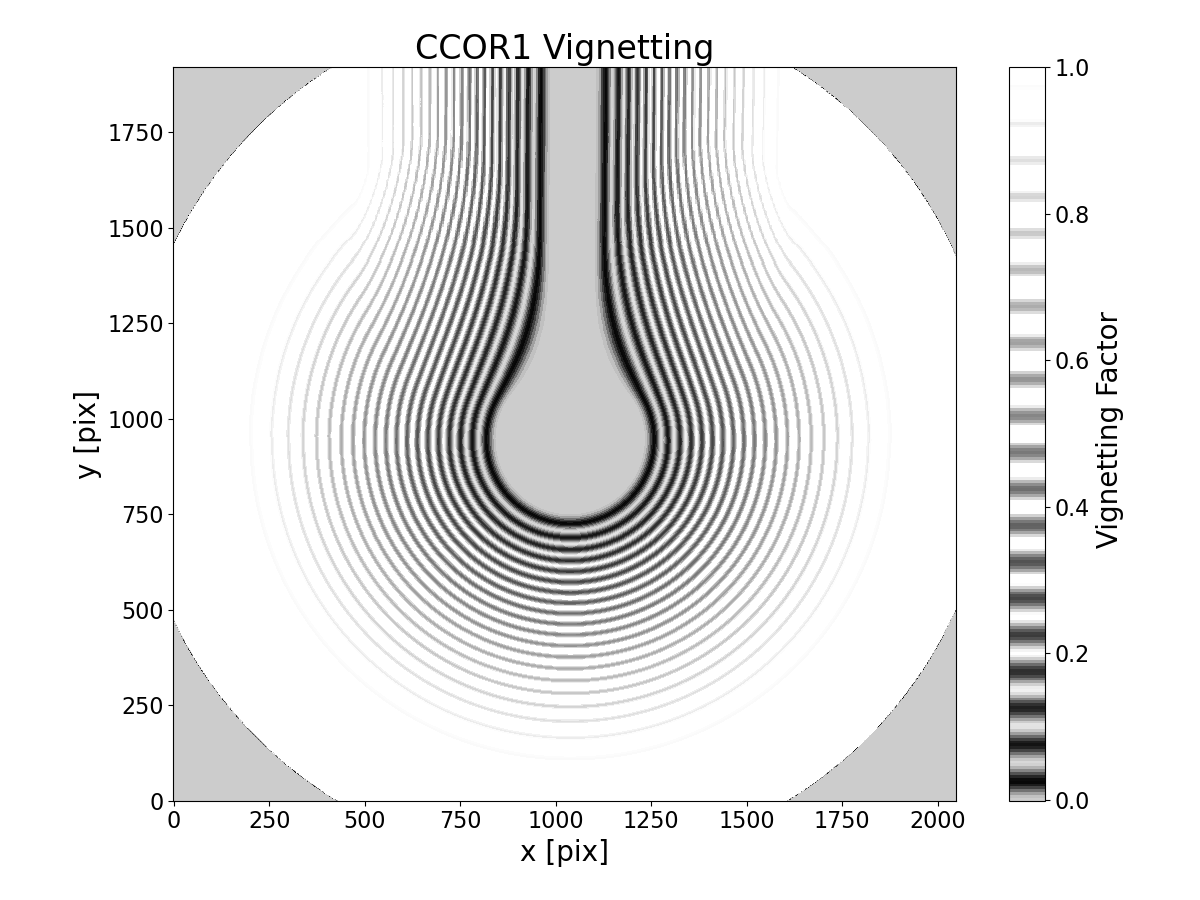}
\includegraphics[width=0.49\textwidth, trim={50 0 50 20}, clip]{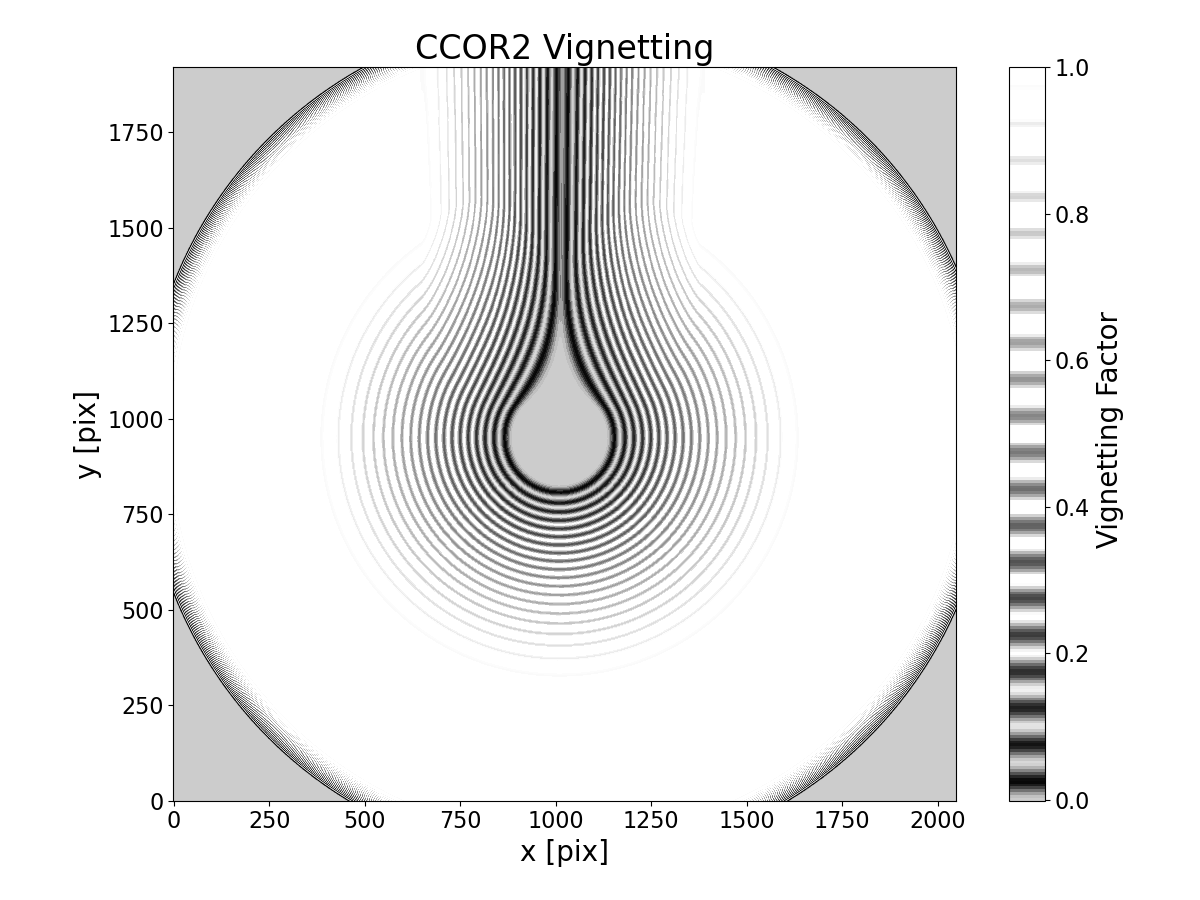}
\caption{Vignetting of CCOR-1 and CCOR2.}\label{fig_ccorVignetting}
\end{figure}

The vignetting images are shown in Figure \ref{fig_ccorVignetting}. The rotationally symmetric part away from the pylon was calculated using the analytical expression given in \cite{Bayanna_2011ExA....29..145B}. The area around the pylon was calculated using ray tracing. The center and rotation of the vignetting were then adjusted using the flat-field images of the flight instruments.

\subsection{Lens Design}
\begin{figure}[h]
\centering
\includegraphics[width=0.49\textwidth]{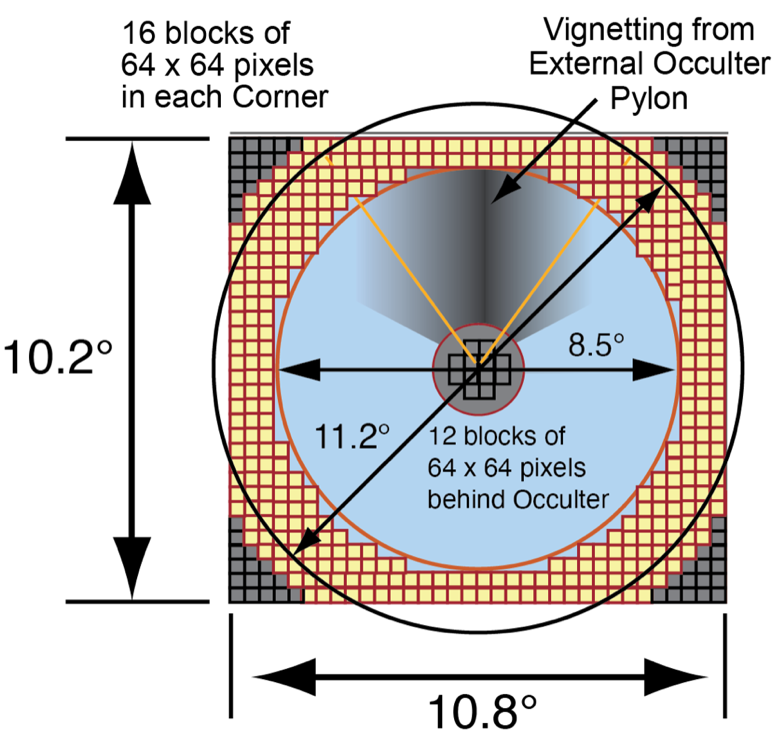}
\includegraphics[width=0.49\textwidth]{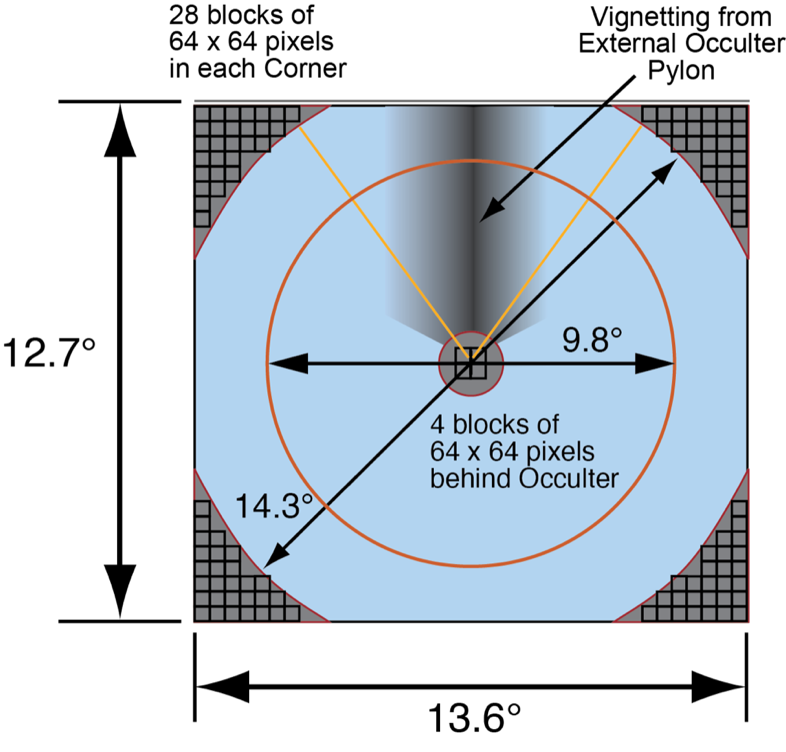}
\caption{CCOR-1 and CCOR-2 Fields of View. The orange circle, at $8.5\degree$ for CCOR-1 and $9.8\degree$ for CCOR-2, show the limit of the lossy and lossless sub frames; this is discussed in more details in Section \ref{sec_onboardprocessing}.}\label{fig_ccorFOVandBlocks}
\end{figure}

\begin{figure}[h]
\centering
\includegraphics[width=0.9\textwidth]{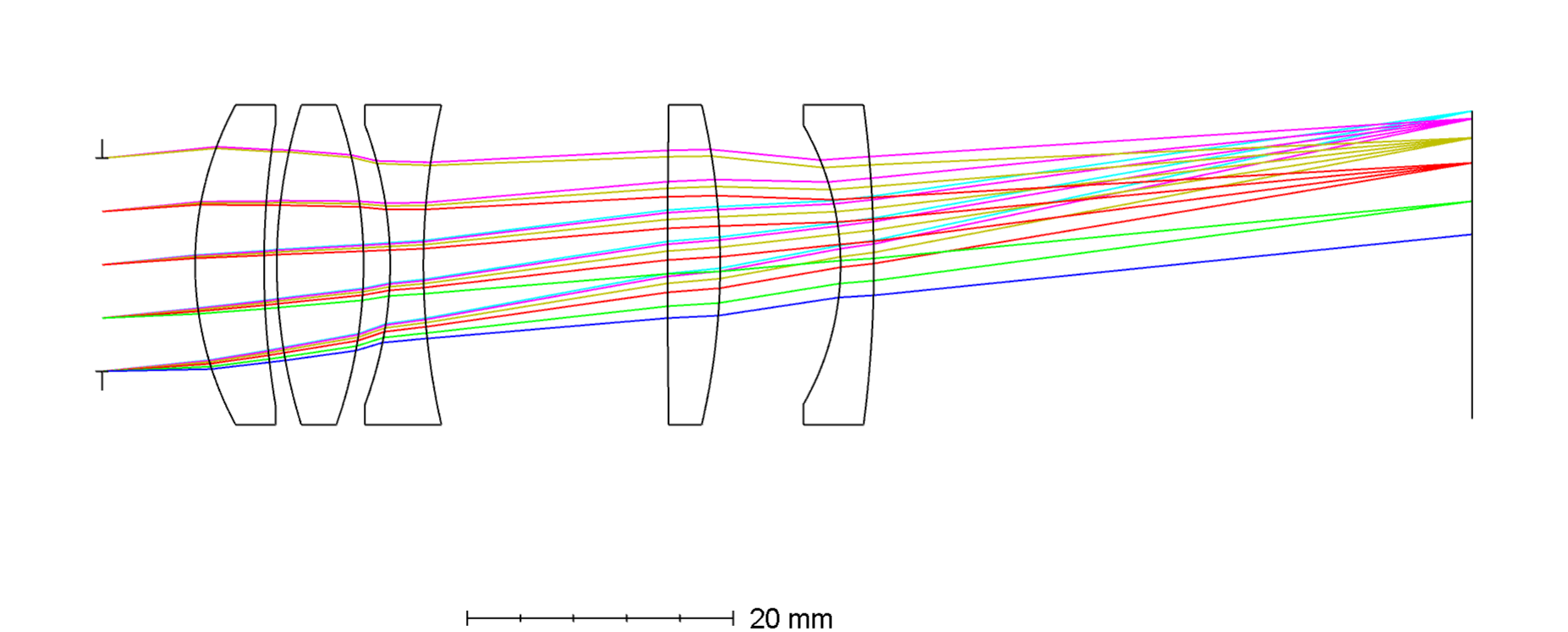}
\includegraphics[width=0.9\textwidth]{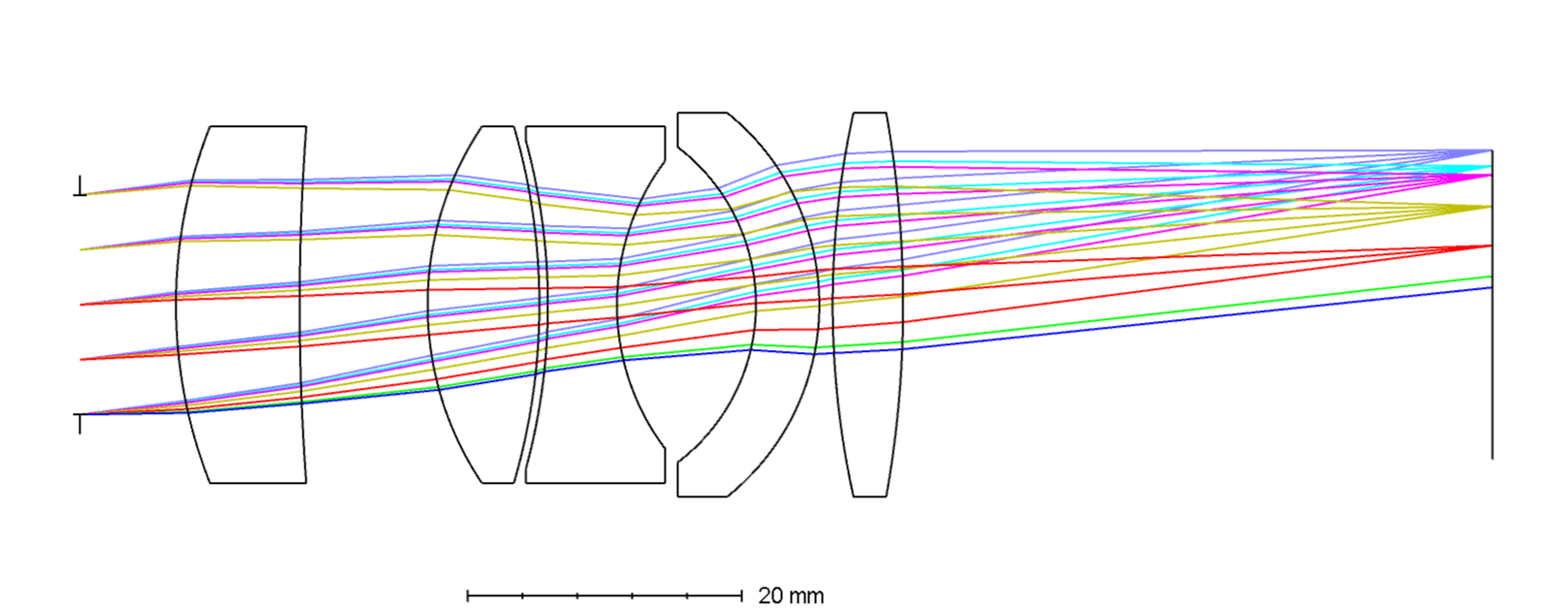}
\caption{Lens ray bundle for CCOR-1 and CCOR-2.}\label{fig_lensraybundle}
\end{figure}

\begin{figure}[h]
\centering
\includegraphics[width=0.45\textwidth]{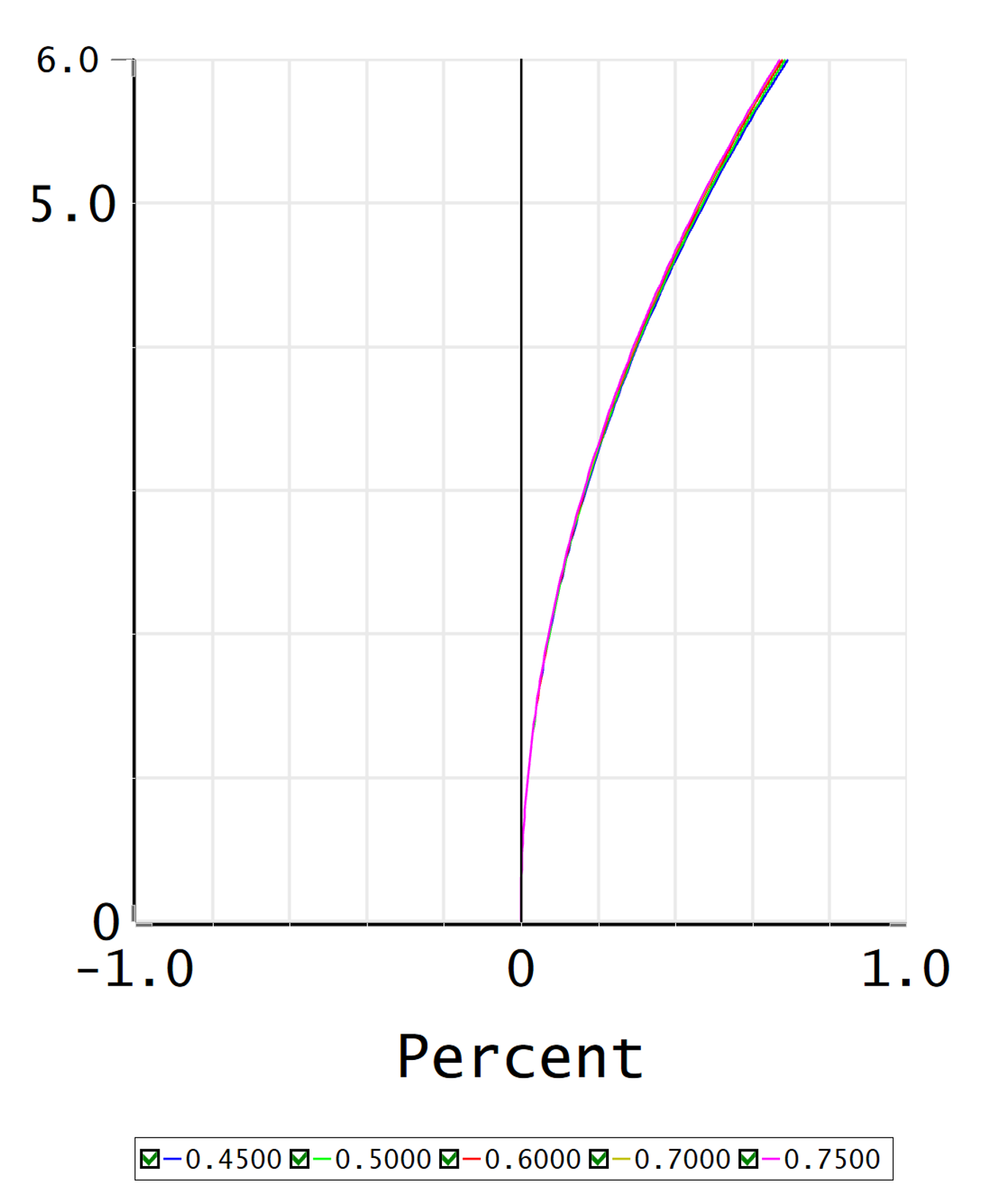}\includegraphics[width=0.45\textwidth]{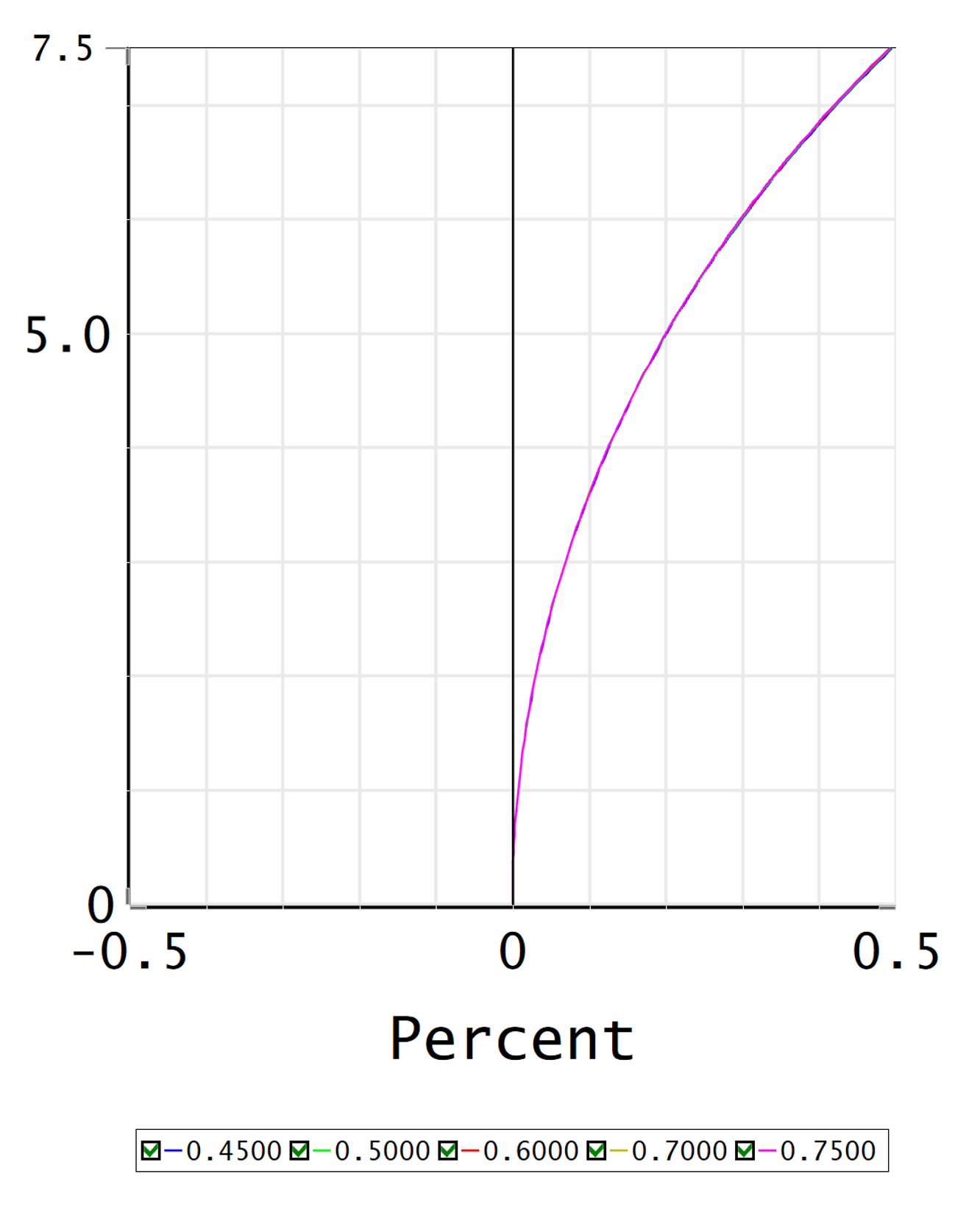}
\caption{F-Tan(Theta) distortion for CCOR-1 (left) and CCOR-2 (right). Maximum field distortion for CCOR-1 is 0.678\% at $6\degree$ field. Maximum field distortion for CCOR-2 is 0.493\% at $7.5\degree$ field.}\label{fig_lensdistortion}
\end{figure}

\begin{figure}[h]
\centering
\includegraphics[width=0.9\textwidth]{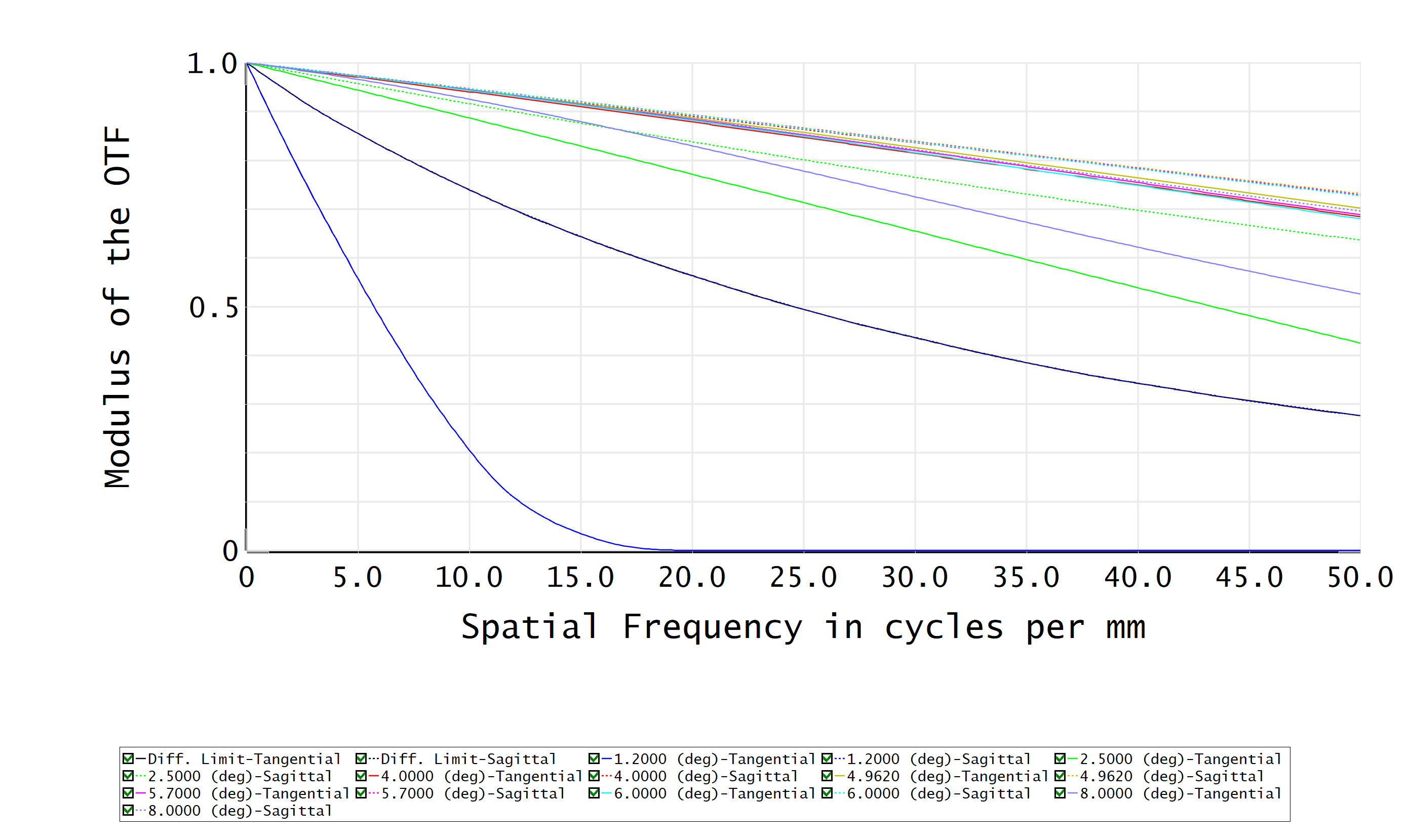}
\includegraphics[width=0.9\textwidth]{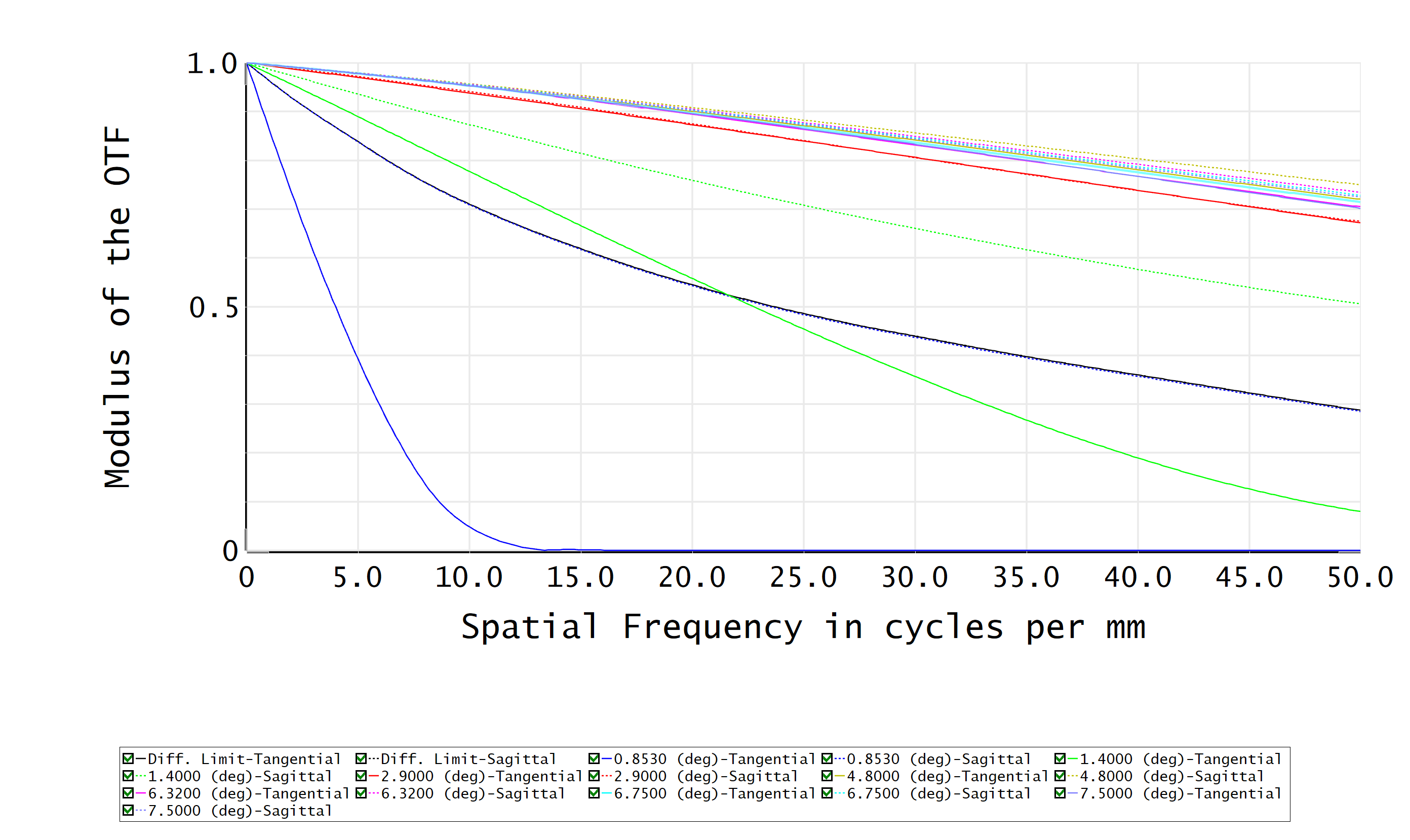}
\caption{Lens Modulation Transfer Function (MTF) for CCOR-1 and CCOR-2, with the occulter present.}\label{fig_lensMTF}
\end{figure}

The lens design is driven by the required fields of view, the detector size, the required spatial resolution, and the required signal-to-noise ratio. Both CCORs use a five-element lens design. The ray bundles are shown in Figure \ref{fig_lensraybundle} and the baseline Modulation Transfer Function (MTF) performances, which includes the presence of the occulter, are shown in Figure \ref{fig_lensMTF}. The field distortion is also minimal, as shown in Figure \ref{fig_lensdistortion}. The FOV achieved for both CCORs is shown in Figure \ref{fig_ccorFOVandBlocks}. The spatial resolution numbers given in Table \ref{tab:optical_param} also take into account the jitter and stability of the spacecraft.

\begin{figure}[h]
\centering
\includegraphics[width=0.7\textwidth]{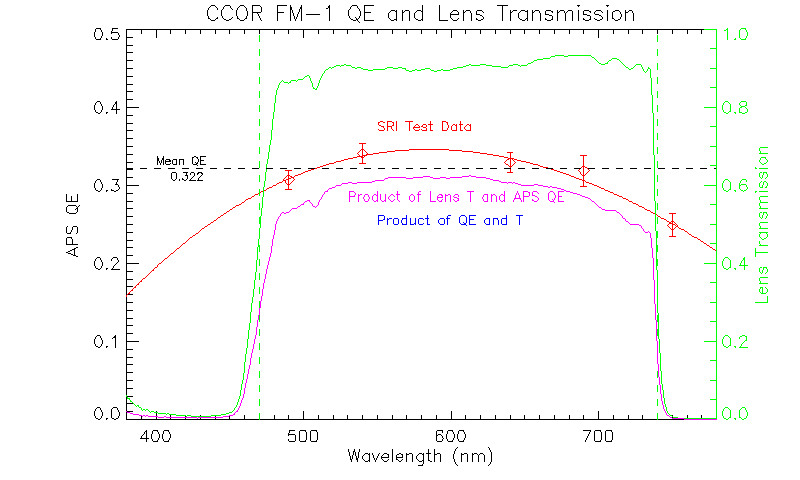}
\includegraphics[width=0.7\textwidth]{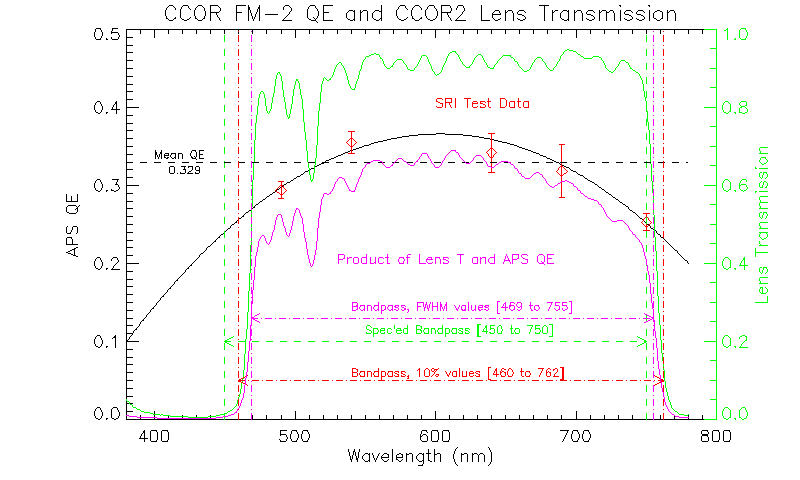}
\caption{Measured lens transmission of the flight lens assemblies and detector QE, for CCOR-1 and CCOR-2.}\label{fig_lenstransmission}
\end{figure}

Ghosting performances were also taken into account in the optimization, especially for minimizing the stray light due to the bright diffraction ring around the occulter and bright planets and comets that will regularly transit the FOV. An anti-reflecting coating was used on each of the surfaces, except on the surfaces that received thin-film high- and low-pass filters. The first surface of the first element was instead coated with a magnesium fluoride AR coating, as it is more resistant to electron radiation compared to more elaborate AR coatings \citep{Boise_10.1117/12.190938}. The darkening of the glass material as a result of the Total Ionizing Dose at the end of life was estimated. The measured bandpass transmission of the flight lens assemblies is shown in Figure \ref{fig_lenstransmission}.

The lens barrel assembly is made of titanium, to minimize any thermal variation effects on the optical performances. Thermal effects, as well as fabrication tolerances and errors, were taken into account with the help of the lens assembly manufacturer. Off-axis stray light rejection was also considered in the design of the lens retainers and lens barrel interior. A Point Source Transmittance (PST) was computed before fabricating the lens assembly and it was measured on the as-built flight assemblies.

\clearpage

\subsection{Stray Light}
Coronagraph design is all about stray light management. The instrument is aimed straight at the Sun, which has a radiance of 1 Mean Solar Brightness (also noted Bsun, $B_\odot$, or $B/B_\odot$), and 3 or 4 solar radii from the sun center we want to observe the K-corona signal, which is 10 to 12 orders of magnitude dimmer than the Sun's surface. As direct sunlight hits the instrument, it is diffracted, reflected, scattered, and refracted. A fraction of that light ends up on the detector, which creates a veiling background on the images and degrades the signal-to-noise ratio. If not properly managed, it can totally swamp the signal and saturate the detector, resulting in total loss of any signal.

\begin{figure}[h]
\centering
\includegraphics[width=0.9\textwidth]{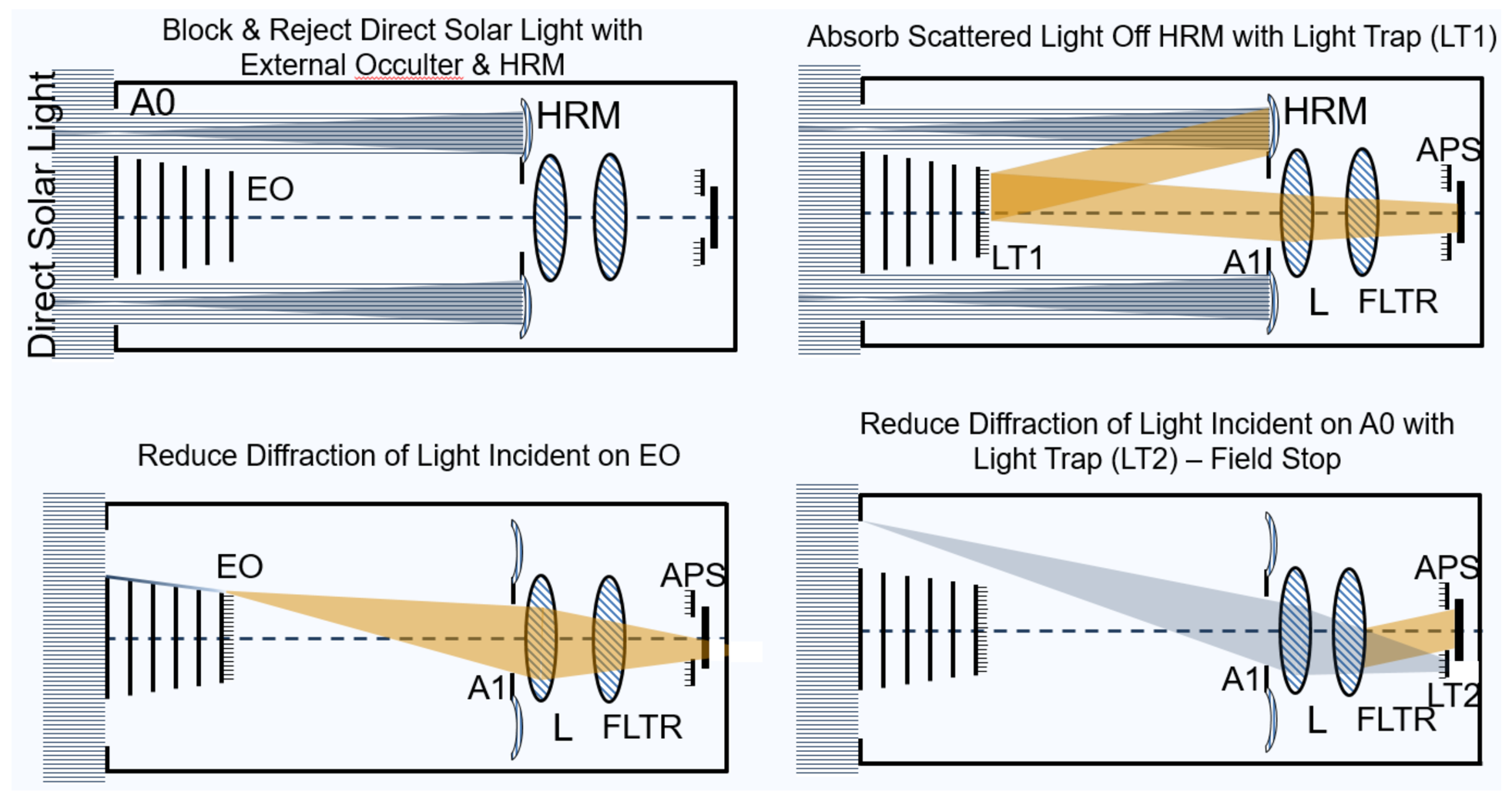}
\caption{Main stray light paths in a compact coronagraph.}\label{fig_strayLightDiagram}
\end{figure}

Figure \ref{fig_strayLightDiagram} describes the main stray light paths in the instrument. In the first panel, the occulter blocks direct sunlight and prevents it from entering the A1 pupil of the telescope. The sunlight passing around the occulter and through the A0 is reflected by the Heat Rejecting Mirrors (HRMs) located at the bottom of the stray light tube. On the second panel, most of the sunlight reflected off the HRMs leave the telescope, however, a small fraction then reflects off the back of the occulter. In the third panel, the occulter does not fully block the sunlight; a residual diffracted light still makes it to the detector and contributes to the stray light background. Finally, the last panel shows that A0, which also sits in full sun, will also contribute to the stray light background. We address these stray light paths in the next sections.

\subsubsection{Diffraction}\label{sec_diffraction}
The CCORs use a multiple-disk occulter (19 for CCOR-1, 24 for CCOR-2) that is optimized on the basis of the target IFOVCO and the diffracted stray light level that the instrument needs to achieve to detect the CMEs. Detector dynamic range, fabrication tolerances, alignment (internal to the instrument and instrument alignment on the spacecraft), vibration loads, thermal distortion, and spacecraft pointing accuracy are all considered in the design. The disk size design follows what is described in \cite{Buffington_10.1117/12.330266}, \cite{Thernisien_2005}, \cite{Thernisien_2018SPIE10698E..0ET}. The occulter disk lateral cross section profile has an ogive shape, as described in \cite{Deforest_2025ApJ...982...58D}. Each disk sits in the shadow of the disk it precedes. The angle between the first two disks, noted $\alpha_{12}$, must be $> 1 R_\odot$ ($\approx0.266 \degree$ if at 1 AU). The angle between the last disk and A1 defines the Inner Field Of View Cutoff (IFOVCO), as explained in Section \ref{subsec:IFOVCO} and Figure \ref{fig_IFOVCO}. For our design, the angular step between the first two disks and the IFOVCO is simply $(IFOVCO - \alpha_{12})/(n-1)$, with $n$ being the number of disks.

\cite{Buffington_2000ApOpt..39.2683B} demonstrated by test that adding more disks on the occulter always improved the diffraction rejection, up to the point where you have a smooth surface, equivalent to an infinite number of disks. Details about the theory can be found in \cite{1988SSRv...47...95K_koutchmy}, \cite{Aime_2020A&A...637A..16A}, and in \cite{Deforest_2025ApJ...982...58D}. \cite{Gong_Socker_2004} studied the rejection of a three-disk occulter using the General Laser Analysis and Design (GLAD) diffraction code. This was used to design and optimize the SECCHI COR-2 occulters. As the CCOR design was derived from the COR-2 design, we attempted to use the same software and code to predict the performances of the CCOR occulter but found that it was unpractical for CCOR because of the number of disks (19 for CCOR-1); the computation took several hours per case, and the results seemed too good when compared to test results. We found that an experimental and semi-empirical approach was more suitable, as it also allowed us to learn about the alignment procedure and side effects that are not fully predicted by the perfect-world diffraction theory. 

During the development of CCOR we have demonstrated by test that more occulter disks always yielded better rejection. We tested occulters that had up to 60 disks, but found that the improvement on the rejection becomes marginal at some point. \cite{Deforest_2025ApJ...982...58D} showed that indeed the rejection tends towards a limit and is not infinite when the number of disks becomes infinite (smooth occulter). \cite{Bout_2000ApOpt..39.3955B} tested a multi-disk occulter shaped as a truncated cone (frustum) for LASCO C2 (160 disks, or threads), which was then used as the external occulter for the flight model of LASCO C2. They also measured that a smooth occulter (infinite number of disks) yields the best rejection. For the CCORs, we selected a number of disks that were sufficient to achieve the desired performances.

\begin{figure}[h]
\centering
\includegraphics[width=0.82\textwidth]{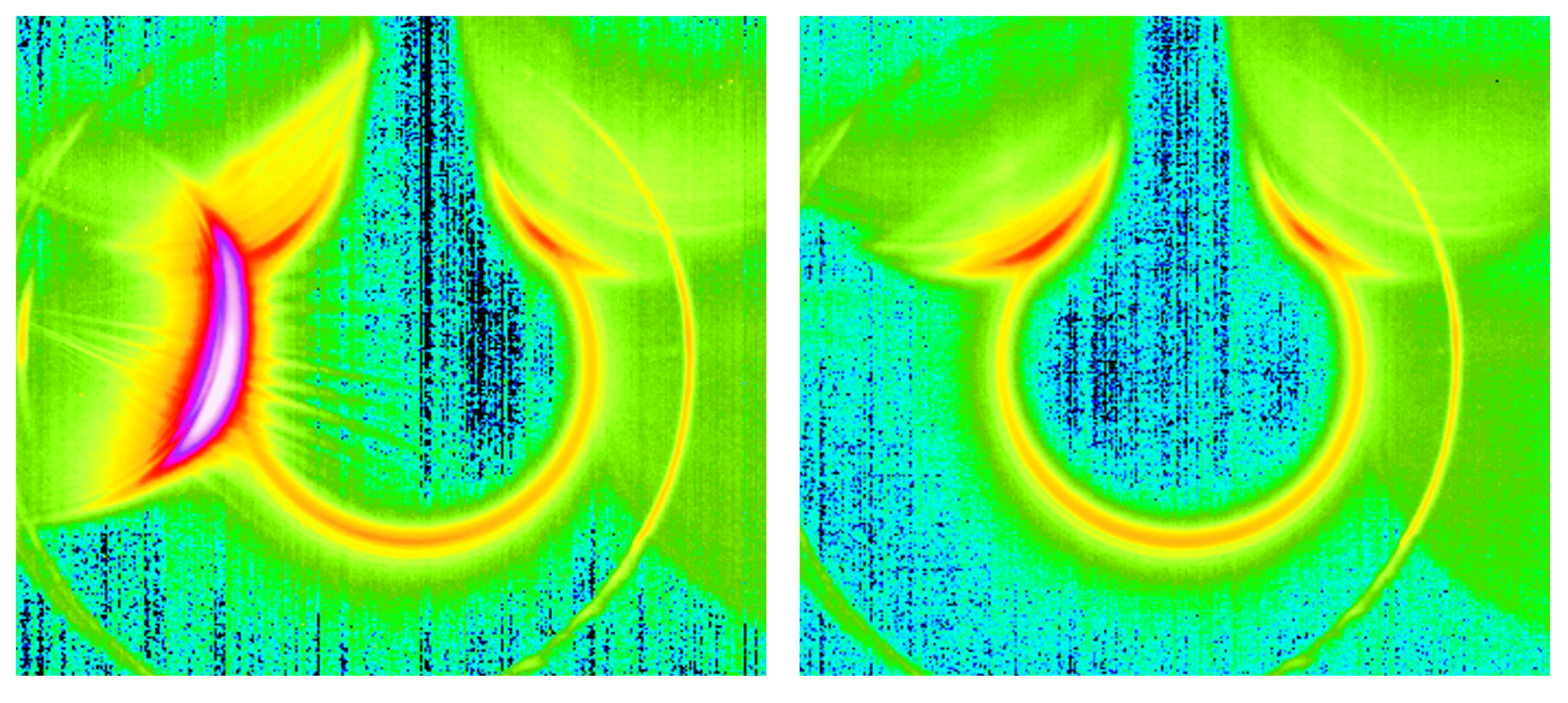}\includegraphics[width=0.08\textwidth]{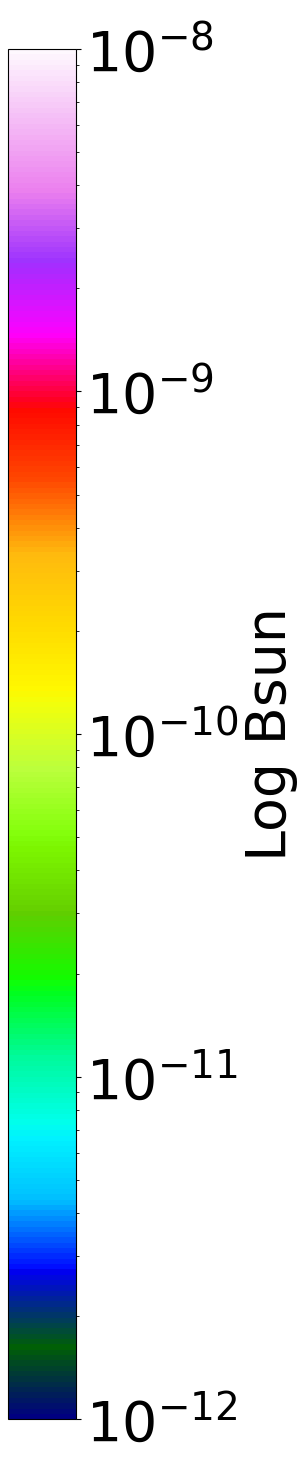}
\caption{Diffraction pattern measured on the ground for CCOR-1. On the left, there was a contamination clearly visible on the left of the occulter. This only showed up in vacuum. On the right, after cleaning, the stray light pattern disappeared.}\label{fig_glint}
\end{figure}

A point we also considered in the occulter design is the susceptibility to dust contamination. The probability of dust contamination increases as you increase the number of disks. A single particle, about the size of a few wavelengths, stuck on one of the disks and sticking out in full sunlight can create a strong stray light signature on the detector. This kind of contamination occurred during the testing of CCOR-1, as shown in Figure \ref{fig_glint}. A small particle stuck on one of the occulter disks created a strong stray light pattern when illuminated by the solar simulator. This only appeared in vacuum; the pattern was not present in air. After local cleaning of the occulter, the pattern disappeared.

Returning to the occulter design, we should also note the innovative design of the METIS coronagraph \citep{Romoli_metis_2017SPIE10563E..1MR, METIS_2020A&A...642A..10A}. Based on the Babinet's principle of reciprocity of diffraction of an obscuration and a clear aperture, METIS is using an inverted external occulter. \cite{Landini_2012_10.1117/12.926225} optimized the occulter design, which is a polished hollow frustum, equivalent to an infinite number of disks.

ASPIICS \citep{shestov_aspiics_2021A&A...652A...4S} is yet another notable innovative coronagraph, not so much by its design, which follows the traditional Lyot design, but by its size. It is a formation flying coronagraph in which an external occulter is held by a spacecraft posted 150 meters from a mother/camera spacecraft. The theory and numerical computation of a 2-disk occulter case related to ASPIICS are treated by \cite{Aime_2020A&A...637A..16A}.

\begin{figure}[h]
\centering
\includegraphics[width=0.82\textwidth]{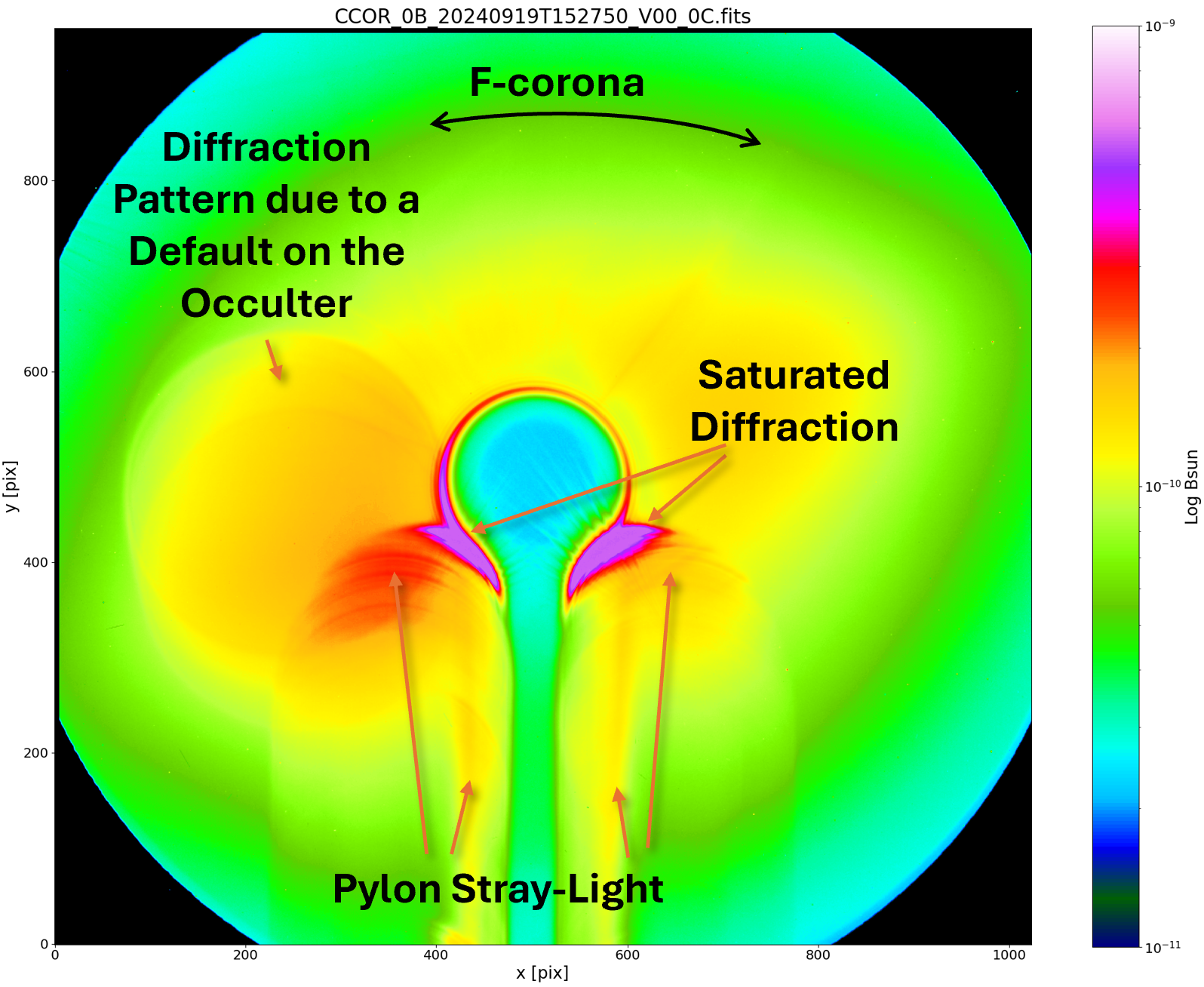}
\caption{CCOR-1 flight image ($2 \times 2$ binned) showing the various stray light features observed in the raw data.}\label{fig_circulardefault}
\end{figure}

Looking now at the measured performances of the flight instruments. For CCOR-1, there is a known imperfection on the last disk of the occulter that creates a circular pattern of stray light, as shown in Figure \ref{fig_circulardefault}. This was detected during the first stray light test in vacuum. Although the stray light is higher in that region, the requirements are still met, so no corrective action was taken. We believe that this is due to the optical coating that is applied on that last disk. This is a sprayed paint and, although the edge was masked, there is probably a very small imperfection of that paint that sticks out. The diameter of the circular pattern it produces on the detector is consistent with the blur diameter of a point source located on the edge of the occulter last disk. Note that with a traditional Lyot design, this would probably be intercepted by the internal occulter. Because there is no internal occulter in CCOR, particular attention is needed to the fabrication process and cleanliness of the occulter and pylon.

\subsubsection{Pylon}\label{subsec:pylon}
The function of the pylon is to keep the occulter in the center of A0 and to keep it aligned with A1. This alignment is optimized during instrument integration and test, but it needs to survive launch and the flight thermal environment. The pylon is designed to ensure these two functions, but it also sits in full sunlight and is imaged, although de-focused, on the detector. As a result, there is residual sunlight diffracted on the edges of the pylon that is imaged on the detector, which contributes substantially to the stray light in that region.

\begin{figure}[h]
\centering
\includegraphics[width=0.99\textwidth]{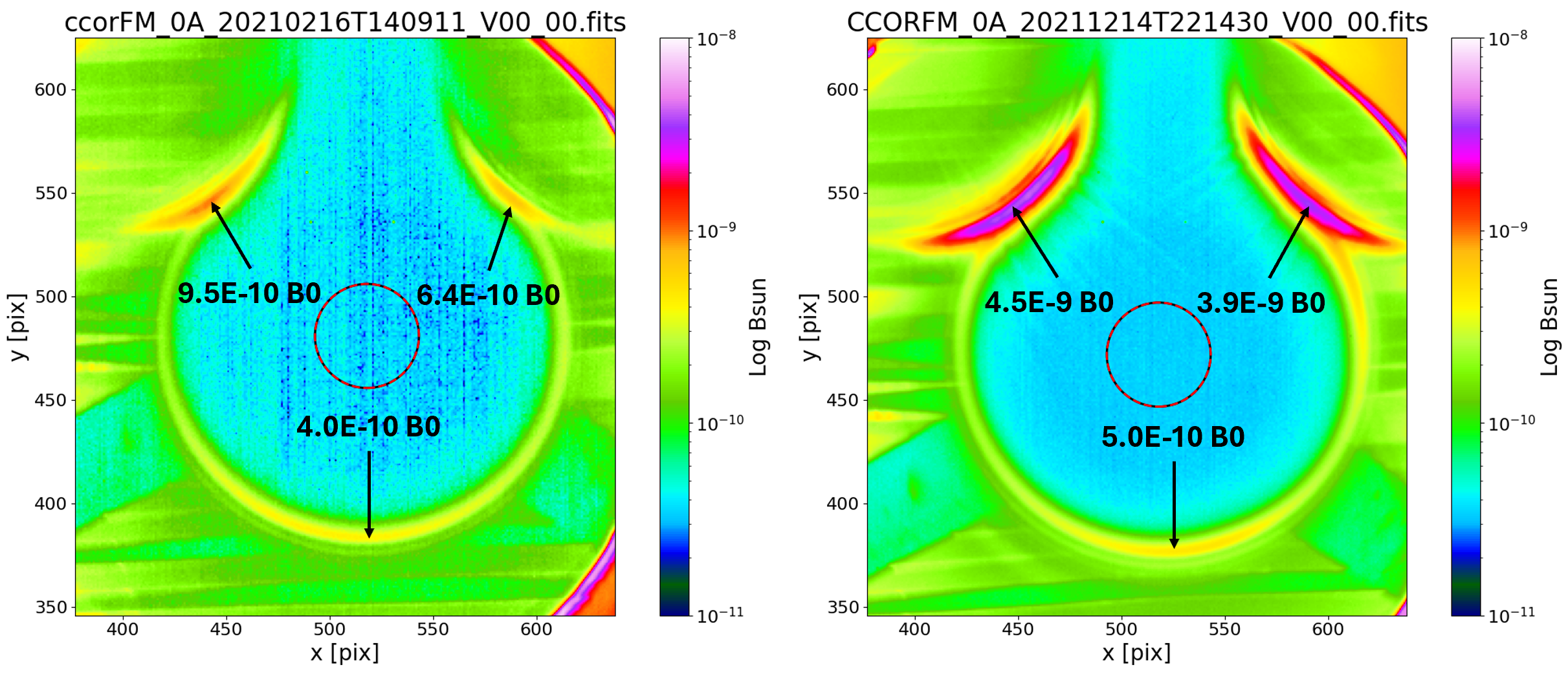}
\caption{Diffraction pattern measured on the ground for CCOR-1, before and after the vibration test. The horizontal patterns are features of the test chamber. Note that these images are $2 \times 2$ binned. The circle at the center is 1 Rsun radius.}\label{fig_ccor1diffractionpattern}
\end{figure}

For CCOR-1, the occulter moved during the vibration test. This resulted in an increase in the level of the diffraction pattern around the occulter, on the pylon side. This is shown in Figure \ref{fig_ccor1diffractionpattern}. The brightness of that pattern increased by a factor of 5.5 on average, which is sufficient to saturate the detector in that region. As a result, there will be no signal in this region. The optimal sun-vector pointing of the instrument was slightly corrected, but this was not sufficient to remedy the problem. Since this is a small area of the field of view, it was decided to use the instrument as is and not attempt a time-consuming and costly repair at that point.

\begin{figure}[h]
\centering
\includegraphics[width=0.55\textwidth]{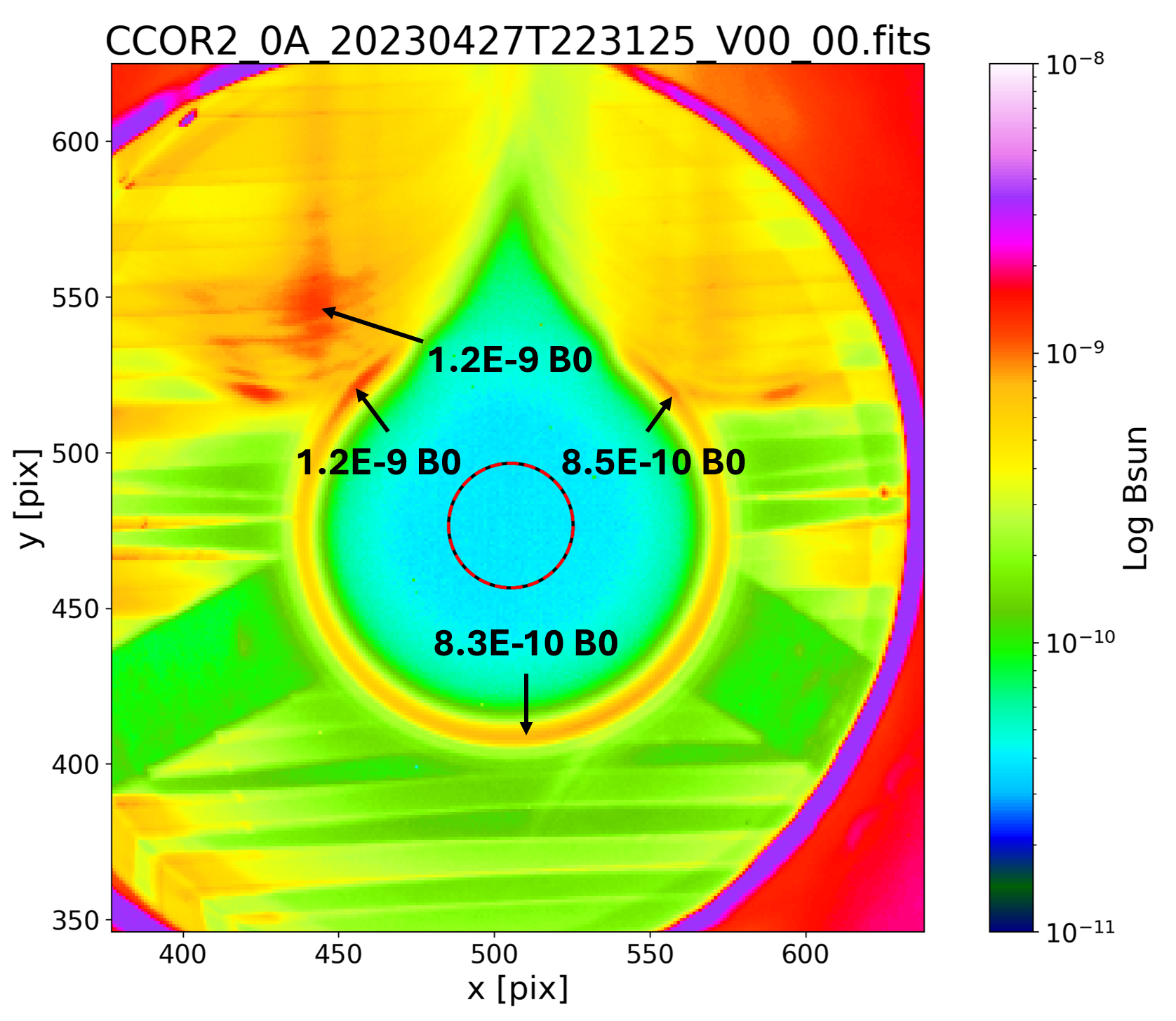}

\caption{Diffraction pattern around the occulter for CCOR-2, measured on the ground. The circle at the center is 1 Rsun radius. The image is 2x2 binned. The red areas in the corners are reflections inside the test chamber. The horizontal lines are also features of the test chamber. }\label{fig_ccor2diffractionpattern}
\end{figure}

For CCOR-2, there was an error in the process applied on the edges of the pylon, which also resulted in a strong stray light pattern next to the occulter on the pylon side. This is shown in Figure \ref{fig_ccor2diffractionpattern}. Similarly to CCOR-1, this region will be saturated, but as it is only a small portion of the field of view, there is no impact on the ability of the instrument to detect CMEs, so it was decided to proceed as is, since it would have been too costly to wait for the pylon to be reworked.

The pylon creates strong vignetting, as shown in Figure \ref{fig_ccorVignetting}. There was no requirement for CCOR to meet the S/N in a sector angle of $\pm 35 \degree$ around the pylon. Although this region presents a higher stray light level, the corona can still be imaged there, with a slightly degraded S/N. Behind the pylon, there is a fully vignetted zone where no coronal signal is imaged, especially on CCOR-1.

\subsubsection{Heat Rejection Mirrors}\label{subsubsec:HRM}
The sunlight that passes around the occulter and through the A0 ends up on the Heat Rejection Mirrors (HRMs) and is then reflected outside the instrument. Approximately 98\% of the incoming light energy is reflected back into space. The focal of the HRMs is optimized to keep the specular reflected sunlight as far as possible from the occulter. Compared to LASCO C3, SECCHI COR2 or PUNCH NFI, CCORs use a three-segment HRM instead of a single torus-shaped HRM, as shown in Figures \ref{fig_ccor1cad} and \ref{fig_ccor2cad}. This was necessary due to the size of the occulter, the distance of the last occulter disk to A1, and the diameter of A0. Specular rays would hit the occulter if a single toroidal HRM was used. This would illuminate it with full sunlight, which would be disastrous in terms of stray light level.

The top right of Figure \ref{fig_strayLightDiagram} illustrates the scattering of sunlight on the HRMs. A fraction of that scattered light hits the back of the occulter, which is imaged, although defocused, on the detector. Note that this stray light pattern is only present in the vignetted zone of the field of view. To minimize the light level reaching the detector on that path, HRMs are polished down to 10 \AA{} RMS roughness \citep{Dittman_K-correl_2006SPIE.6291E..0RD}. In addition, more HRMs were fabricated than needed, which allowed a part selection to be performed based on the measured Bidirectional Reflectance Distribution Function (BRDF) \citep{Nicodemus_BRDF_1965ApOpt...4..767N} of each HRM.

As the light scattered off the HRMs illuminates the back of the occulter and pylon, there are light traps implemented in the back of these two parts. The light trap in the back of the occulter is an inverted cone coated with a specular black paint. The light trap in the back of the pylon has a W-shape and is also coated with specular black paint.

\subsubsection{A0}
The A0 has a serrated shape to divert diffracted light away from the A1. If it were circular, it would concentrate the diffracted sunlight in the A1. This is a typical design for the A0 of coronagraphs, as described in \cite{1988SSRv...47...95K_koutchmy} and \cite{Bout_2000ApOpt..39.3955B}. The edges of the A0 are polished to reduce residual diffraction that would be due to the surface roughness. 

The diameter of the lens assembly is slightly larger than what the field of view requires, to allow the ray bundle coming from the A0 to pass through the lens without interfering with the lens barrel or the lens retainers, which would create additional stray light. This is especially visible in the CCOR-2 lens design (see Figure \ref{fig_lensraybundle}), as the two last elements have a larger diameter than the previous ones.

Finally, a mask on the detector is sized to intercept the rays coming from the A0, to avoid any of these rays being imaged on the detector. This mask acts as a field stop.

\subsubsection{Stray Light Tube}
The stray light tube, or Stray-Light Assembly Structure, connects the A0 to the telescope. It has a series of baffles that are sized so that the A1 only sees the back side of these baffles, but never the cylindrical tube surface interior. In addition, the door aperture only sees the front of the baffles. This is a typical design that is described in \cite{Breault_10.1117/12.948095} and \cite{JungSun_2000}, for example. The interior of the stray light tube is coated with a dark matte coating.

The HRMs scatter a fraction of the sunlight on the back of the stray light tube baffles, but these are sufficiently outside the field of view that this contribution gets trapped in the lens barrel; only a very small fraction of that path contributes to the overall stray light budget on the detector.

\subsubsection{End Cap and Mitigation of the Earthshine on CCOR-1}\label{sec::earthshine}
CCOR-1 is on GOES-19, a geostationary satellite. In that orbit, the Earth passes near or inside the CCOR field of view every day (see Section \ref{sec::GOES_accommodations} and Figure \ref{fig_EclipseTiming}). It will induce strong and variable stray light patterns in the images. To mitigate this, an extension of the instrument front, called end cap, was implemented. The end cap connects the A0 to the door rim (see Figure \ref{fig_ccor1cad}). It prevents any light coming from angles of elongation larger than $45\degree$ from entering directly the A0. 

\begin{figure}[h]
\centering
\includegraphics[width=0.49\textwidth]{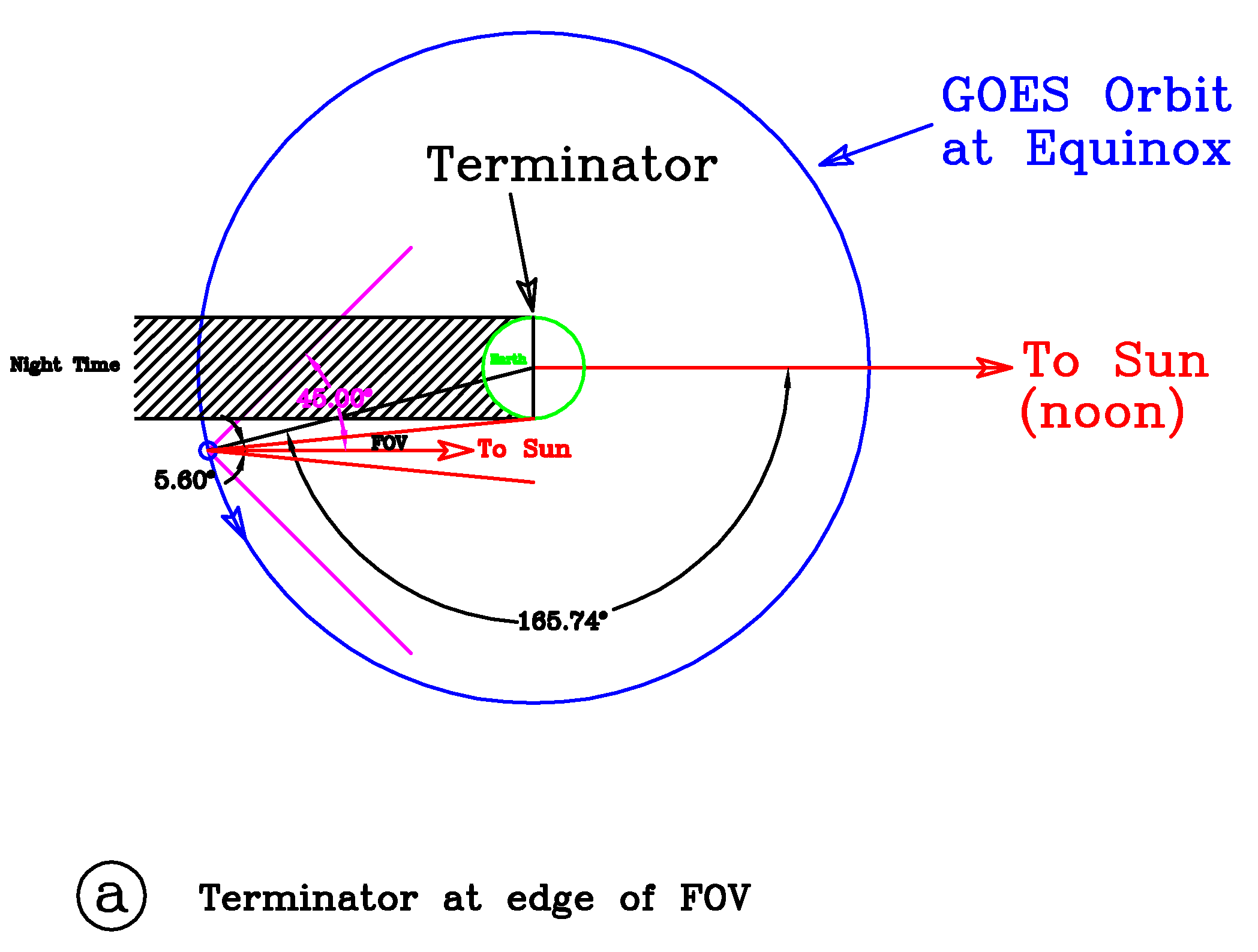}
\includegraphics[width=0.45\textwidth]{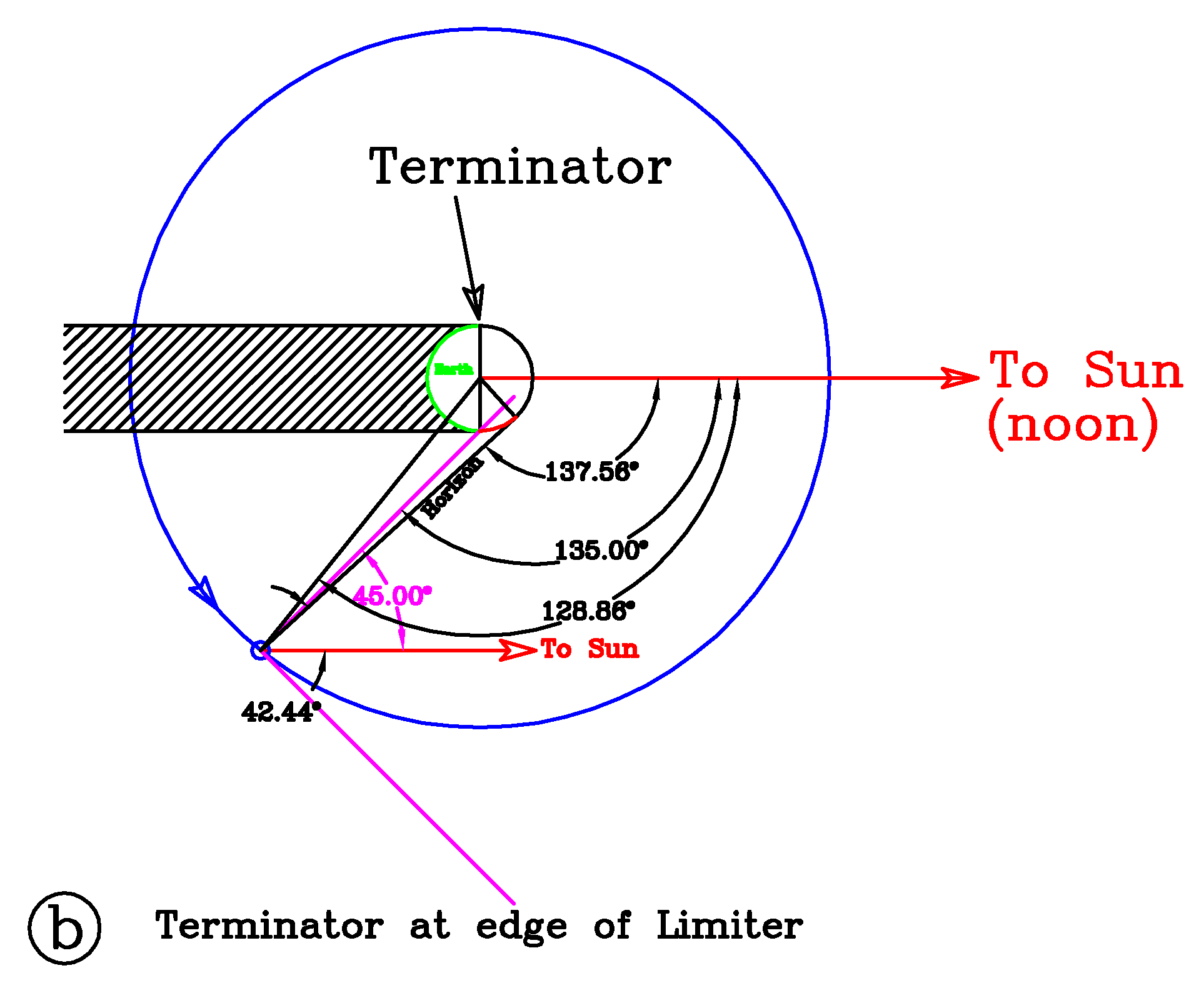}
\includegraphics[width=0.49\textwidth]{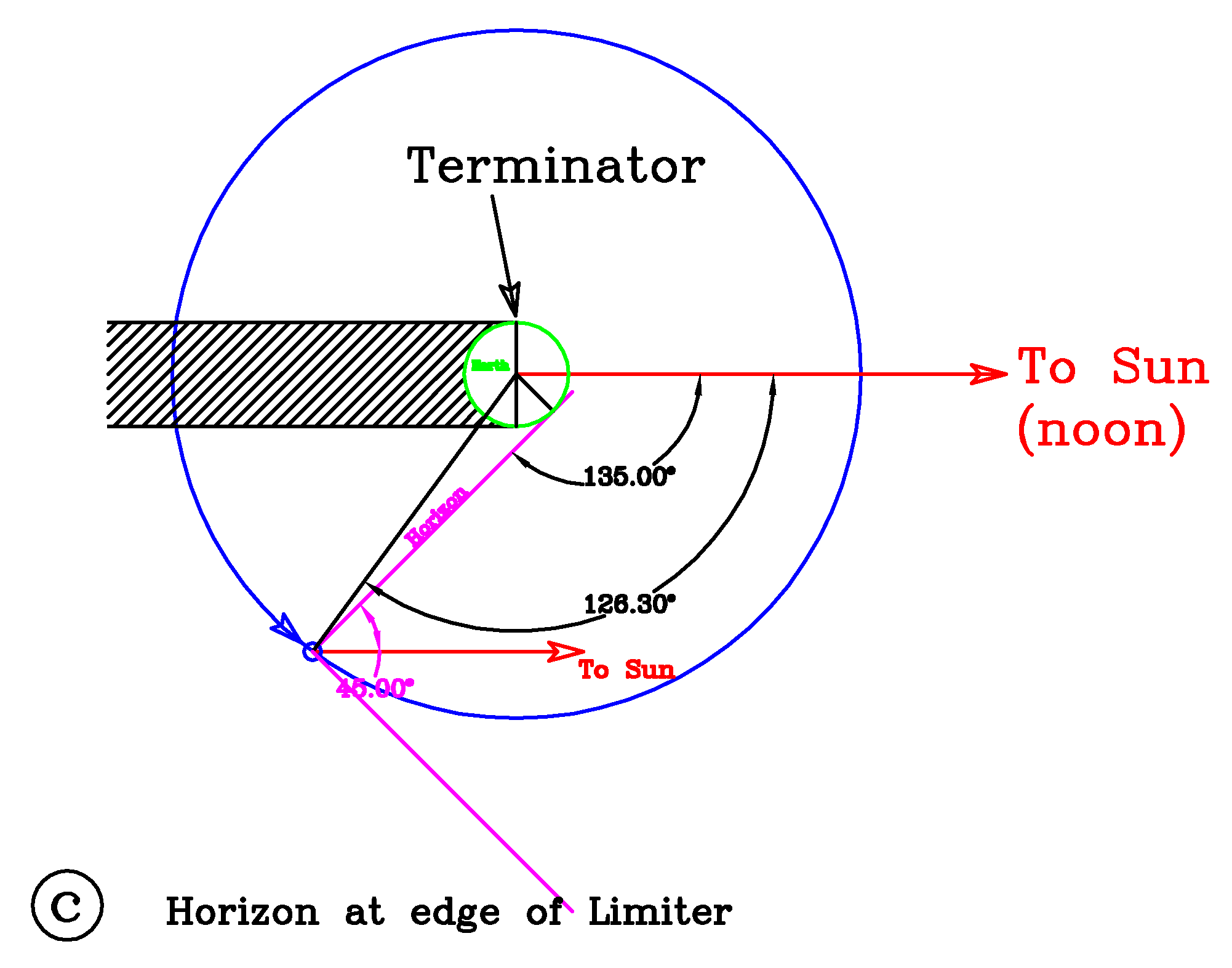}
\caption{Limits of the earthshine entering CCOR-1 function of the three GOES-19 orbital positions.}\label{fig_goes_limits}
\end{figure}

\begin{figure}[h]
\centering

\includegraphics[width=0.9\textwidth]{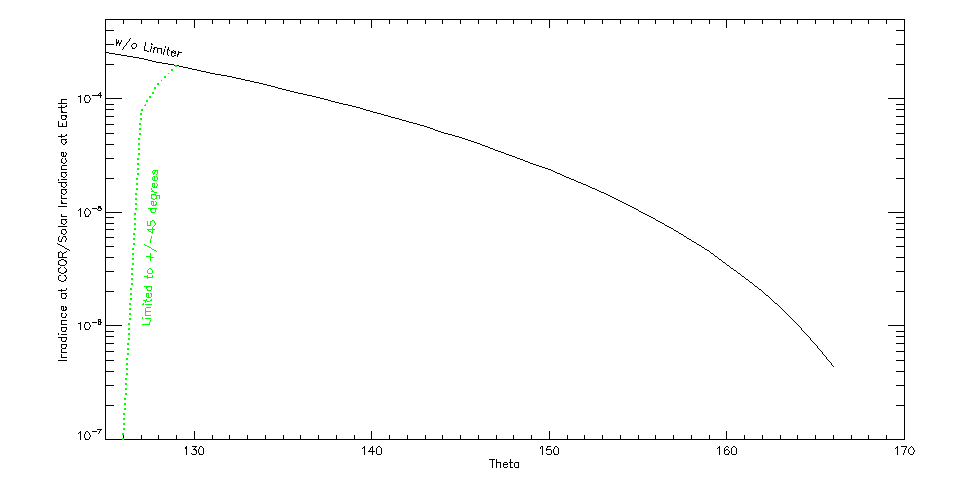}
\caption{Earth irradiance received in front of CCOR-1 function of the angle to orbital noon ($0 \degree$). The green line shows the attenuation provided by the end cap compared to the no end cap case (labeled W/O Limiter), below $126\degree$. When CCOR gets closer to orbital midnight ($180\degree$), the Earth radiance decreases up until it would drop to zero during the eclipse. }\label{fig_EarthirradianceatCCOR}
\end{figure}

Figure \ref{fig_goes_limits} shows the position of the Earth with respect to the end cap / limiter and the CCOR FOV, versus three orbital positions. Figure \ref{fig_EarthirradianceatCCOR} shows the earthshine irradiance received onto CCOR's A0 aperture function of the orbital position angle. In the two figures, the orbital position angle origin is taken at orbital noon. The irradiance is calculated assuming that the Earth is a Lambertian sphere with an albedo of 40\%. The vertical green line in Figure \ref{fig_EarthirradianceatCCOR} demonstrates the attenuation provided by the end cap compared to the case without the end cap (labeled W/O Limiter), below $126\degree$. Above $126\degree$, earthshine enters the A0 and stray light levels in the image plane become substantial, even though the earthshine irradiance decreases as the spacecraft approaches orbital midnight. Around orbital midnight, the earthshine drops to zero as the Earth fully eclipses the Sun. Note that from the geostationary altitude the angular radius of Earth is $8.7\degree$.

Modeling using Monte Carlo ray tracing of the earthshine was done and allowed estimating the impact of this effect before flight. The first flight data confirmed the results of the modeling, although in some cases the levels were an order of magnitude higher than initially modeled. Accurately modeling the earthshine is a difficult task: the Earth is never at the same place, the Earth brightness depends on the cloud coverage and sunlight angle of incidence on the oceans and lands, it creates some diffraction effects that are not modeled in the ray tracing, the optical properties of the instrument, although well known, have some uncertainties, and the geometry of the model might slightly differ from the actual as-built flight instrument. The Lambertian model with 40\% albedo of Earth we used for this study was probably too simplistic also.

Flight data show that earthshine starts to be noticeable on the pylon when Earth is $\leq 65\degree$ from the CCOR boresight, and is noticeable inside the rest of the FOV when $\leq 25\degree$. When the angle becomes $\leq 10.9\degree$, the Earth limb starts to directly illuminate A1; this is when the stray light level of the earthshine reaches levels similar or higher than those of the K + F coronae. When the Earth is inside the FOV, its limb has a radiance level of $ > 10^{-8}$ Bsun, which saturates the detector and creates strong stray light patterns throughout the FOV.

\begin{figure}[h]
\centering
\includegraphics[width=0.9\textwidth]{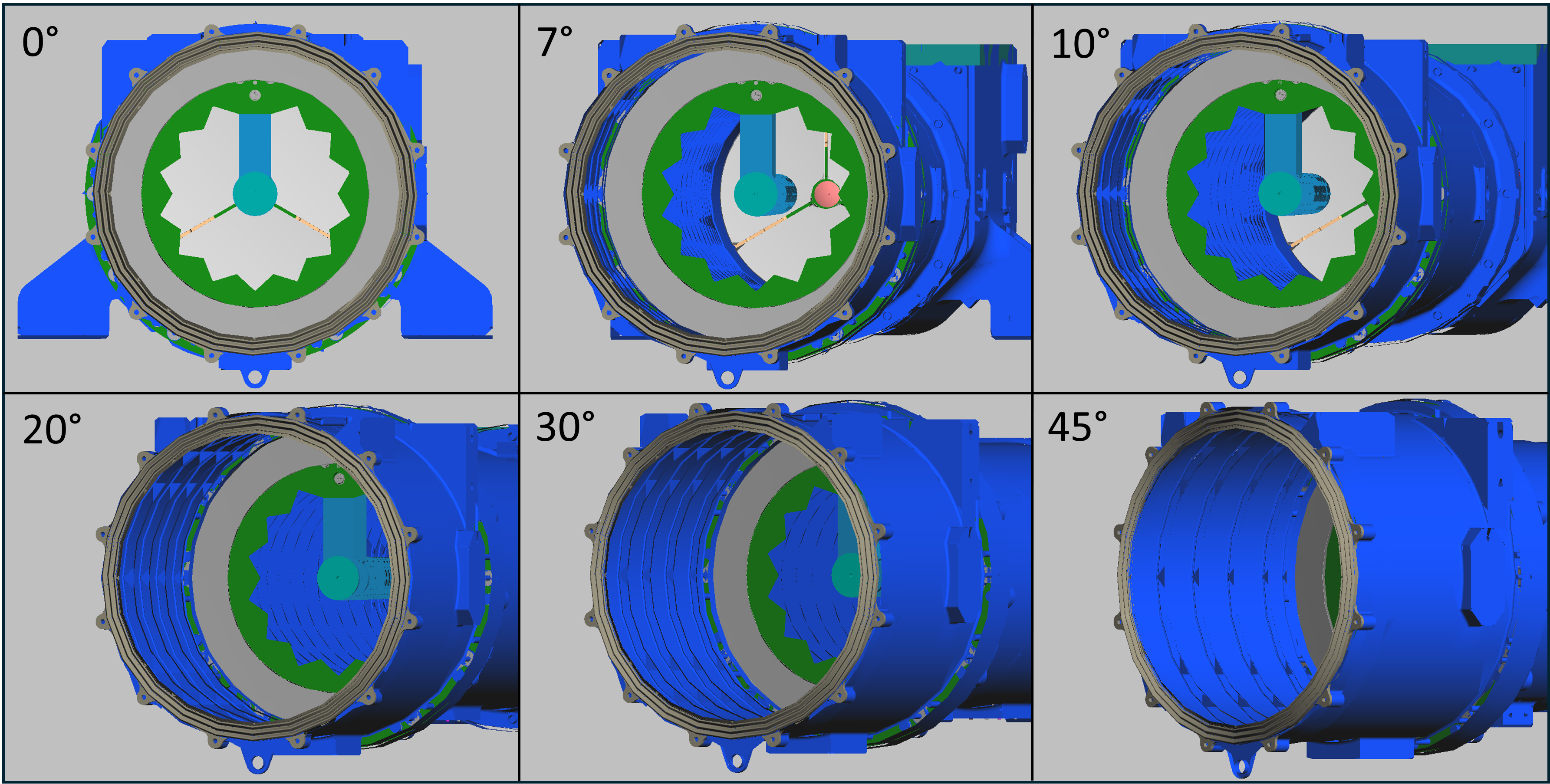}
\caption{Shadowing effect of the end cap function of the elongation from the CCOR-1 boresight/optical axis. }\label{fig_endcapVsEarth}
\end{figure}

One way to visualize how earthshine interacts with CCOR is to look at the CAD model from the perspective of the Earth limb. This is shown in Figure \ref{fig_endcapVsEarth}, in the case of the equinox.

\begin{description}
    \item[$0\degree - 7\degree$:] The Earth limb is imaged on the detector. The Earth disk masks substantial portions of the image. The Earth limb saturates.
    \item[$7\degree - 10\degree$:] $7\degree$ is slightly outside the field of view, but still all internal CCOR structures are illuminated, including the A1 entrance pupil and the objective lens. Although Earthshine is not directly imaged, it bounces inside the objective lens and creates substantial stray light on the detector.  
    \item[$10\degree - 20\degree$:] No more direct illumination of the lens.  Earthshine still illuminates the occulter and the pylon. A glint off the last disk of the occulter will appear on the detector.
    \item[$20\degree - 30\degree$:] The Earthshine mostly illuminates the occulter and pylon, except the last disk.
    \item[$30\degree - 45\degree$:] Illumination of some fraction of the stray light tube.
    \item[$45\degree - 90 \degree$:] No more light passes through A0, but reflections off the end-cap onto the occulter and pylon are still present. This is observed in the flight data.
    \item[$\geq 90\degree$:] No more earthshine as the Earth is behind the door aperture.
\end{description}

The current plan is to use flight data to build an earthshine model function of the orbital angle (in other words, the time of the day) and the angle between the Sun vector and the Earth equator (in other words, the time from the vernal equinox). A model would need to be accurate on the order of the K-corona signal, which is less than 10\% of the total signals. Knowing that the earthshine stray light is highly variable and has high spatial frequencies, this is a very challenging task, but we believe it can be reduced enough to unveil any CMEs that could be occurring during the transits.

CCOR-2 will be at the Lagrange point L1, just like SOHO-LASCO, so there will be no issue with the earthshine.

\subsection{Spacecraft Induced Stray Light}
CCOR-1 is located on the Solar Pointing Platform (SPP) of the GOES-U spacecraft. GOES-U has several structures that could potentially reflect a significant fraction of the sunlight inside CCOR-1; see Figure \ref{fig_goes19}. We modeled the following cases:

\begin{itemize}
    \item Solar Array Wing Assembly scattering sunlight when the SPP is oriented $23 \degree$ upward, away from the deck.
    \item Geostationary Lightning Mapper (GLM) instrument radiator and other deck structures when the SPP is oriented $23 \degree$ upward, away from the deck.
    \item GLM instrument radiator and other deck structures when the SPP is parallel ($0 \degree$) to the deck.
\end{itemize}

None of these paths presented a significant contribution to the overall CCOR-1 stray light budget.

For CCOR-2, we looked at the contributions of the Solar Array top edge, SWiPS, and STIS, which are all in the vicinity of the CCOR-2 front aperture. There is still an end cap on CCOR-2, but it is shorter than that on CCOR-1. Although there is no earthshine concern for CCOR-2, the end cap still helps shadow any reflections from the nearby spacecraft structures and instruments. In the end, the modeling showed that none of these structures contributed significantly to the stray light budget.

\subsection{Stray Light Summary}
\begin{figure}[h]
\centering
\includegraphics[width=0.9\textwidth]{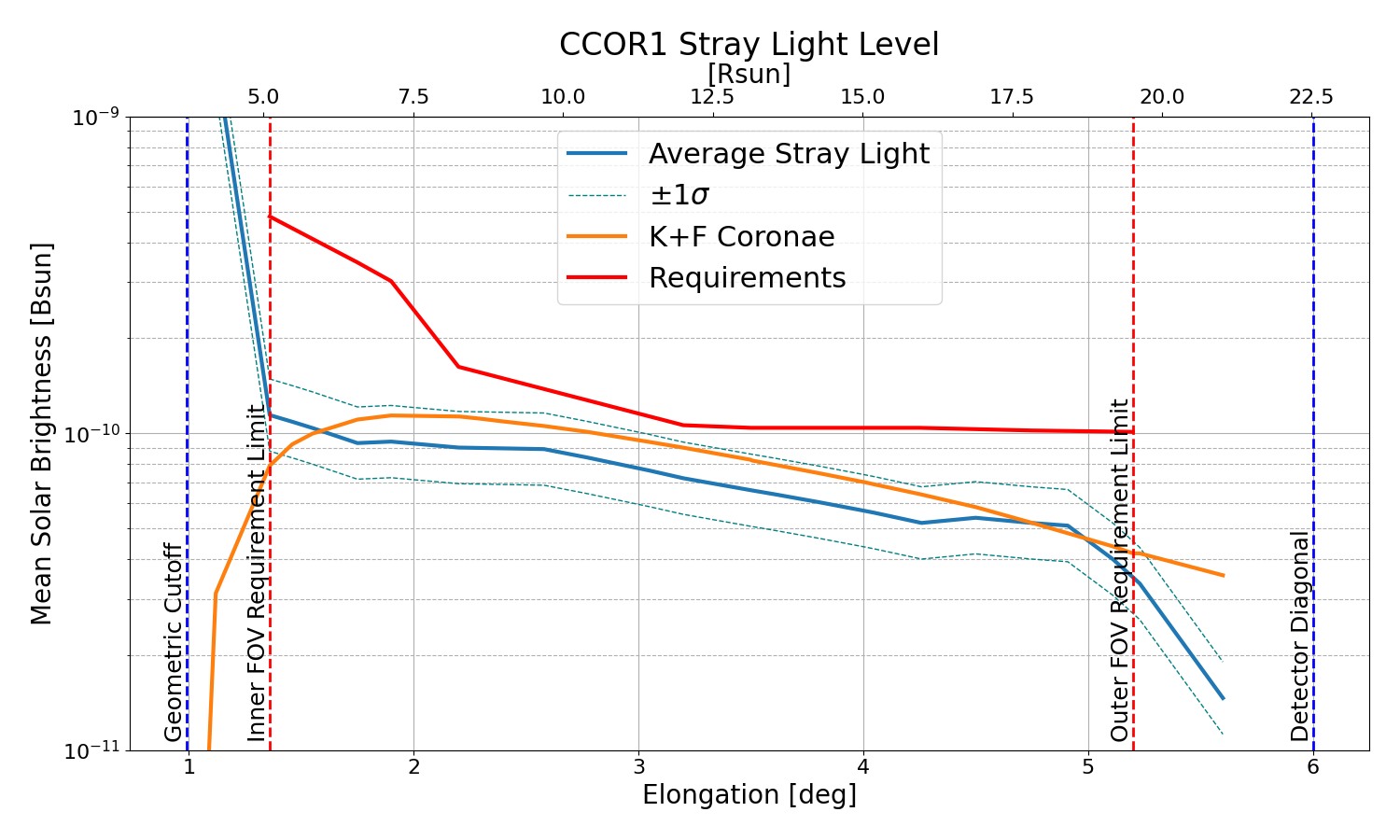}
\includegraphics[width=0.9\textwidth]{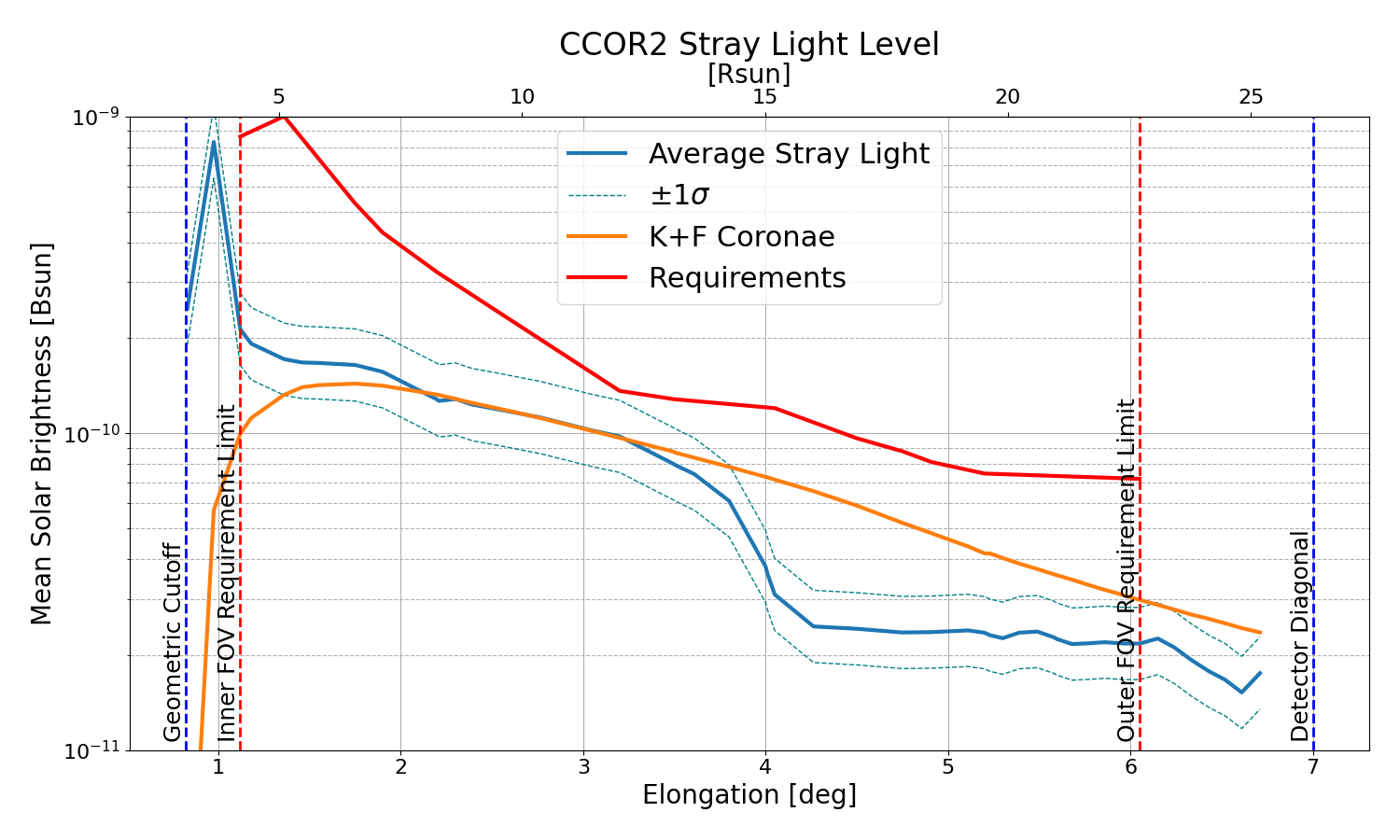}
\caption{Expected stray light level on orbit for CCOR-1 and CCOR-2.}\label{fig_stray lightprofiles}
\end{figure}

We show profiles of the expected stray light level in orbit versus the elongation from the Sun center in Figure \ref{fig_stray lightprofiles}, for both CCOR-1 and CCOR-2. The plots include all the stray light components described above, diffracted and scattered, with the exception of the pylon as we had a requirement exclusion zone of $\pm 35\degree$ around the pylon. The diffraction was measured under vacuum with a solar simulator, as shown in Figures \ref{fig_ccor1diffractionpattern} and \ref{fig_ccor2diffractionpattern}. The scattered components were obtained using a Monte Carlo ray-tracing model of the instrument. The model was correlated on the basis of the vacuum test of the flight instruments fully illuminated by the solar simulator source. The solid red line shows the stray light level requirements. We see that the expected stray light level meets the requirements at all elongations within the required FOV depicted by the vertical red dashed lines.

\clearpage

\section{Mechanical Design}
\subsection{Door}\label{subsec:door}

\begin{figure}[h]
\centering
\includegraphics[width=0.75\textwidth]{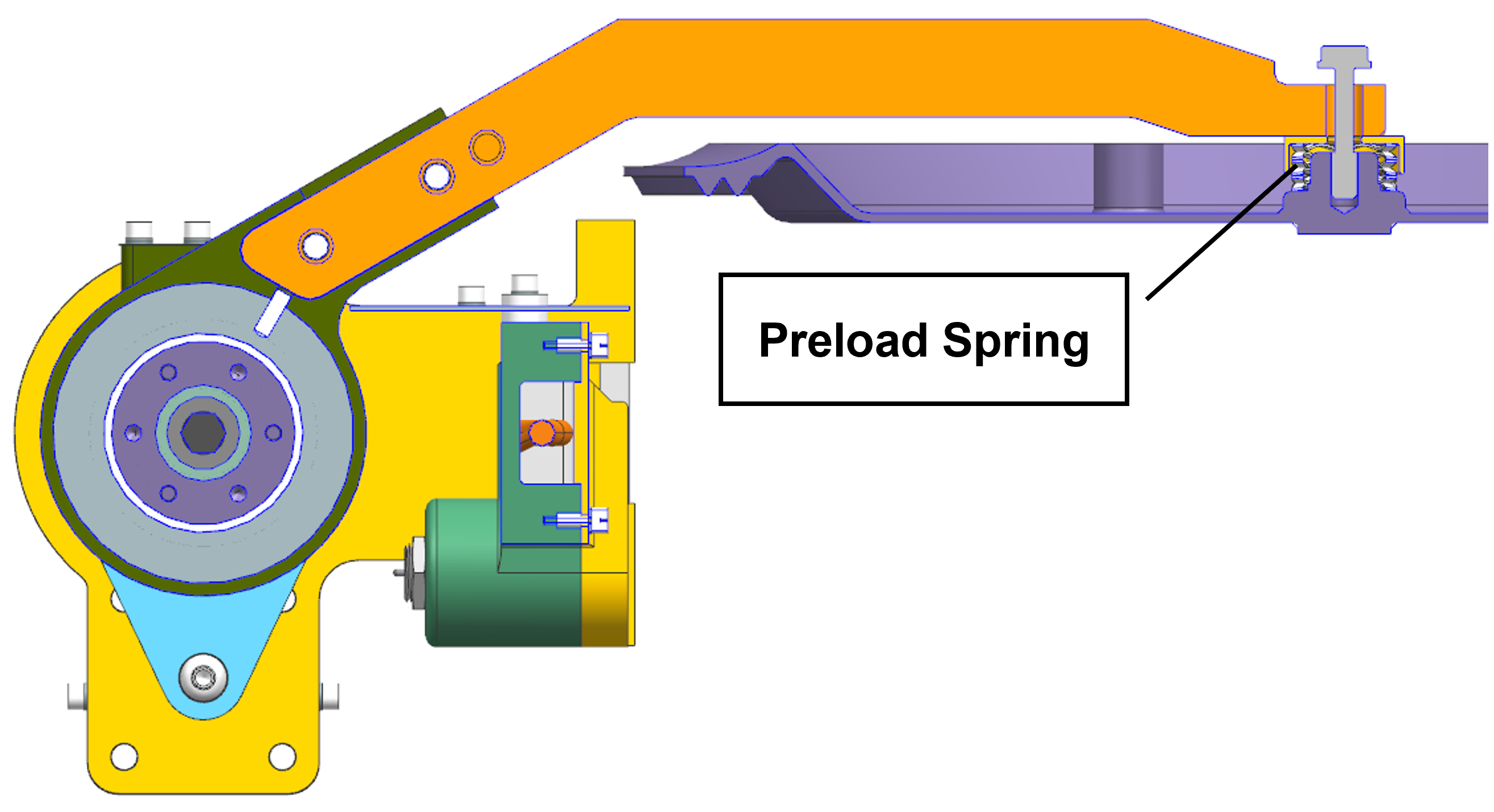}
\centering\includegraphics[width=0.9\textwidth]{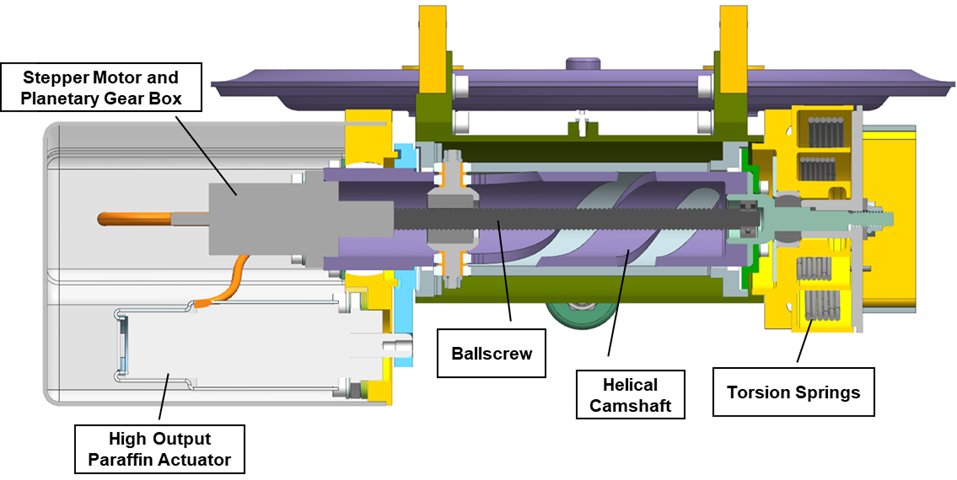}
\caption{Overview of the door mechanism. Top: cross section of one of the door hinge and the lid with location of the door pre-load spring. Bottom: principle of the door mechanism, showing both the re-closable aspect and the fail safe.}\label{fig_door1}
\end{figure}

The CCOR instruments utilize a redundant, re-closable door mechanism (see Figure \ref{fig_door1}). The re-closable aspect of the design only applies to the primary method of actuation. The redundant method of actuation disables the primary and can only be used once. A manual reset of the redundant system is required to re-enable actuation of the primary and redundant actuators. 

The primary, re-closable actuation system consists of a bipolar 2-phase stepper motor, a planetary gear box, a ball screw, and a helical camshaft. This system generates torque, which is then converted to linear motion, and then converted back to torque to actuate the door. The door is preloaded into the structure with springs when closed in order to prevent gaping and/or motion during launch and transportation. In a purely rotational system, this preload force would be reflected back into the motor and a brake would be required in order to maintain the preload. The need for a brake was eliminated by adding the ballscrew and helical camshaft to the system. There is a portion of the cam profile that prevents rotation of the door. When in this portion of the cam the door preload force is reacted into the structure rather than back to the motor. 

The redundant single-use actuation system consists of a high-output paraffin actuator and a set of torsion springs. The primary actuation system is mounted on bushings and has a rotational degree of freedom on the rotation axis of the door. Torque is applied to the primary actuation system and the door using torsion springs. This torque is then retained using the paraffin actuator. When the paraffin actuator is released, the torsion springs drive the door open. Since the paraffin actuator also pins the primary system to the structure, once it is released, the primary system no longer has a way to react forces into the structure, thus rendering it inoperable.

\section{Thermal}
The CCOR instruments have six operational heater zones at five CCOR Instrument Module (CIM) locations to maintain the operational temperature limits for the CCOR detector, optics, electronics, and the metering structure between optical elements. 5 operational thermal modes will be used:

\begin{enumerate}
    \item Nominal Operations: Hot Door Open
    \item Nominal Operations: Cold Door Open
    \item Nominal Operations: Door Closed
    \item Eclipse (except for CCOR-2)
    \item Detector Anneal
\end{enumerate}

The CCOR flight software controls the heaters to maintain a set-point temperature using a Proportional-Derivative (PD) heater controller with a Duty Cycle Estimator. The controller applies heater power over a 64 second time period using pulse width modulation (PWM) such that the average power over this time period will converge on the temperature setpoint and then maintain this temperature at each sample time for the temperature telemetry.

The anneal heater allows the detector to be heated to 30C for evaporating molecular contamination and repair of lattice displacement induced by accumulated radiations through the mission. Detector anneal will be performed at the start of the mission and possibly after 5 years of operation.

The GOES-U and SWFO spacecraft provide survival power to the CCOR instrument and capture the CIM exterior temperature measurements at the CIM survival heater locations, at the CIM conductive interface with the spacecraft deck, and on the exterior of the PSB. There are 3 survival heater zones on the CIM: Endcap, Telescope Tube, and ICB zones. Each survival heater location is controlled by thermostats wired in series to provide redundancy against thermostats failing closed.  Neither CCOR-1 nor CCOR-2 provide redundant survival heaters.  However, CCOR-1 uses separately controlled Bias and Trim heaters in the ICB survival heater zone for more stable control over the range of required survival heater power. The separately controlled Bias and Trim heaters provide limited power redundancy if one of the two heaters fails. CCOR-2 uses Primary and Backup heaters in the Telescope Tube and ICB survival heater zones. The separately controlled Backup heater provides full power redundancy, although at a lower set point, and would augment the Primary heater power if the Primary does not fail but additional power is needed. 

\begin{figure}[h]
\centering\includegraphics[width=0.9\textwidth]{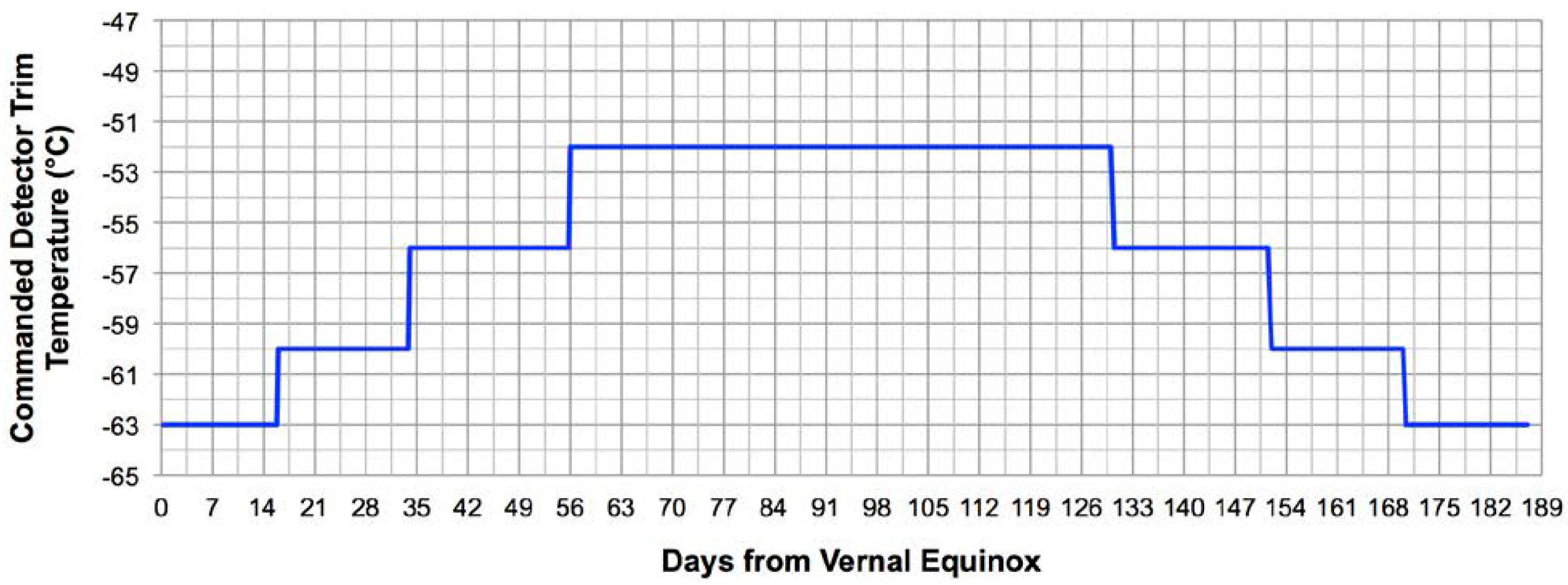}
\caption{Predicted CCOR-1 detector setpoint versus days from vernal equinox. This will be updated based on the actual flight performances. }\label{fig_tempsetpoint}
\end{figure}

There are seasonal variations in the detector temperature of CCOR-1 due to the geostationary orbit of GOES. The predicted temperature set point is shown in Figure \ref{fig_tempsetpoint}. Note that each time the set-point is changed, the detector bias has to be updated, which can result in a slight flicker in the data.

CCOR-2 detector temperature set-point is expected to be fixed. The heat input only depends on the distance to the Sun, which varies from 0.974 to 1.006 AU.

\section{Detector}\label{sec_detector}
The detector is a Complementary Metal-Oxide Semiconductor (CMOS) Active Pixel Sensor (APS) designed and manufactured by SRI International \citep{janesick_10.1117/12.678867}. The same detector is used in the SoloHi Heliospheric Imager \citep{Howard_2020} and the WISPR imager \citep{Vourlidas_2016}. 

The development and qualification of the device to the required Technology Readiness Level (TRL) of 6 is summarized in \cite{Korendyke_2013SPIE.8862E..0JK}. Testing especially included exposure to proton radiation. Quantum efficiency (QE) and gain of the tested devices showed little change after exposure to 100 Krad (Si). The post-radiation dark current showed a significant increase, but it improved after a temperature annealing. 

The specifications of the detector and the performance measured during ground tests of CCOR-1 and CCOR-2 are summarized in Table \ref{tab:detectorparameters}.

The detector is used in high gain, which yields a conversion factor from electrons to Digital Numbers (DN) of about $\approx 2.1$ e-/DN. High gain yields a lower read noise than the low gain mode, and allows a shorter exposure time, although the linear full well is about 22,000 e-, compared to 100,000 in low gain. 

The linear region of the pixels electron well goes from 0.1k e- to 22k e-. The linearity is better than 2.2\% in that region. From 22k e- to about 30k e-, there is still some usable signal, but the linearity is degraded, although it can be partially corrected. Above 30k e-, the signal becomes too close to the saturation level to be properly calibrated. More details on the detector are discussed by \cite{2021SoPh..296...94H} who performed its characterization during the WISPR in-flight calibration. Figure 5 of that article shows the typical linearity of these detectors. 

The detector is passively cooled to operational temperatures ranging from -70C to -40C, yielding a dark current at the Beginning of Life (BOL) of about 2.4 e-/s/pix. Radiation tests were performed for the SoloHI program. Based on the expected Total Ionizing Dose at the End of Life (EOL) of both missions, the expected dark current at EOL is 20 e-/s/pix.

The Detector Readout Board (DRB) readouts the detector row by row. Each row takes 1.28 ms to read and reset. There are 1920 rows, so the whole detector takes 2.4576 s to be fully readout. This means that there is a time delay between the first rows that are read and the last rows. 2.4576 s is the minimum exposure time that the camera can achieve. A shorter exposure time can be achieved if fewer rows are read. A double exposure observing program using a normal full frame exposure for the outer FOV and a shorter exposure for only the inner FOV is actually considered to possibly mitigate the saturation due to the pylon stray light present on CCOR-2. See Section \ref{subsec:pylon} and Figure \ref{fig_ccor2diffractionpattern}.

The first 10 rows and columns are opaque. This allows us to measure the detector bias and column-to-column variations. In operation mode, this bias and column-to-column variation is measured during flight and removed from each of the images that are transmitted to the ground. This permits reducing the data volume and can be reversed on the ground if needed. The detector bias depends on the temperature. For CCOR-1, the detector temperature setpoint will be adjusted based on the seasons (see Figure \ref{fig_tempsetpoint}). Each time the temperature setpoint is changed, a new bias image is taken. This could result in small jumps in the column-to-column variation when looking at movies on the ground. On CCOR-2, no seasonal variations are expected so the detector temperature setpoint should not change, unless some aging effects are noticed after months or years of operation.

\begin{table}[h]
    \centering
    \begin{tabular}{|c|c|c|c|}
        \hline
        Parameter & CCOR-1 & CCOR-2 & BoE\\
        \hline
        Pixels & 2048 x 1920 & 2048 x 1920 & D\\
        Imaging Pixels & 2038 x 1910 & 2038 x 1910 & M\\
        Pixel Size & 10 $\mu$m & 10 $\mu$m & D\\
        Gain & 2.08 e-/DN & 2.06 e-/DN ($\pm$0.06)& M\\
        Gain fine setting & 8 & 8 & S\\
        Gain setting & high & high & S\\
        Linear full well ($\leq$2.2\%) & 25,600 e- & 23,700 e- & M\\
        Saturated full well & $\approx$33,000 e- & $\approx$33,000 e- & M \\
        Read Noise & 8.2 e-/pix & 15.4 e-/pix & M\\
        Dark Current BOL & 2.07 e-/s/pix & 2.37 e-/s/pix & M \\
        Dark Current EOL & 20 e-/s/pix & 20 e-/s/pix & E\\
        Average QE & 32.2\% (470-740nm FWHM) & 32.9\% (469-755nm FWHM)\% & M \\
        Readout rate & 2 Mpix/s & 2 Mpix/s & D\\
        \hline
    \end{tabular}
    \caption{Detector specifications and as tested performances. BoE: Base of Estimate. Legend for the BoE column: D: as designed target, M: measured, S: set point, E: estimated.}
    \label{tab:detectorparameters}
\end{table}

The devices Modulation Transfer Function (MTF) was measured by SRI. They are very close to the theoretical function $sinc$ for a $10\mu m$ pixel matrix whose MTF is 63\% at a spatial frequency of 50 lines/mm. The measured spatial resolution of Table \ref{tab:optical_param} takes into account the detector.

\section{Flight Software}
The CCOR flight software (FSW) runs on a Cobham Gaisler LEON3FT (Field-Programmable Gate Array) FPGA on the Instrument Control Box (ICB) Processor Card (PC). The LEON3 FT has 256MB of RAM, 4MB of magnetic bubble Non-Volatile Random Access Memory (NVRAM), and 64 kB of Programmable Read-Only Memory (PROM) with Error Detection and Correction (EDAC). The LEON3FT FPGA is responsible for all CCOR computer functions, including image scheduling, image processing, command processing, housekeeping, and camera control. The PC provides the SpaceWire interfaces to the spacecraft, the Low Voltage Differential Signaling (LVDS) interface to control and read out the CCOR camera, and a Joint Test Action Group (JTAG) interface and serial ports that are used as test ports on the CCOR Engineering Model and Emulator. The Camera/Mechanism Card (CMC) controls the camera and controls the stepper motor and interfaces with the door encoder to open and close the CIM re-closeable door.

On both CCORs, the Power System Box (PSB) receives power from the satellite power system. It is controlled by the LEON3FT and provides operational power, thermistor readings, door power and heater power. The heaters are controlled by a Proportional-Derivative (PD) algorithm that uses periodic thermistor readings to sense temperature.

\begin{figure}[h]
\centering
\includegraphics[width=0.75\textwidth]{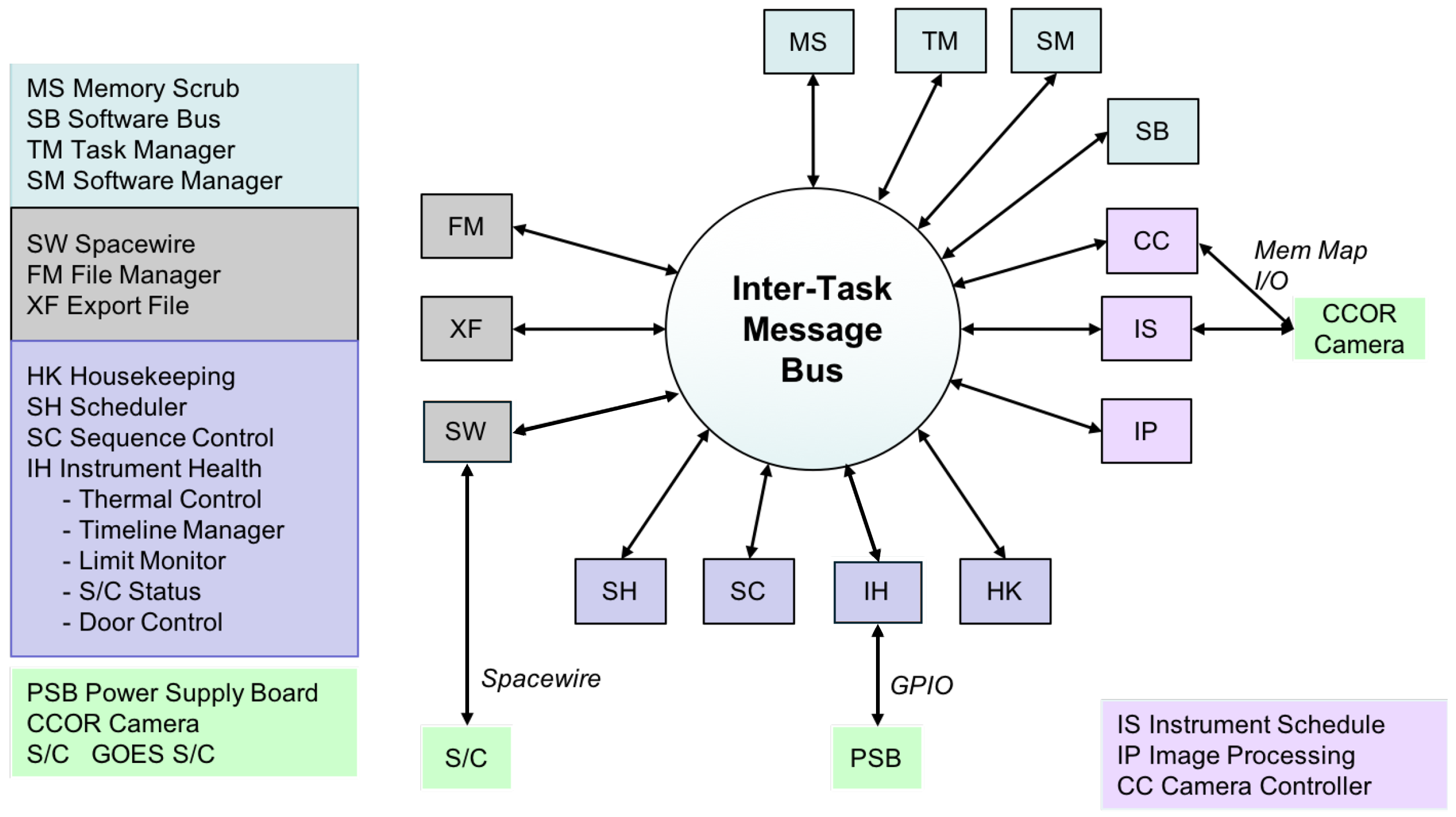}
\caption{CCOR flight software block diagram.}\label{fig_FSBlockDiagram}
\end{figure}

The CCOR flight software is derived from the SoloHI instrument flight software \citep{Howard_2020}, which has a heritage of STEREO/SECCHI. The Flight Software block diagram shown in Figure \ref{fig_FSBlockDiagram} is identical to that of SoloHI. Primary and redundant copies of the flight software are stored in the 4 MB Magnetoresistive Random-Access Memory (MRAM, non-volatile). There is a small 'data vault' in MRAM used to store system parameters that need to be maintained through a power cycle. These hard-coded parameters include: SpaceWire interface (primary or redundant), housekeeping telemetry conversion parameters, heater settings, checksums, and autonomy rule enable status. The 64kB PROM contains software to perform self-tests, checksum calculation, initialization, and a boot loader.

\section{On-Board Processing}\label{sec_onboardprocessing}
The CCOR Nominal Observing program will capture Full-Frame averaged images of the coronal scene with brightness measurement within the CCOR detector linear range at a nominal image cadence of 15 min (except during the scheduled Earth Eclipse periods and Spacecraft yaw-flip maneuvers, specific to GOES-19).

Full frame images are the average of several single exposure images. Multiple exposures are used to perform a Solar Energetic Particle (SEP) or Cosmic Rays (CR) scrub, which is described in \cite{crump_2019RNAAS...3..183C}. Note that detector bias is subtracted for each individual exposure before being processed (see Section \ref{sec_detector}). Each pixel of the sequence of multiple exposures is compared, and the pixels that saturate are removed from the averaging. Based on the number of SEP hits detected, the algorithm can use more single exposure images to perform the scrubbing; the goal is to keep a sufficient number of uncorrupted pixels to maintain the nominal signal-to-noise ratio. The number of single exposures versus storm level is given in Tables \ref{tab:CCOR1_StormIntegTime} and \ref{tab:CCOR2_StormIntegTime}.

\begin{table}[h]
    \centering
    \begin{tabular}{|c|c|c|}
        \hline
        Case &  Number of Images& Integration Time [s]\\
        \hline
        Single image & 1 & 8.5 \\
        \hline
        No Storm & 3 & 25.5 \\
        \hline
        Minor Storm & 5 & 42.5 \\
        \hline
        Major Storm & 7 & 59.5 \\
        \hline
    \end{tabular}
    \caption{CCOR-1 number of averaged exposures and integration time function of the SEP storm level. These numbers are subject to change during flight.}
    \label{tab:CCOR1_StormIntegTime}
\end{table}

\begin{table}[h]
    \centering
    \begin{tabular}{|c|c|c|}
        \hline
        Case &  Number of Images& Integration Time [s] \\
        \hline
        Single image & 1 & 2.7 \\
        \hline
        No Storm & 10 & 27 \\
        \hline
        Minor Storm & 16 & 43.2 \\
        \hline
        Major Storm & 22 & 59.4 \\
        \hline
    \end{tabular}
    \caption{CCOR-2 number of averaged exposures and integration time function of the SEP storm level. These numbers are subject to change during flight.}
    \label{tab:CCOR2_StormIntegTime}
\end{table}

The number of cosmic-ray pixels replaced by the CMC cosmic-ray scrub routines is read from the camera electronics and placed in the image header. This includes pixels scrubbed for all exposures used in the CCOR averaged image.

\begin{figure}[h]
\centering
\includegraphics[width=0.6\textwidth]{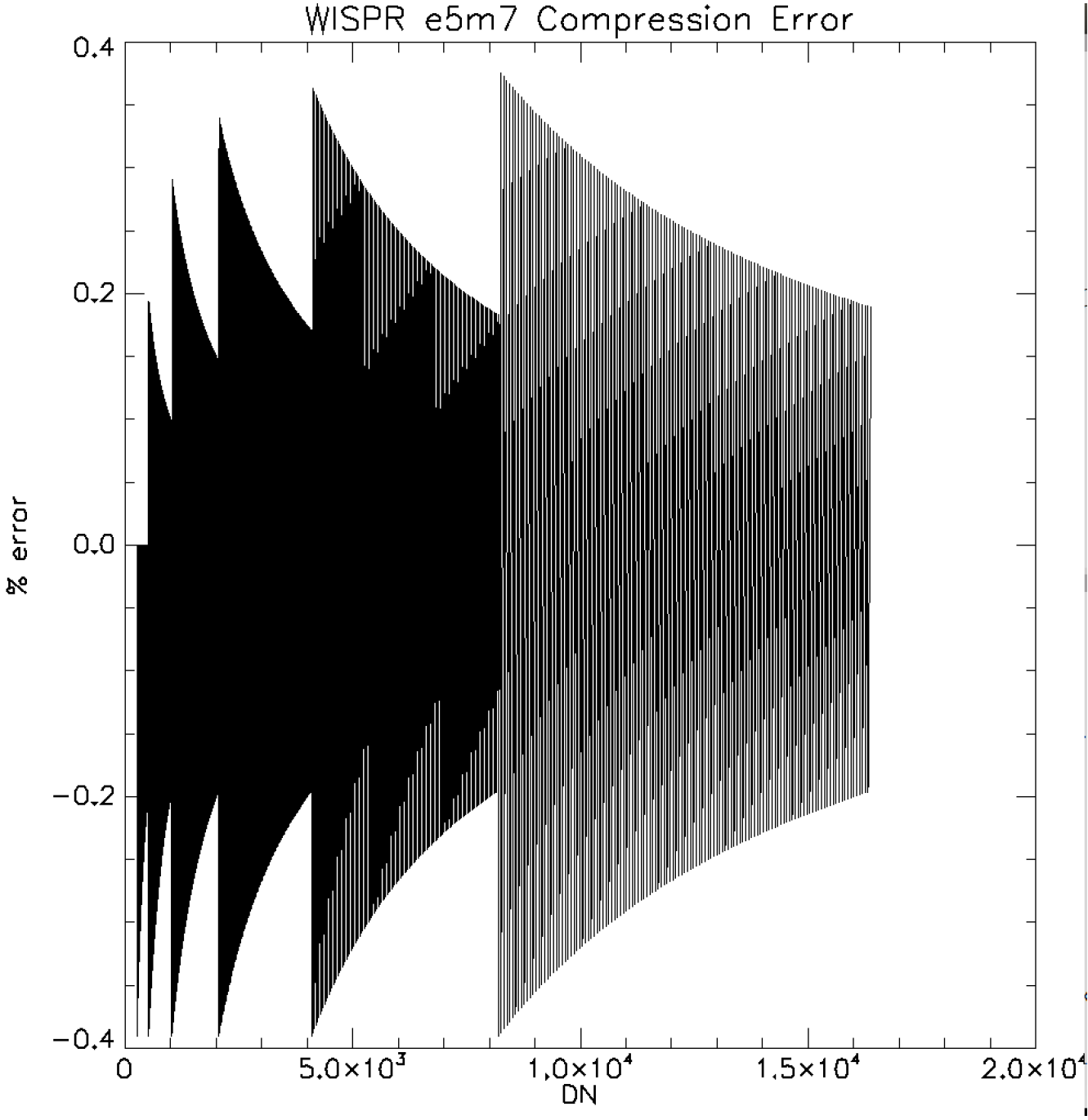}
\caption{Compression Error for the E5M7 exponent and mantissa lossy compression used on both CCORs.}\label{fig_compressionerror}
\end{figure}

After SEP scrubbing and averaging, the Full Frame image is divided into an Inner Field image that will be compressed using Base2Log lossy compression (using an algorithm similar to what is described in \cite{Yang_LogCompress_8660478}), and an Outer Field image that will be compressed using Base2Log lossless compression. The boundaries of the Inner and Outer fields are indicated in Figure \ref{fig_ccorFOVandBlocks}. The compression error versus DN value is shown in Figure \ref{fig_compressionerror}. The lossy compression maximum error is 0.4\%. Figure \ref{fig_ccorFOVandBlocks} shows the Inner Field pixels highlighted in blue, the Outer Field pixels highlighted in yellow, and the Image corner/Occulter pixels that will not be downlinked highlighted in gray. The Full Frame image is reassembled on the ground by the Ground Processing Algorithm, and saved as a FITS file (Level 0b).

\section{Signal to Noise Ratio}
The required signal-to-noise ratio (S/N) is 10, from 5.1 $R_\odot$ to 21 $R_\odot$ radial distances for CCOR-1, and from 4.4 $R_\odot$ to 22.7 $R_\odot$ for CCOR-2. The S/N is evaluated taking into account all signals, vignetting, noises, and errors. The signal of interest is the K-corona, as described in Section \ref{sec_coronalesignals}. The other main signals are the F-corona and the instrument stray light. Both K and F signals are vignetted, but not the instrument stray light. The noises and errors taken into account in the S/N evaluation are: the photon shot noise, the background subtraction error, the compression error (see Figure \ref{fig_compressionerror}), the read noise, and the pixel-to-pixel gain variation and residual bias. 

CMEs are not detected by using a single pixel, but by using large groups of contiguous pixels. The K-corona/CME signal decreases by almost two orders of magnitude in the 5.1 $R_\odot$ - 21 $R_\odot$ range. The CMEs are occupying more pixels in the outer field compared to the inner field. Therefore, we used a pixel binning factor that ranges from 10 pixels at 4.1 $R_\odot$ to 90 pixels at 22.7 $R_\odot$, to compensate for the CME expansion and brightness dimming as it propagates away from the Sun.

In addition to using the pixel averaging, the Solar Energetic Particle (SEP) scrubbing algorithm, described in the previous section (Section \ref{sec_onboardprocessing}) averages several images based on the level of the storm detected. The range of single images averaged are given in Tables \ref{tab:CCOR1_StormIntegTime} and \ref{tab:CCOR2_StormIntegTime}. The image averaging is taken into account in the evaluation of the S/N.

\begin{figure}[h]
\centering
\includegraphics[width=0.9\textwidth]{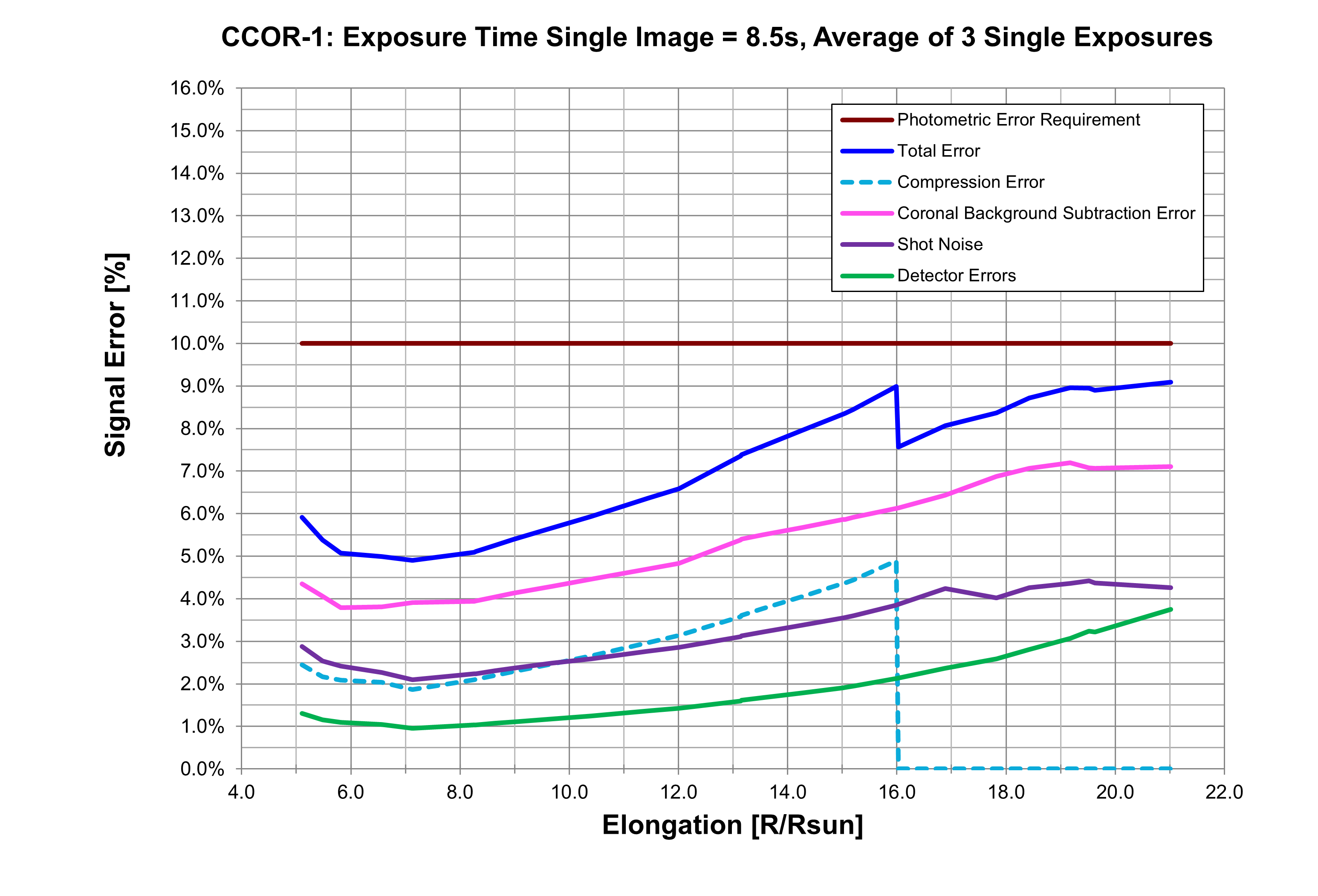}
\caption{Signal error break down for CCOR-1.}\label{fig_CCOR1SNRError}
\end{figure}

\begin{figure}[h]
\centering
\includegraphics[width=0.9\textwidth]{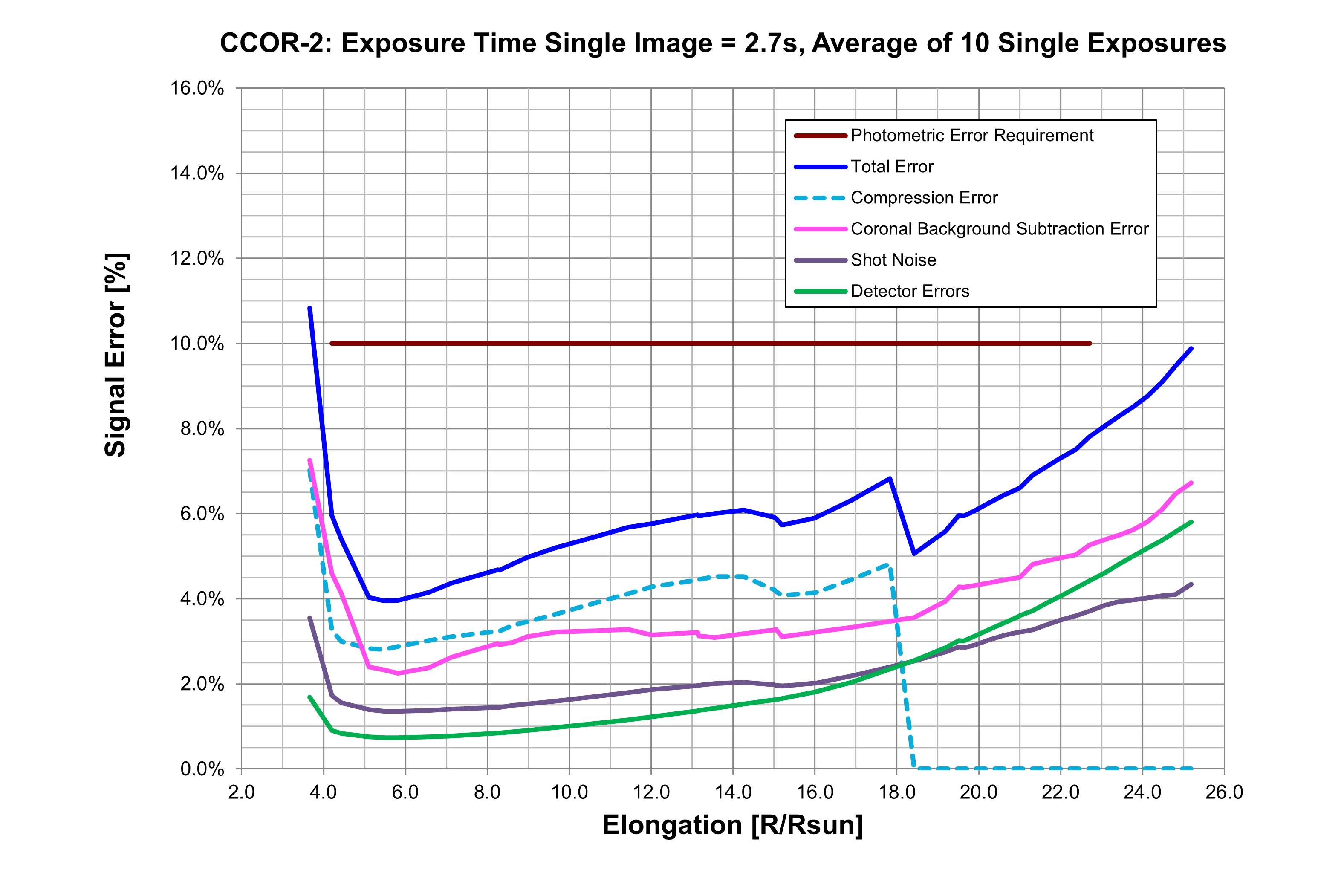}
\caption{Signal error break down for CCOR-2.}\label{fig_CCOR2SNRError}
\end{figure}

Instead of plotting the S/N, we use here the Signal Error, which is simply the inverse of the S/N expressed in percent. The plot of the Signal Errors are given in Figures \ref{fig_CCOR1SNRError} and \ref{fig_CCOR2SNRError}. The plots are broken down into 4 components: shot noise, detector errors, compression error, and background subtraction error. The root sum square of these components gives the total error, shown in blue in the plots. Again, the inverse of the total error is the S/N, and an S/N of 10 corresponds to a Signal Error of 10\%.

The shot noise is the photon statistics that follows a Poisson distribution. The detector errors include read noise, pixel-to-pixel gain variation, and bias variation. The compression error, described in Section \ref{sec_onboardprocessing}, is self-explanatory. Note that the compression error drops to 0 above 16 $R_\odot$ for CCOR-1 and 18 $R_\odot$ for CCOR-2 as a lossless compression is used in the outer FOV; see Figure \ref{fig_ccorFOVandBlocks}. The background subtraction error accounts for the fact that the error is twice the variance of the signal when an image running difference is computed, which is a commonly used technique to detect CMEs in a sequence of images.

\section{Operations, Data Processing, and Products}
When received on the ground, the image packets are decompressed and reassembled into a FITS file. The following levels will be produced in real time by SWPC. NCEI will reprocess the data and retrospectively distribute them to the scientific community.

Level 0: Level 0 data are created by the SWFO Command and Control (C2) and the GOES-19 Ground Segment (GS) and delivered to the Product Generation/Product Distribution (PG-PD) element as packets or frames, which consist of instrument science and housekeeping data and observatory housekeeping data. Frames are in the time-ordered Consultative Committee for Space Data Systems (CCSDS) format, with duplicates removed.

For CCOR-1/GOES-19, frames are broken down and packets containing only CCOR-1-related information are delivered to GeoCloud facilitated by Amazon Web Services (AWS) Console in the form of Network Common Data Form (NetCDF) files.

For CCOR-2/SWFO-L1, the frames from the antenna stations go to C2 and then are delivered to the AWS console in the form of NetCDF files for SWPC. Frames are also processed by SWPC (passed from C2) and delivered to NRL via SWPC sftp server in the form of NetCDF files and binary (.bin) files.

Level 0a: Level 0a is an intermediate data level between Level 0 data and combined images. CCSDS science packets and spacecraft telemetry packets, optionally compiled in NetCDF-format files, are ordered
and assembled into individual CCOR images. These images are decompressed where necessary. The flags and the number of missing and bad pixel blocks (of size 64 by 64 pixels for the full-resolution image) are created and recorded in FITS-format header. The output is in the form of FITS-format image files. Telemetry values required for higher-level processing are extracted from spacecraft telemetry packets and added to the FITS files. In normal observing mode, two Level 0a FITS files are produced, one containing only the outer FOV image and the other one containing only the inner FOV image.

\begin{figure}[h]
\centering
\includegraphics[width=1\textwidth]{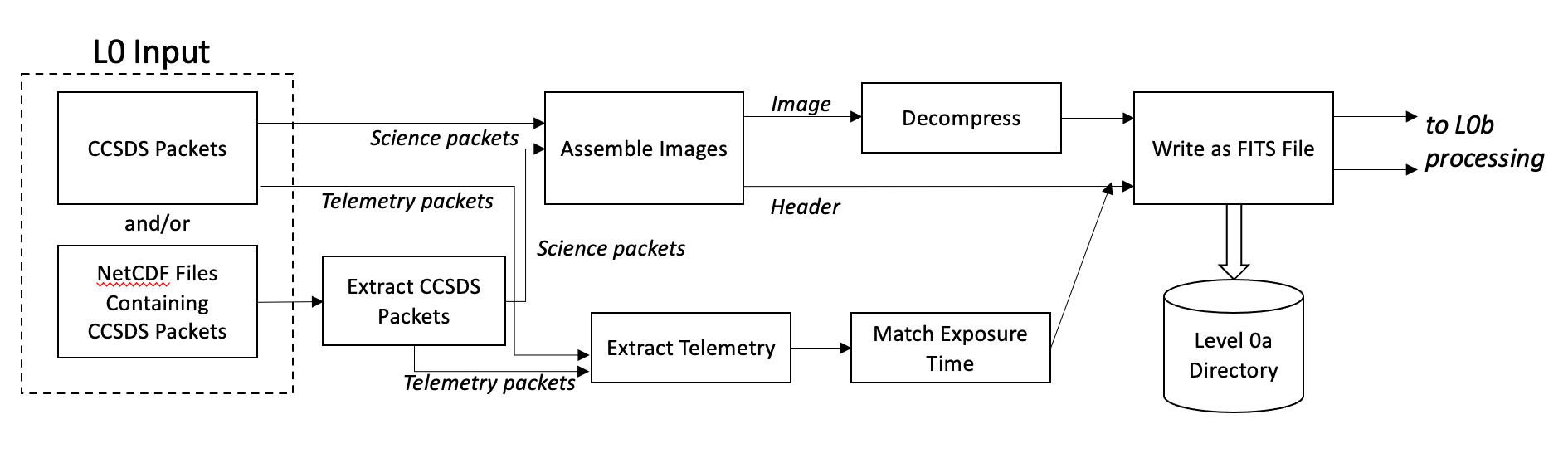}
\caption{Steps in Level 0a processing}\label{fig_l0aprocessing}
\end{figure}

Level 0b: Level 0b processing reconstructs individual FITS files for each full resolution image. The pipeline ingests Level 0a data, either in the form of processed FITS files or "save" data from Level 0a processing, and combines partial images as necessary. In normal observing mode, after the inner and outer FOV images are combined, the pixel values of the right- and left-side of the image are swapped, and the image is rotated by increments of 90 degrees to bring the solar North vector into the upper half of the image. Housekeeping and ancillary data required for higher-level processing (e.g., instrument mode, whether compression is applied, etc.) are included. The World Coordinate System (WCS) \citep{greisen_fits_2002A&A...395.1061G, calabretta__FITS_2002A&A...395.1077C} keywords and Solar Ephemeris Data keywords for the FITS header are populated. The FITS header uses the Helioprojective Cartesian Coordinates (HPC, default header) \citep{Thompson_2006A&A...449..791T} and Geocentric Equatorial Inertial (GEI, header A) to identify stars and small bodies.

Level 1a: Derived from the L0b image at full resolution, L1a images are column-to-column bias corrected (normally done onboard, but possibly reprocessed for retrospective data), optionally corrected for detector nonlinearity \citep{2021SoPh..296...94H}, divided by the exposure time, corrected for image binning, corrected for vignetting, and converted to Mean Solar Brightness (MSB).

Level 1b (CCOR-1 only): Derived from the L1a image, L1b images are planned to have the earthshine model removed. The earthshine model could have two components: the stray light produced by the earthshine rays hitting various components of the instrument and the earthshine itself which is superposed onto the background scene brightness (e.g., F-corona).

Level 2: Derived from L1b data for CCOR-1 and from L1a data for CCOR-2, L2 images are produced by subtracting the minimum background images. The algorithm described here is based on the algorithm used for the LASCO data. Minimum background images are produced by the following steps. First, a daily median image is created using all normal-mode images (of L1b for CCOR-1 and of L1a for CCOR-2) taken in one day and taking the median over time for each pixel in the image. Second, minimum background image is created using all median images over a certain number of days and taking minimum over time for each pixel in the image. The number of days for minimum background is different for real-time (SWPC) and retrospective (NCEI) processing. Currently, for the former, it is 7 counting from the day before the date processed to L2, and for the latter, it is $\pm$14 centered on the date that is processed to L2 (subject to change in the future).

Data and documentation will be available online on the NOAA website, links provided in the Appendix \ref{secA1}. The files are formatted in the (Flexible Image Transport System) FITS standard with (World Coordinate System) WCS compliant information, which will allow the use of the usual solar physics tools available in SolarSoft for IDL (Interactive Data Language) users, or SunPy for Python users. Calibration data and tools will be made available in SolarSoft (IDL) and possibly as a Python package.

\section{Summary and Conclusions}
CCOR-1 and CCOR-2 are the first operational solar coronagraphs, as opposed to science coronagraphs which are designed to address science questions. Among other sensors, they will be used by SWPC to detect and track CMEs for Space Weather forecast. They are designed to provide full-resolution images of the corona every 15 minutes and are robust enough to survive radiation and SEP storms to reduce the probability of missing images that could present geoeffective CMEs. Their sensitivity and field of view are similar to LASCO C3 and SECCHI COR-2, which have been used by SWPC so far.

The CCORs are also the first compact coronagraphs; all the solar occultation is done by a multi-disk external occulter. There is no second-stage internal occulter as in the traditional Lyot design. This allowed us to reduce the length of the instrument by half compared to SECCHI COR-2, and significantly save on mass as well, allowing only a minor degradation on the overall signal-to-noise ratio.

The ground testing and design analysis of the CCORs show that their performance meets the requirements defined by NOAA. The pylon design was more challenging than initially anticipated, resulting in additional stray light in that region, although there was a requirement exclusion zone in that area. We believe that we understand the issues and have tested solutions that will be implemented for CCOR-3.

At the time this article was written, CCOR-1 already collected its first light and was commissioned. The GOES-19 spacecraft replaced GOES-16 on April 2025 and became fully operational. CCOR-1 started delivering synoptic data from then on. CCOR-2 is planned to be launched in the fall of 2025. CCOR data are, or will be, available on NOAA's website (see Appendix \ref{secA1}).

Although CCOR-1 images will be impacted by Earth-induced stray light and transits for a couple of hours a day, CCOR-2 will be placed at L1, which will allow an uninterrupted view of the corona. The CCORs will deliver synoptic and regular image cadence with consistent point of view and resolution. Overall, the CCORs should be useful for the scientific community in continuing to monitor the corona the same way it was done with LASCO and SECCHI.

Finally, sponsored by NOAA and under the scientific and technical management of the NOAA NESDIS Office of Space Weather Observation (SWO), the CCOR team at NRL started working on CCOR-3, which will be on board the European Space Agency's (ESA) Vigil spacecraft, scheduled for launch in 2031, with operations starting in 2032. Vigil will be in the ecliptic at the Lagrange point L5, located $60\degree$ from the Sun-Earth line, on the trailing side. Proven by the NASA STEREO mission, in combination with the CCORs or CCOR-like coronagraphs located at L1 or near Earth, this additional vantage point will allow much improvement in the detection and accuracy of estimating the kinematics of solar wind transients that can affect the Space Weather at Earth or everywhere in the solar system.

\bmhead{Acknowledgments}
The authors thank Dr. Robin Colaninno for the initial review and helpful comments.

The CCOR team thanks the SWFO team, Jeff Kronenwether, and the GOES team.

\section*{Declarations}

Distribution Statement A. Approved for public release: distribution is unlimited.

\begin{itemize}
\item Funding: This work was funded by the National Oceanic and Atmospheric Administration.
\item Conflict of interest/Competing interests: not applicable
\item Ethics approval and consent to participate: not applicable
\item Consent for publication: approved by the National Oceanic and Atmospheric Administration.
\item Data availability: see appendices
\item Materials availability: see appendices
\item Code availability: see appendices
\end{itemize}


\begin{appendices}

\section{Data Distribution}\label{secA1}

CCOR data are open access following  \href{https://www.noaa.gov/information-technology/open-data-dissemination}{NOAA's Open Data Dissemination} here \url{https://www.noaa.gov/information-technology/open-data-dissemination}.

Real-time data can be seen on the \href{https://www.swpc.noaa.gov/}{SWPC} website: \url{https://www.swpc.noaa.gov/}.

Operational and retrospective science quality data can currently be accessed from here: \url{https://noaa-nesdis-swfo-ccor-1-pds.s3.amazonaws.com/index.html\#SWFO/GOES-19/CCOR-1/}.

NRL CCOR website \url{https://ccor.nrl.navy.mil/}.

\end{appendices}


\end{document}